\documentclass[tighten,iop,apj,numberedappendix,twocolappendix]{emulateapj}

\usepackage{amsmath}
\usepackage{txfonts}
\usepackage{graphicx}
\usepackage{url}
\usepackage{color}

\newcommand{\pd}{\partial}
\newcommand{\ud}{\ensuremath{\mathrm{d}}}

\begin{document}

\title{The Last Minutes of Oxygen Shell Burning in a Massive Star}

\author{Bernhard~M\"uller\altaffilmark{1,2,6},
Maxime~Viallet\altaffilmark{3},
  Alexander~Heger\altaffilmark{2,4,5,6},
  Hans-Thomas Janka\altaffilmark{3}}
\affil{
\\
\altaffilmark{1}{Astrophysics Research Centre, School
of Mathematics and Physics, Queen's University
Belfast, Belfast, BT7~1NN, United Kingdom;
  \href{mailto:b.mueller@qub.ac.uk}{b.mueller@qub.ac.uk}}
\\
\altaffilmark{2}{Monash Centre for Astrophysics, School of
  Physics and Astronomy, Monash University, Victoria
  3800, Australia}
\\
\altaffilmark{3}{Max-Planck-Institut f\"ur Astrophysik,
  Karl-Schwarzschild-Str. 1, 85748 Garching, Germany}
\\
\altaffilmark{4}{School of Physics \& Astronomy,
  University of Minnesota, Minneapolis, MN 55455, U.S.A.}
\\
\altaffilmark{5}{Center for Nuclear Astrophysics, Department of
  Physics and Astronomy, Shanghai Jiao-Tong University, Shanghai
  200240, P. R. China.}
\\
\altaffilmark{6}{Joint Institute for Nuclear Astrophysics, 1 Cyclotron
  Laboratory, National Superconducting Cyclotron Laboratory, Michigan
  State University, East Lansing, MI 48824-1321, U.S.A.}
}

\begin{abstract}
We present the first $4 \pi$-3D  simulation of the last minutes of oxygen shell
burning in an $18 M_\odot$ supernova progenitor up to the onset of
core collapse. A moving inner boundary is used to accurately model the
contraction of the silicon and iron core according to a 1D stellar
evolution model with a self-consistent treatment of core
deleptonization and nuclear quasi-equilibrium.  The simulation covers
the full solid angle to allow the emergence of large-scale convective
modes. Due to core contraction and the concomitant acceleration of
nuclear burning, the convective Mach number increases to
$\mathord{\sim}0.1$ at collapse, and an $\ell=2$ mode emerges shortly
before the end of the simulation. Aside from a growth of the oxygen
shell from $0.51 M_\odot$ to $0.56 M_\odot$ due to entrainment from
the carbon shell, the convective flow is reasonably well described by
mixing length theory, and the dominant scales are compatible with
estimates from linear stability analysis.  We deduce that artificial
changes in the physics, such as accelerated core contraction, can have
precarious consequences for the state of convection at collapse.  We
argue that scaling laws for the convective velocities and eddy sizes
furnish good estimates for the state of shell convection at collapse
and develop a simple analytic theory for the impact of convective seed
perturbations on shock revival in the ensuing supernova. We predict a
reduction of the critical luminosity for explosion by $12 \ldots 24\%$
due to seed asphericities for our 3D progenitor model relative to the
case without large seed perturbations.
\end{abstract}


\section{Introduction}
\label{sec:intro}
It is well known that core and shell burning in massive stars
typically drives convective overturn \citep{kippenhahn}.  Although
convective heat transport and mixing are inherently multi-dimensional
phenomena, the dynamical, convective, Kelvin-Helmholtz, and nuclear
time-scales are typically too disparate for modeling convection in
three dimensions (3D) during most phases of stellar evolution.
Spherically symmetric (1D) stellar evolution models therefore need to
rely on mixing-length theory (MLT; \citealp{biermann_32,boehm_58}) or some generalization
thereof \citep{kuhfuss_86,wuchterl_98,straka_08}. Such an effective
1D treatment of convection in stellar evolution is bound to remain
indispensable even with the advent of modern, implicit hydrodynamics
codes \citep{viallet_11,viallet_16,miczek_15} that permit multi-D simulations
over a wider range of flow regimes and time-scales.

The final stages of a massive star before its explosion as a supernova
(SN) are among the notable exceptions for an evolutionary phase where
the secular evolution time-scales are sufficiently short to remain
within reach of multi-D simulations (see, e.g.,
\citealp{mocak_08,mocak_09,stancliffe_11,herwig_14} for other examples
in the case of low-mass stars). There is also ample motivation for
investigating these final stages in 3D. Aside from the implications of
multi-D effects in convective shell burning for pulsar kicks
\citep{burrows_96,goldreich_97,lai_00,fryer_04,murphy_04} and their
possible connection to pre-SN outbursts \citep{smith_14}, they have
recently garnered interest as a means for facilitating shock revival
in the ensuing supernova \citep{couch_13,mueller_15a,couch_15}, which
has been the primary impetus for this paper.  While
the idea that progenitor asphericities arising from convective motions
with Mach numbers $\mathord{\sim} 0.1$ can aid shock revival by
boosting turbulent motions in the post-shock regions appears
plausible, it would be premature to claim that this new
idea is a decisive component for the success of the neutrino-driven
mechanism. Major questions
about this so far undervalued ingredient remain unanswered; and in this
paper we shall address some of them.

To evaluate the role of pre-SN seed perturbations in the explosion
mechanism, we obviously need multi-D simulations of shell burning up
to the onset of core collapse. As shown by the parametric study of
\citet{mueller_15a}, the typical Mach number and scale of the
convective eddies at this stage determine whether the seed
asphericities can effectively facilitate shock revival. None of the
available multi-D models can reliably provide that information yet.
While there is a large body of 2D and 3D simulations of earlier phases
of shell burning
\citep{arnett_94,bazan_94,bazan_98,asida_00,kuhlen_03,meakin_06,meakin_07,meakin_07_b,arnett_11,jones_16} ,
a first, exploratory attempt at extending a model of silicon shell
burning up to collapse has only been made recently by
\citet{couch_15}, albeit based on a number of problematic
approximations. \citet{couch_15} not only assumed octant symmetry,
which precludes the emergence of large-scale modes, but also
artificially accelerated the contraction of the iron core due to
deleptonization, which leads to a gross overestimation of the
convective velocities in the silicon shell as we shall demonstrate in
this paper.  Moreover, convective silicon burning often (though not
  invariably)  terminates minutes before collapse in stellar evolution
models (see, e.g., Figures~22 and 23 in \citealt{chieffi_13} and
Figure~16 in \citealt{sukhbold_14}), as it apparently also does in the
1D model of \citet{couch_15} calculated with the \textsc{MESA} code
\citep{paxton_11,paxton_13}. Obviously, simulations covering the full
solid angle ($4 \pi$) with a more physical treatment of the core
contraction are required as a next step.

Moreover, the efficiency of progenitor asphericities in triggering shock
revival in supernova simulations varies considerably
between different numerical studies.  The models of
\citet{couch_13,couch_14} are compatible with a small or moderate 
reduction of the critical luminosity \citep{burrows_93} for runaway
shock expansion. \citet{couch_15} observe shock revival in
their perturbed and non-perturbed model alike, i.e.\, the perturbations
are not crucial for shock revival at all in their study (which appears somewhat
at odds with their claims of a significant effect).  On the other
hand, a much stronger reduction of the critical luminosity of the
order of tens of percent has been inferred by \citet{mueller_15a} for
dipolar or quadrupolar perturbation patterns based on 2D models with
multi-group neutrino transport. These claims may not be in conflict
with each other, but could simply result from the different scale and
geometry of the pre-collapse velocity/density perturbations, the
different progenitor models, and the different treatment of neutrino
heating and cooling in these works. A more quantitative theory about
the impact of progenitor asphericities on shock revival that could
provide a unified interpretation of these disparate findings is still
lacking.

In this paper, we attempt to make progress on both fronts.  We present
the first full-$4\pi$ 3D simulation of the last minutes of oxygen
shell burning in an $18 M_\odot$ star.  The model is followed up to
collapse by appropriately contracting the outer boundary of the excised
(non-convective) core as in the corresponding 1D stellar evolution
model computed with the \textsc{Kepler} code \citep{weaver_78,heger_10}. By focusing on oxygen shell burning, we avoid the intricacies
of deleptonization in the iron core and the silicon shell and the
nuclear quasi-equilibrium during silicon burning, so that nucleon burning
can be treated with an inexpensive $\alpha$-network. Our simulation covers
the last $293.5 \, \mathrm{s}$  before collapse to keep
the last three minutes ($\mathord{\sim} 9$ turnover time-scales) free
of the artificial transients.

In our analysis of the simulation, we single out the properties of the
convective flow that are immediately relevant for understanding
pre-collapse asphericities in supernova progenitors and their role in
the explosion mechanism, while a more extensive analysis of the flow
properties based on a Reynolds decomposition (as in
\citealt{arnett_09,murphy_11,viallet_13,mocak_14}) is left to a future
paper. The key question that we set out to answer in this paper is
simply: \emph{Can we characterize the multi-dimensional structure of
  supernova progenitors (and perhaps their role in the explosion
  mechanism) already based on 1D stellar evolution models?} We shall
argue that this question can be answered in the affirmative, and
demonstrate that the typical velocity and scale of the convective
eddies comport with the predictions of mixing length theory (MLT) and
linear stability analysis.  In preparation for future core-collapse
simulations using multi-D progenitors, we develop a tentative theory
for the effects of pre-collapse seed perturbations on shock revival
that allows one to single out promising models for such
simulations. Aside from some remarks on convective boundary mixing,
  we largely skirt the much more challenging question whether
  deviations from MLT predictions have a long-term effect on the
  evolution of supernova progenitors during earlier phases. 

Our paper is structured as follows: In Section~\ref{sec:numerics}, we
describe the numerical methods used for our 3D simulation of oxygen
shell burning and briefly discuss the current version of the
\textsc{Kepler} stellar evolution code and the $18 M_\odot$ supernova
progenitor model that we consider. In Section~\ref{sec:results}, we
present the results of our 3D simulation, compare them to the 1D
stellar evolution model, and show that the key properties of the
convective flow are nicely captured by analytic scaling laws.  We
point out that these scaling laws impose a number of requirements on 3D
simulations of shell burning in Section~\ref{sec:requirements}.
In Section~\ref{sec:theory} we formulate a simple estimate for the effect of
the pre-collapse asphericities with a given typical convective
velocity and eddy scale on shock revival. The broader implications of
our findings and questions that need to be addressed by 3D stellar
evolution models of supernova progenitors are summarized in
Section~\ref{sec:summary}. Two appendices address different
formulations of the Ledoux criterion
(Appendix~\ref{app:ledoux}) and possible effects of
resolution and stochasticity (Appendix~\ref{app:res}).

\begin{figure}
  \includegraphics[width=\linewidth]{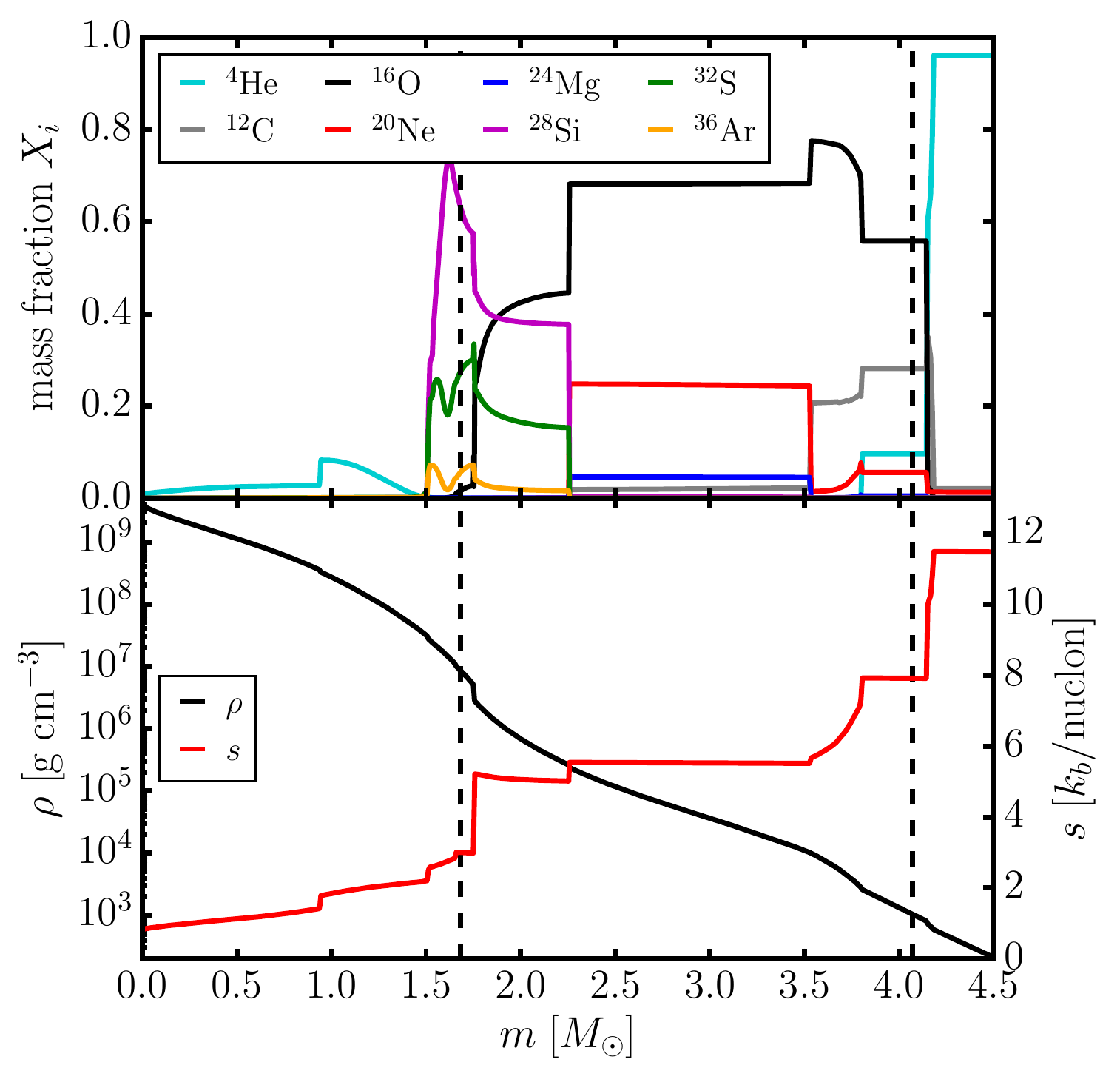}
  \caption{Top panel: Mass fractions $X_i$ of
relevant $\alpha$-elements in the 1D progenitor
model at the onset of collapse as a function
of enclosed mass $m$. Bottom panel:
Profiles of entropy $s$ and density $\rho$ as
a function of $m$.
Dashed vertical lines indicate
the boundaries of the region simulated in 3D.
    \label{fig:composition}}
\end{figure}

\section{Setup and Numerical Methods}
\label{sec:numerics}
\subsection{The \textsc{Kepler} Stellar Evolution Code}
We simulate oxygen shell burning in a non-rotating $18 M_\odot$
solar metallicity star. This stellar model has been evolved to the
onset of core collapse with an up-to-date version of the stellar
evolution code \textsc{Kepler} \citep{weaver_78,woosley_02,heger_10}.   A
19-species nuclear network \citep{weaver_78} is used at low
temperatures (up to oxygen burning); at higher temperatures, we switch
to a quasi-equilibrium (QSE) approach that provides an efficient and
accurate mean to treat silicon burning and the transition to a nuclear
statistical equilibrium (NSE) network after silicon depletion.

The mixing processes taken into account in this model include
convective mixing according to MLT, thermohaline mixing according to
\citet{heger_05}, and semiconvection according to \citet{langer_83},
but modified for a general equation of state as derived in \citet{heger_05}. All
mixing is modeled as a diffusive process with appropriately
determined diffusion coefficients. Since we will compare the
predictions of MLT and the results of our 3D simulation in some
detail, we elaborate further on the numerical implementation of MLT in
fully convective (Ledoux-unstable) regions in \textsc{Kepler},
 which has been outlined in a more compact form
in previous papers \citep{woosley_88,woosley_04}.  For
the implementation of semiconvection and thermohaline convection
(which are not immediately relevant for this paper), we refer the
reader to \citet{heger_00,heger_05}.

MLT assumes that the relative density contrast $\delta \rho/\rho$ between
convective updrafts/downdrafts and the spherically averaged
background state is related to the deviation of the
spherically averaged stratification from convective neutrality
and hence to the Brunt-V\"ais\"al\"a frequency $\omega_\mathrm{BV}$.  If
the Ledoux criterion for convection is used (as in \textsc{Kepler}),
one obtains
\begin{equation}
\label{eq:mlt1}
\frac{\delta \rho}{\rho}
=
\Lambda_\mathrm{mix}
\left(
\frac{1}{\rho} \frac{\pd  \rho}{\pd r}
-\frac{1}{\rho c_s^2}\frac{\pd P}{\pd r}\right)
=
\frac{\Lambda_\mathrm{mix}
\omega_\mathrm{BV}^2}{g},
\end{equation}
for $\delta \rho/\rho$, where both
entropy and composition gradients are implicitly taken into
account (see Appendix~\ref{app:ledoux}). Here, $\rho$, $P$, and $c_s$ denote the
spherically averaged density, pressure, and adiabatic sound speed, and $g$ denotes
the local gravitational acceleration.
$\omega_\mathrm{BV}$ is
the Brunt-V\"ais\"al\"a frequency\footnote{Note the sign
convention used in this paper:
$\omega_\mathrm{BV}^2>0$ corresponds to convective instability.},
and $\Lambda_\mathrm{mix}$ is the mixing length, which is chosen
as one pressure scale height $h_P$ so that we have
\begin{equation}
\Lambda_\mathrm{mix}=h_P
=P \left(\frac{\pd P}{\pd r}\right)^{-1}
=\frac{P}{\rho g},
\end{equation}
under the assumption of hydrostatic equilibrium. The convective velocity
in MLT can then be expressed in terms of
$\omega_\mathrm{BV}$, $\Lambda_\mathrm{mix}$, $g$, and
$\delta\rho/\rho$, and a dimensionless
parameter $\alpha_1$ as
\begin{eqnarray}
  \nonumber
\label{eq:vconv}
v_\mathrm{conv}
&=&
\alpha_1
\omega_\mathrm{BV} \Lambda_\mathrm{mix}
=\alpha_1
\left(\frac{\delta \rho}{\rho} \frac{g}{\Lambda_\mathrm{mix}}\right)^{1/2}
\Lambda_\mathrm{mix}
\\
&=&\alpha_1
\left(g \Lambda_\mathrm{mix}\frac{\delta \rho}{\rho} \right)^{1/2}.
\end{eqnarray}
Note that different normalizations and default values for $\alpha_1$
are used in the literature. Wherever a direct calibration
against observations (as for the solar convection zone, \citealp{christensen_96}) is
not possible, physical arguments can only constrain $\alpha_1$ to
within a factor of a few.

Together with the temperature contrast $\delta T$ between
the convective blobs and their surroundings, $v_\mathrm{conv}$
determines the convective energy flux $F_\mathrm{conv}$,
\begin{eqnarray}
  \label{eq:mlt2}
  \nonumber
  F_\mathrm{conv}
  &=&
  \alpha_2
  \rho c_P \, \delta T \, v_\mathrm{conv}
  =
  \alpha_1 \alpha_2
  \rho c_P \, \delta T\,
  \Lambda_\mathrm{mix} \omega_\mathrm{BV}
  \\
  \nonumber
  &=&
  -\alpha_1 \alpha_2
  \rho c_P \left(\frac{\pd T}{\pd \ln \rho}\right)_P
  \frac{\delta \rho}{\rho}
  \Lambda_\mathrm{mix} \omega_\mathrm{BV}
  \\
  &=&
  -\alpha_1 \alpha_2
  \rho c_P \left(\frac{\pd T}{ \pd \ln \rho}\right)_P
  \frac{\Lambda_\mathrm{mix}^2 \omega_\mathrm{BV}^3}{g}
  ,
\end{eqnarray}
where $c_P$ is the specific heat at constant pressure, and $\alpha_2$
is another dimensionless parameter. Note that the second and third
  line in Equation~(\ref{eq:mlt2}) implicitly assume that the
  contribution of composition gradients to the unstable gradient can
  be neglected inside a convective zone, which is a good approximation
  for advanced burning stages.

For compositional
mixing, \textsc{Kepler} uses a time-dependent diffusion model
\citep{eggleton_72,weaver_78,heger_00,heger_05} for the evolution
of the mass fractions $X_i$,
\begin{equation}
\left(\frac{\pd \rho X_i}{\pd t}\right)_\mathrm{mix}
=
\frac{1}{r^2}\frac{\pd r^2 F_{X_i}}{\pd r}
=\frac{1}{r^2} \frac{\pd }{\pd r} \left(r^2 \rho D \frac{\pd X_i}{\pd r} \right),
\end{equation}
where $F_{X_i}=\rho D \pd X_i/\pd r$ is the diffusive partial
mass flux for species $i$, and the diffusion coefficient is given by
\begin{equation}
\label{eq:mlt_diff}
D
=
\alpha_3 \Lambda_\mathrm{mix} v_\mathrm{conv}
=
\alpha_1 \alpha_3
\omega_\mathrm{BV} \Lambda_\mathrm{mix}^2,
\end{equation}
where we have introduced another dimensionless parameter
$\alpha_3$. If we introduce the composition contrast
$\delta X_i = \Lambda_\mathrm{mix} \pd X_i/\pd r$
 between the bubbles and the background state, the symmetry
to Equation~(\ref{eq:mlt2}) for the convective energy flux becomes manifest:
\begin{equation}
\label{eq:mlt3}
F_{X_i}
=
\alpha_1 \alpha_3\,
\rho \,\delta X_i\,
 \Lambda_\mathrm{mix} \omega_\mathrm{BV}.
\end{equation}

We note that only the products $\alpha_1 \alpha_2$ and $\alpha_1 \alpha_3$
enter the evolution equations, and we are therefore free
to reshuffle an arbitrary factor between $\alpha_1$ and the other
two coefficients. In \textsc{Kepler}, we choose
$\alpha_1 \alpha_2=1/2$ and $\alpha_1 \alpha_3=1/6$,
which is traditionally interpreted as the result of
$\alpha_1=1/2$, $\alpha_2=1$ and $\alpha_3=1/3$, where
the choice of $\alpha_3=1/3$ is motivated by the interpretation
of convective mixing as a random-walk process
in 3D with mean free path $\Lambda_\mathrm{mix}$ and an
average \emph{total} velocity (including the non-radial
velocity components) $v_\mathrm{conv}$.
Setting $\alpha_3=\alpha_2/3$ arguably introduces an asymmetry
in the equations, but we defer the discussion of its effect to
Section~\ref{sec:mixing}. For extracting convective velocities
from the \textsc{Kepler} model, we shall work with the
alternative choice of
$\alpha_1=1, \alpha_2=1/2, \alpha_3=1/6$, however,
as this gives better agreement with the convective
velocity field in our 3D simulation. This is equally
justifiable; essentially this choice amounts
to a larger correlation length for velocity perturbations
and less perfect correlations between fluctuations in velocity
and entropy/composition.

For numerical reasons, $\omega_\mathrm{BV}$ is rescaled
before computing the convective energy and partial
mass fluxes according to Equations~(\ref{eq:mlt2}) and (\ref{eq:mlt3}),
\begin{equation}
\label{eq:rescaling}
\omega_\mathrm{BV} \rightarrow
\omega_\mathrm{BV}
e^{-f /(3 \delta\rho/\rho)},
\end{equation}
where $f$ is an adjustable parameter that is set to $f=0.01$ in our
model. By rescaling $\omega_\mathrm{BV}$ convective mixing and energy
transport are suppressed until a reasonably large superadiabatic
gradient has been established. This procedure avoids convergence
problems due to zones switching too frequently between convective
stability and instability.  The repercussions and limitations of this
numerical approach will be discussed in Section~\ref{sec:comparison},
where we compare the 1D stellar evolution model to our 3D hydrodynamic
simulation.

\subsection{1D Supernova Progenitor Model}
Entropy, density, and composition profiles of the 1D progenitor model
at the onset of collapse are shown in Figure~\ref{fig:composition}.
The progenitor has an extended convective oxygen shell of about $0.5
M_\odot$ with a broader convective carbon burning shell directly on
top of it. The inner and outer boundaries of the oxygen shell are
located at $3000 \, \mathrm{km}$ and $8000 \, \mathrm{km}$ at the
beginning of our 3D simulation and contract considerably until
collapse sets in. The entropy jump between the silicon and oxygen
shell is relatively pronounced, so that no strong overshooting and/or
entrainment at the inner convective boundary is expected because of
the strong buoyancy  barrier at the interface. The boundary between the
oxygen and carbon shell is considerably ``softer'' with only a small
jump of $0.5 \, \mathrm{k}_b/\mathrm{nucleon}$ in entropy.

We note that the balance between energy generation by nuclear burning
and neutrino cooling is broken during the final phase before collapse that
we are considering here. This is due to the acceleration of shell
burning induced by the contraction of the core on a time-scale too
short for thermal adjustment by neutrino cooling. Different from
earlier phases, it is therefore sufficient to follow shell
convection in multi-D merely for several overturn time-scales to reach
the correct quasi-steady state (instead of several Kelvin-Helmholtz
time-scales for earlier phases to ensure thermal adjustment).

\subsection{3D Simulation}
At a time of $293.5 \, \mathrm{s}$ before the onset of collapse, the
stellar evolution model is mapped to the finite-volume hydrodynamics
code \textsc{Prometheus} \citep{fryxell_89}, which is an
implementation of the piecewise parabolic method of
\citet{colella_84}. An axis-free overset ``Yin-Yang'' grid
\citep{kageyama_04,wongwathanarat_10a}, impelemented
  as in \citet{melson_15a} using MPI domain decomposition 
  and an algorithm for conservative advection of
scalars \citep{wongwathanarat_10a,melson_msc} , allows us to retain the advantages
of spherical polar coordinates, which are best suited to the problem
geometry, while avoiding excessive time-step constraints close to the
grid axis. As in \textsc{Kepler}, nuclear burning is treated using a
19-species $\alpha$-network.  The simulations are performed in the so-called implicit large eddy simulations (ILES) paradigm \citep{boris_new_1992,ILES_grinstein}, in which diffusive processes (viscosity, mass diffusion, thermal diffusion) are not explicitly included in the equations. Instead, one relies on the truncation errors of the underlying numerical scheme to mimic the effects of irreversible processes taking place at unresolved scales (truncation errors act as an ``implicit'' sub-grid scale model).

Since there is no convective activity in the Fe core and the Si shell
in the last stages before collapse in the \textsc{Kepler} model, we
excise the innermost $1.68 M_\odot$ of the core and contract
the inner boundary of the computational domain according to the
trajectory of this mass shell in the \textsc{Kepler} run from an
initial radius of $3000 \, \mathrm{km}$ to $1974 \, \mathrm{km}$ at the
onset of collapse. At both the inner and outer boundary, we impose
  reflecting boundary conditions for the radial velocity, and
  use constant extrapolation for the non-radial velocity components.
  The density, pressure and internal energy are extrapolated into
  the ghost zones
assuming hydrostatic equilibrium and constant entropy.
Excising the core  not only reduces the computer time
requirements considerably, but also allows us to circumvent the
complications of deleptonization and Si burning in the QSE regime. The
outer boundary is set to a mass coordinate of $4.07 M_\odot$
(corresponding to a radius of $50,000 \, \mathrm{km}$) so that the
computational domain comprises the outer $0.08 M_\odot$ of the Si shell,
the entire O and C shell, and a small part of the incompletely burnt He shell.
On the other hand, using an inner boundary condition implies that
  we cannot address potential effects of shell convection
  on the core via wave excitation at the convective boundaries,
  such as the excitation of
  unstable g-mode
  \citep{goldreich_97} (whose growth is likely too slow
  to be significant; see \citealt{murphy_04}) or core spin-up
  due to angular momentum transport by internal gravity waves
  \citep{fuller_15}. 

We use a logarithmic radial grid with 400 zones, which implies a
radial resolution of $\Delta r/r=0.7 \%$ at the beginning of the
simulation. Equidistant spacing in $\log r$ is maintained throughout
the simulation as the inner boundary is contracted. $56 \times 148$
angular zones are used on each patch of the Yin-Yang grid, which
corresponds to an angular resolution of $2^\circ$.
A limited resolution study  based on two additional
models with coarser meshes is presnted in Appendix~\ref{app:res}.

\begin{figure*}
  \plottwo{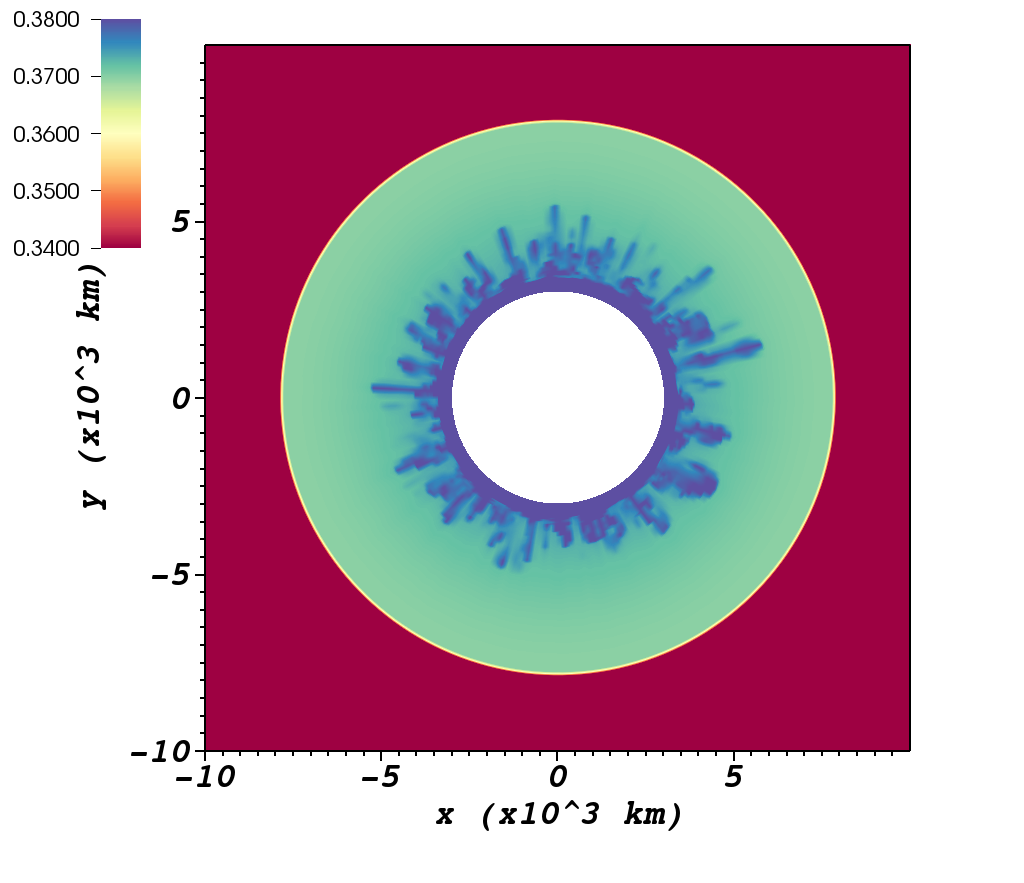}{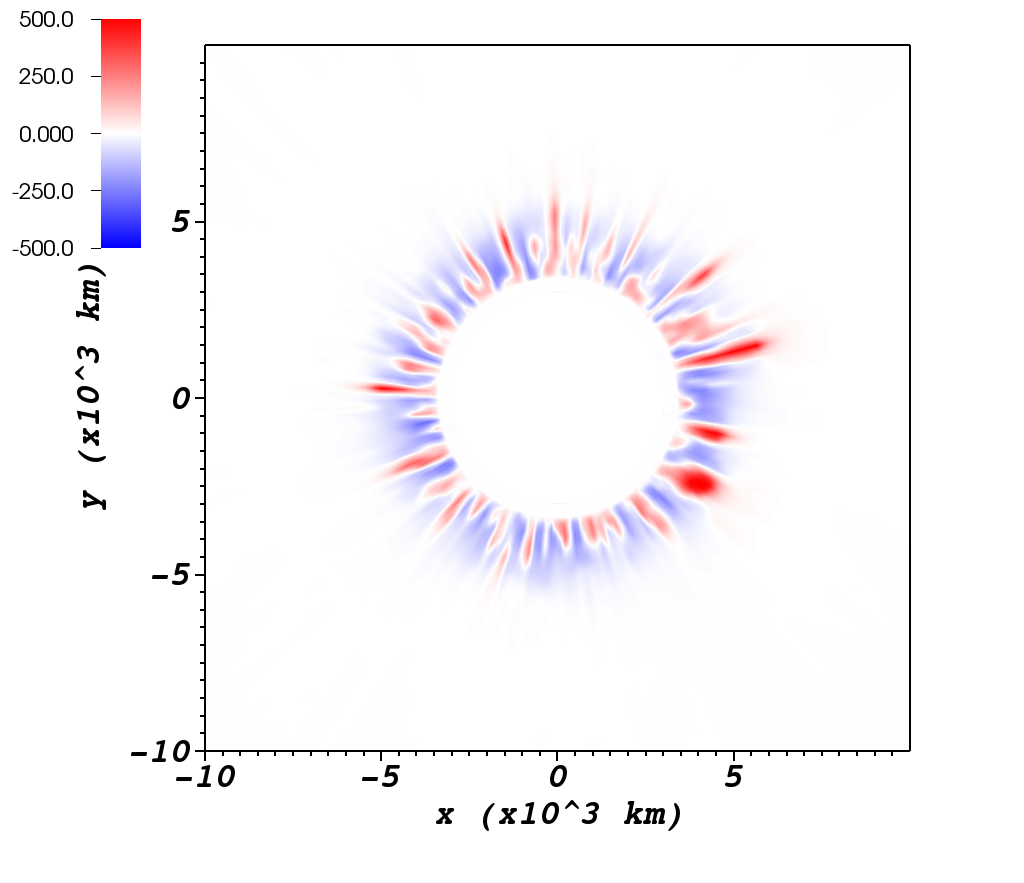}
  \plottwo{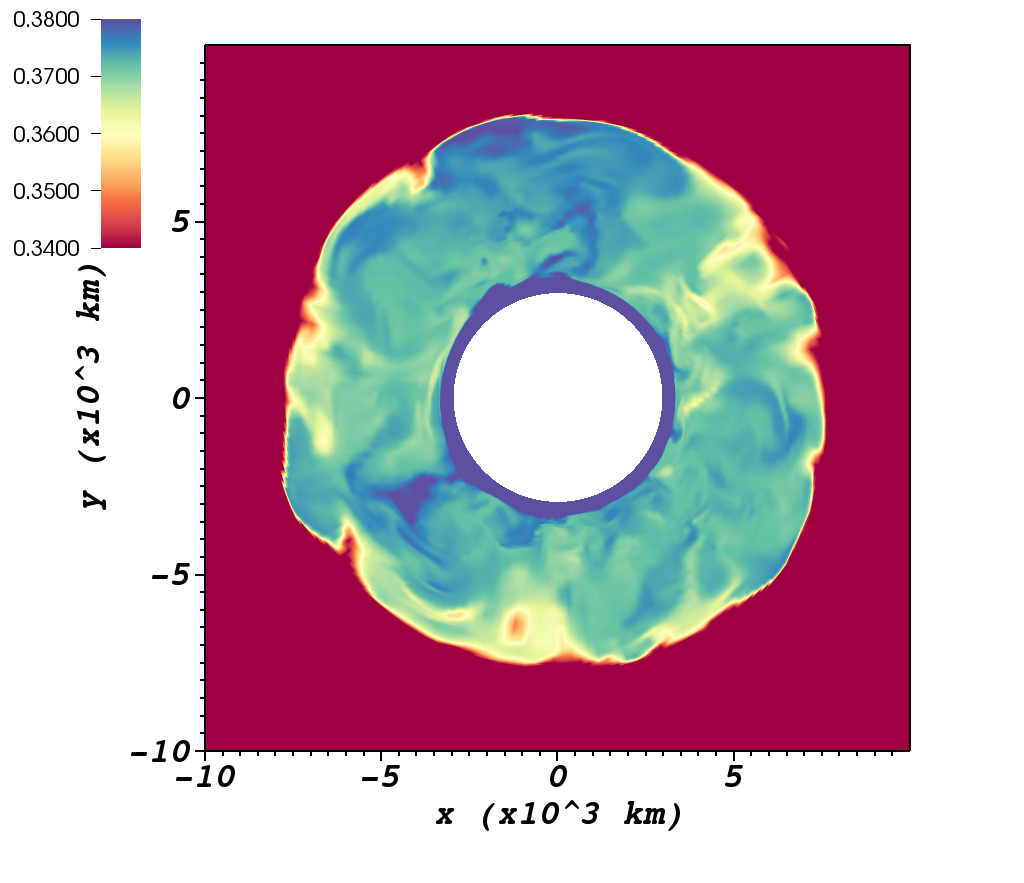}{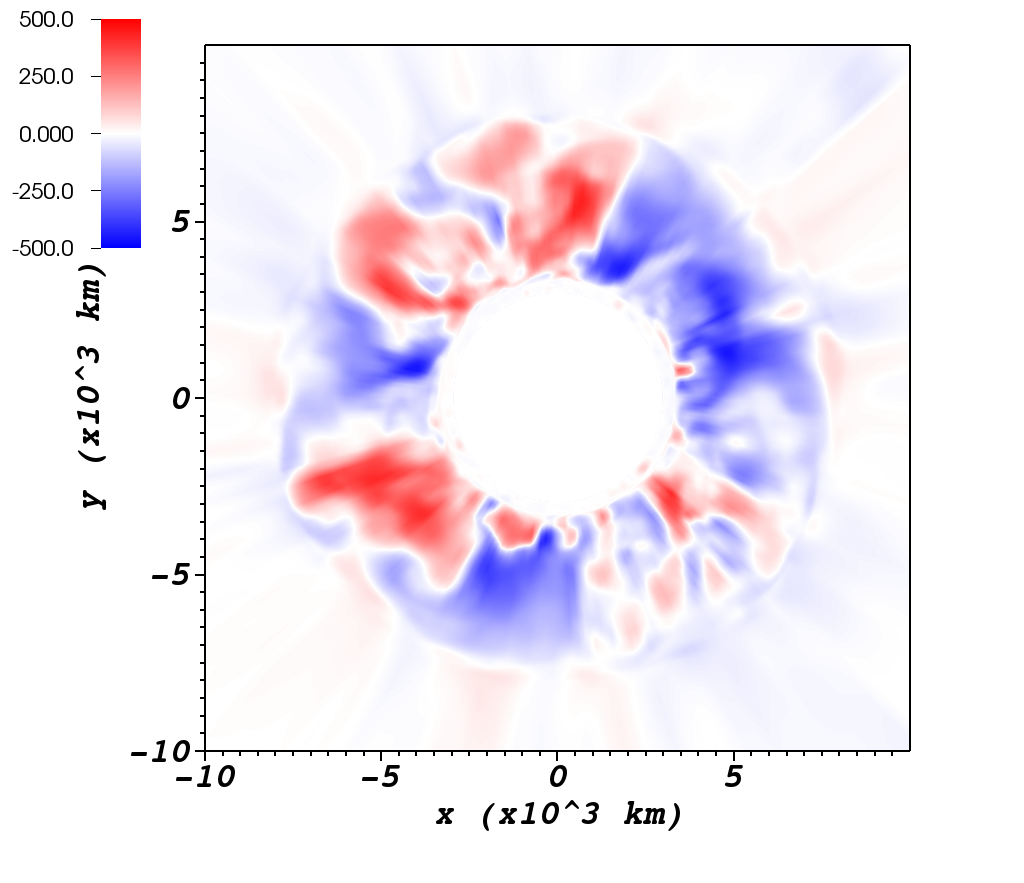}
  \plottwo{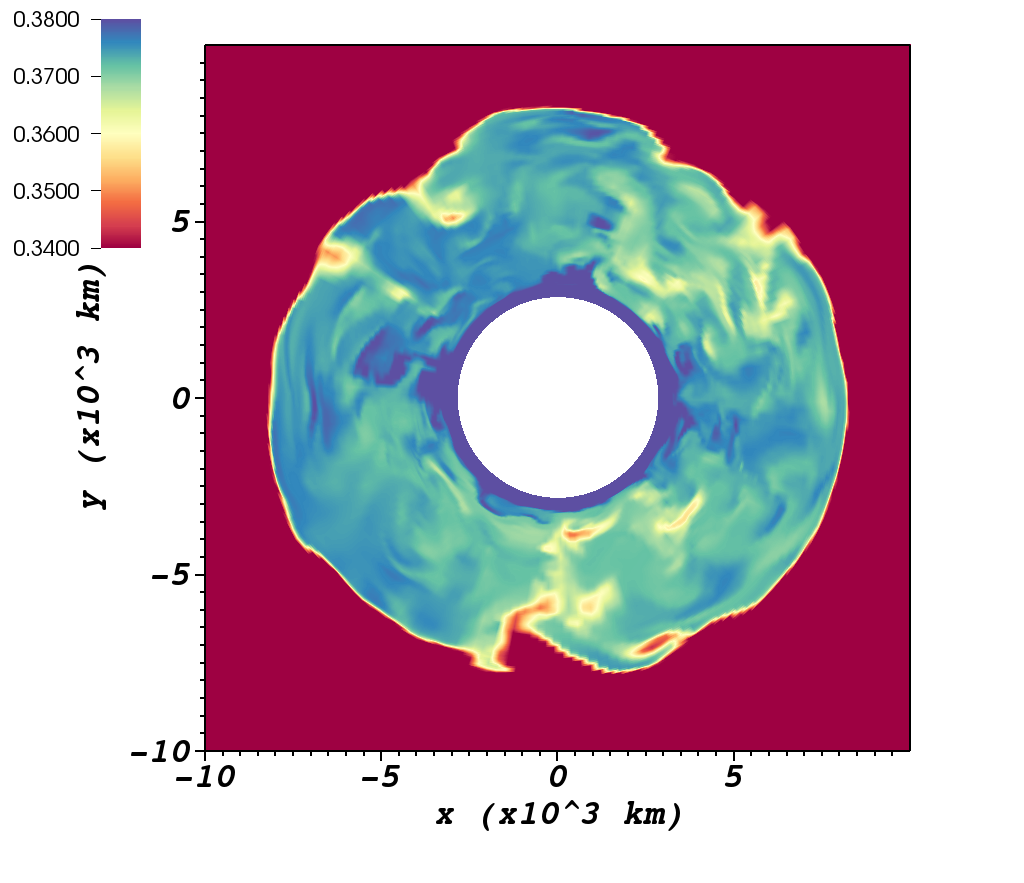}{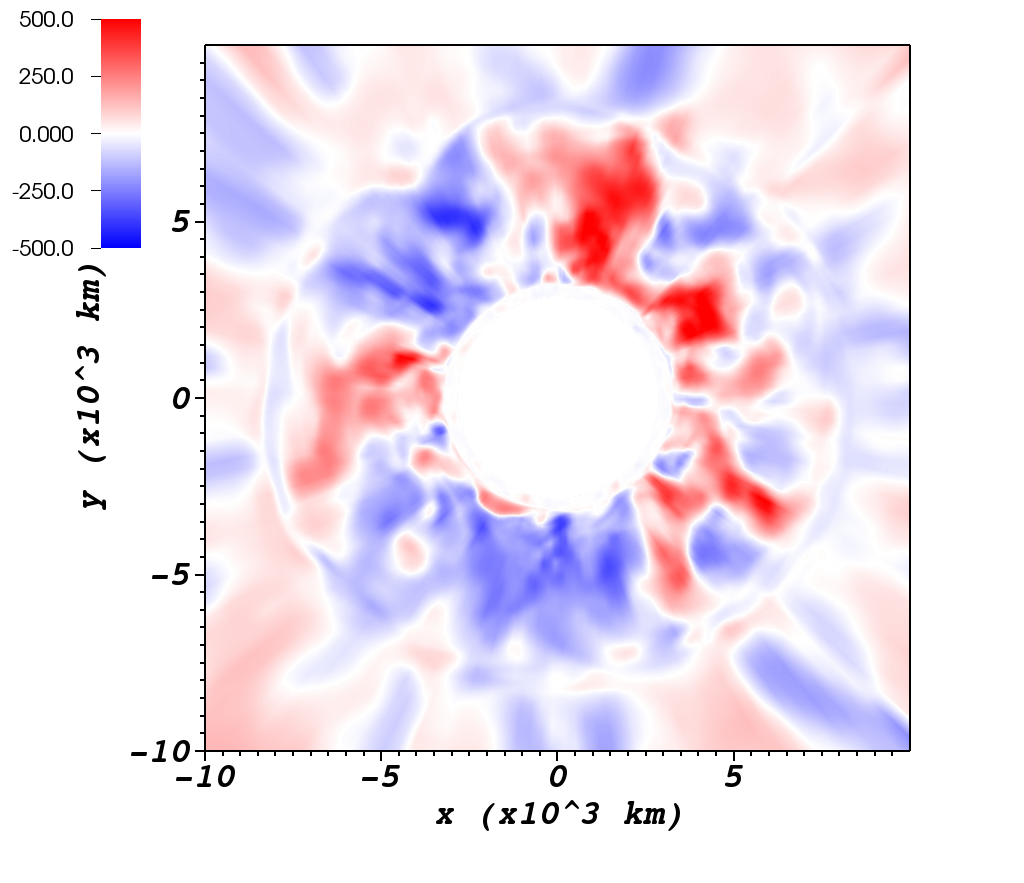}
  \caption{Slices showing the mass fraction
$X_\mathrm{Si}$ of silicon (left column) and the radial velocity $v_r$
(right column)
at times of
$20 \, \mathrm{s}$,
$151 \, \mathrm{s}$, and
$210 \, \mathrm{s}$ after the beginning of the 3D simulation
(top to bottom). $v_r$ is given in units
of $\mathrm{km} \, \mathrm{s}^{-1}$.
The boundary between the patches of the Yin-Yang grid is
  located in the right half of the panels at $45^\circ$ and $135^\circ$ from the vertical direction. 
Note that convection initially develops on small angular scales
after mapping from the 1D stellar evolution as a strongly
superadiabatic gradient builds up in the narrow region of strongest
nuclear burning at the base of the oxygen shell (top row). Once convection
is fully developed, large-scale overturn emerges.
The position of the updrafts and downdrafts shifts freely
  across the boundaries between the grid patches. 
\label{fig:snap2d_a}}
\end{figure*}

\begin{figure*}
  \plottwo{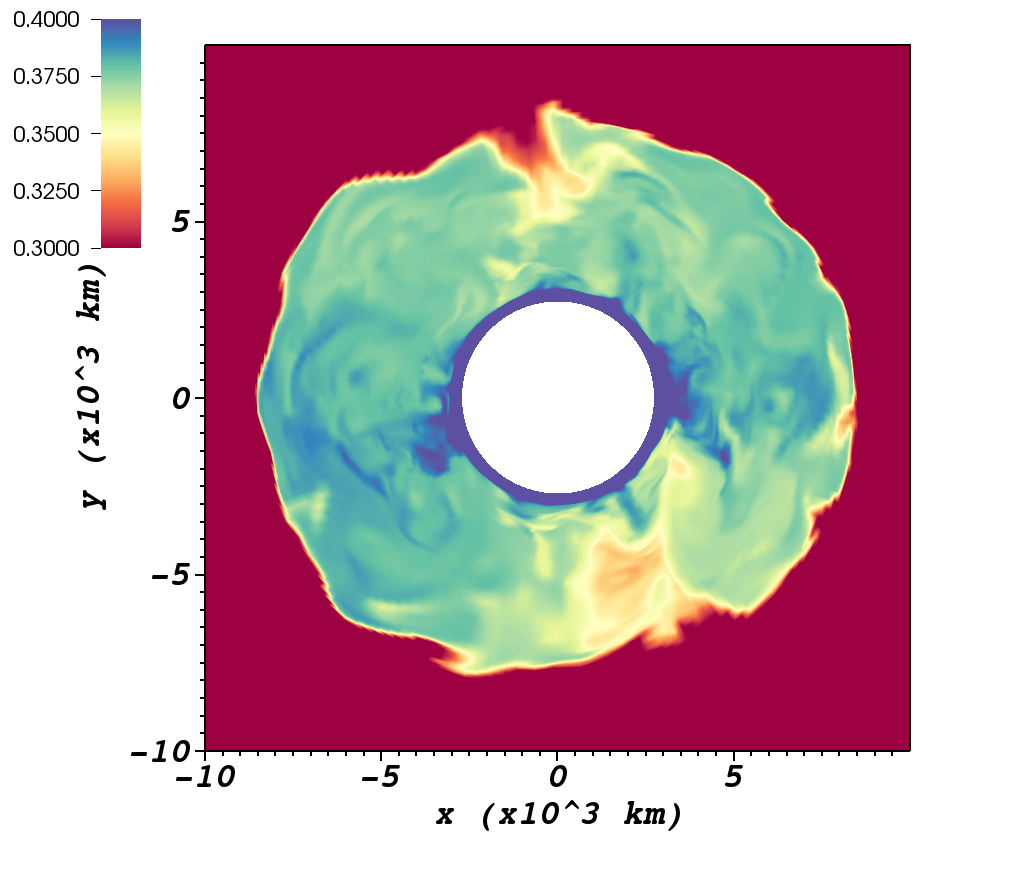}{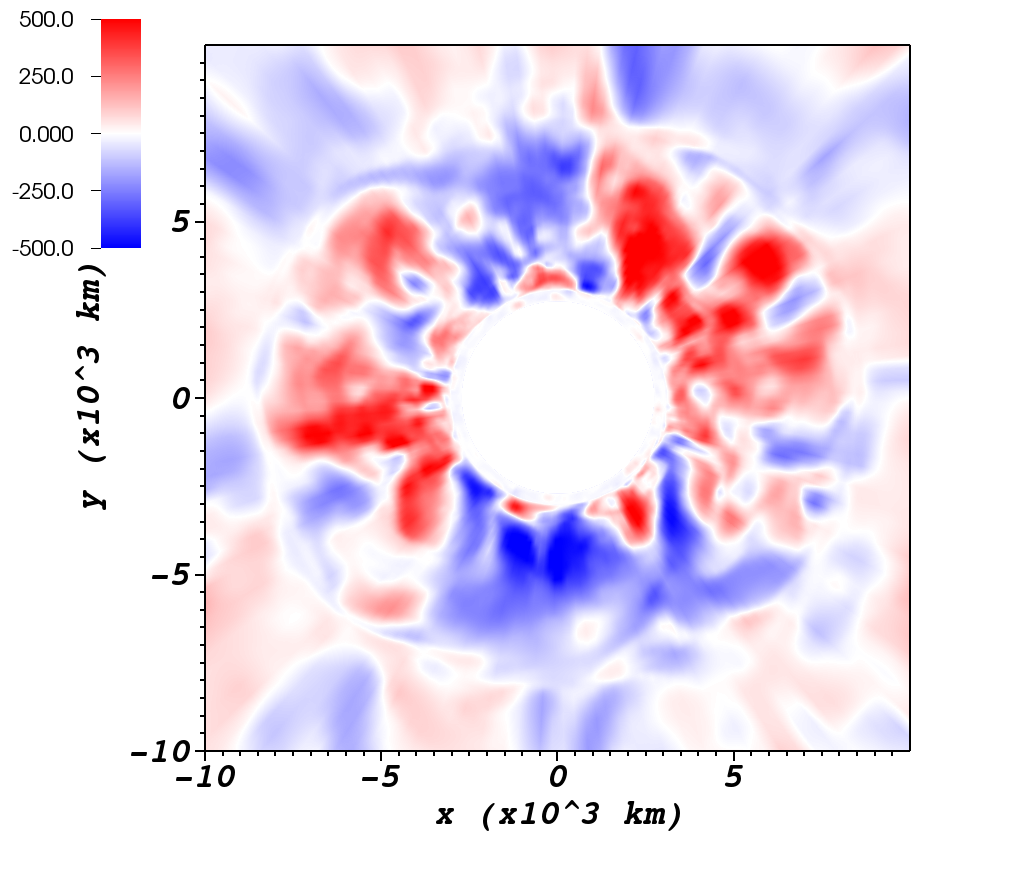}
  \plottwo{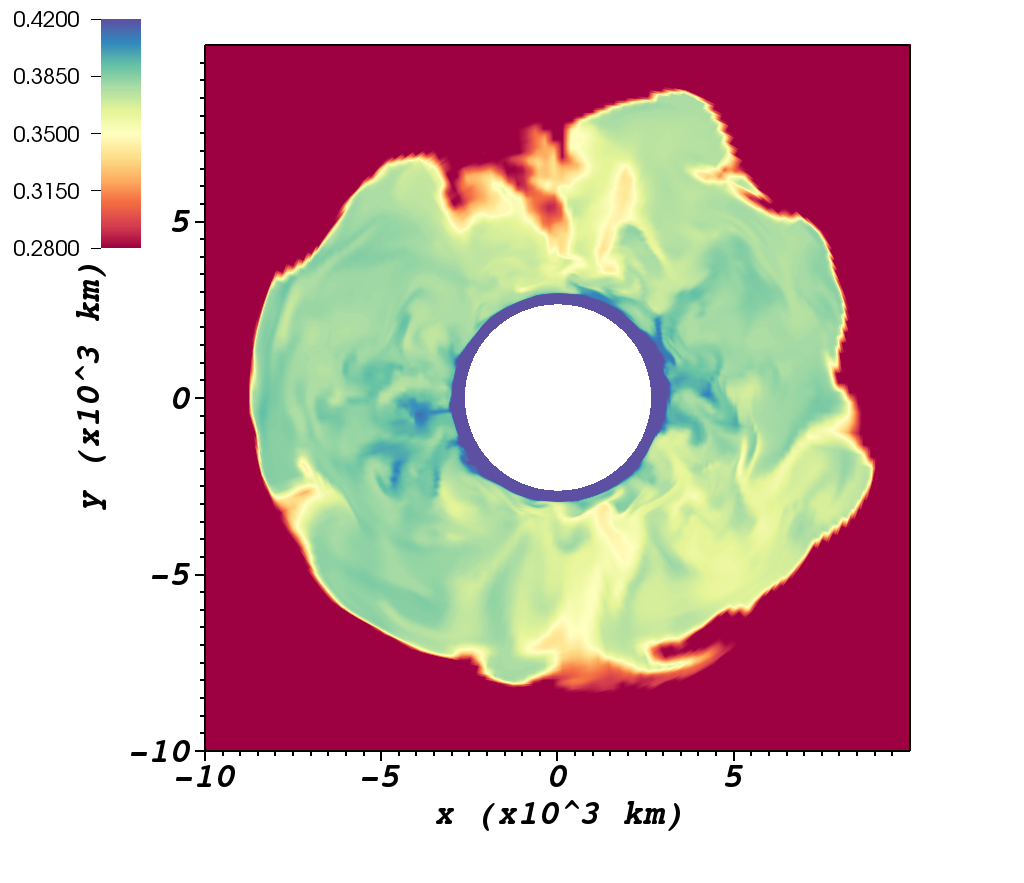}{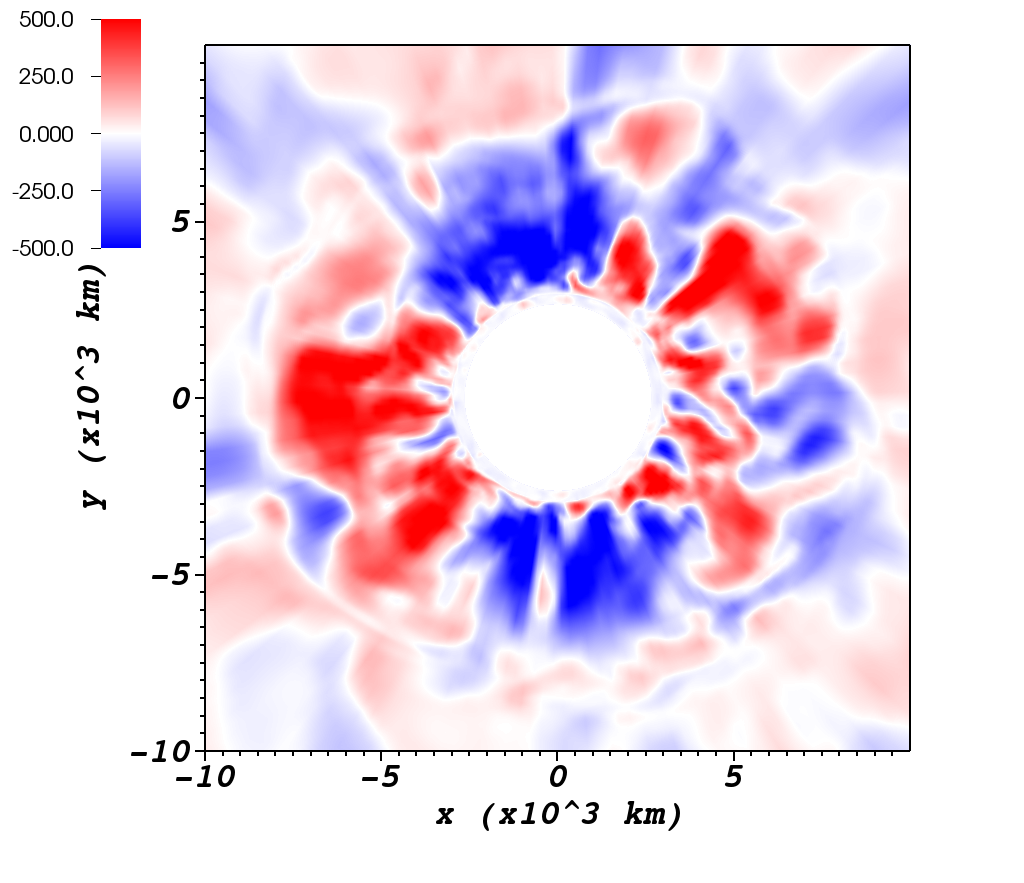}
  \plottwo{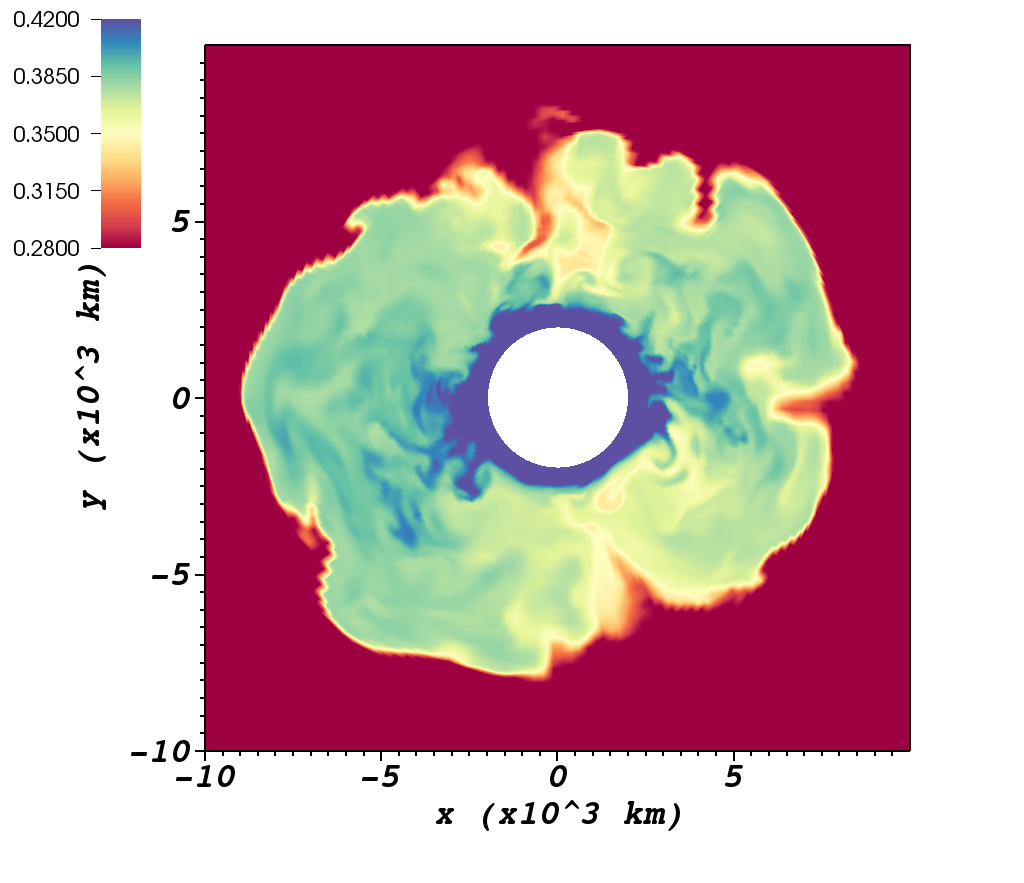}{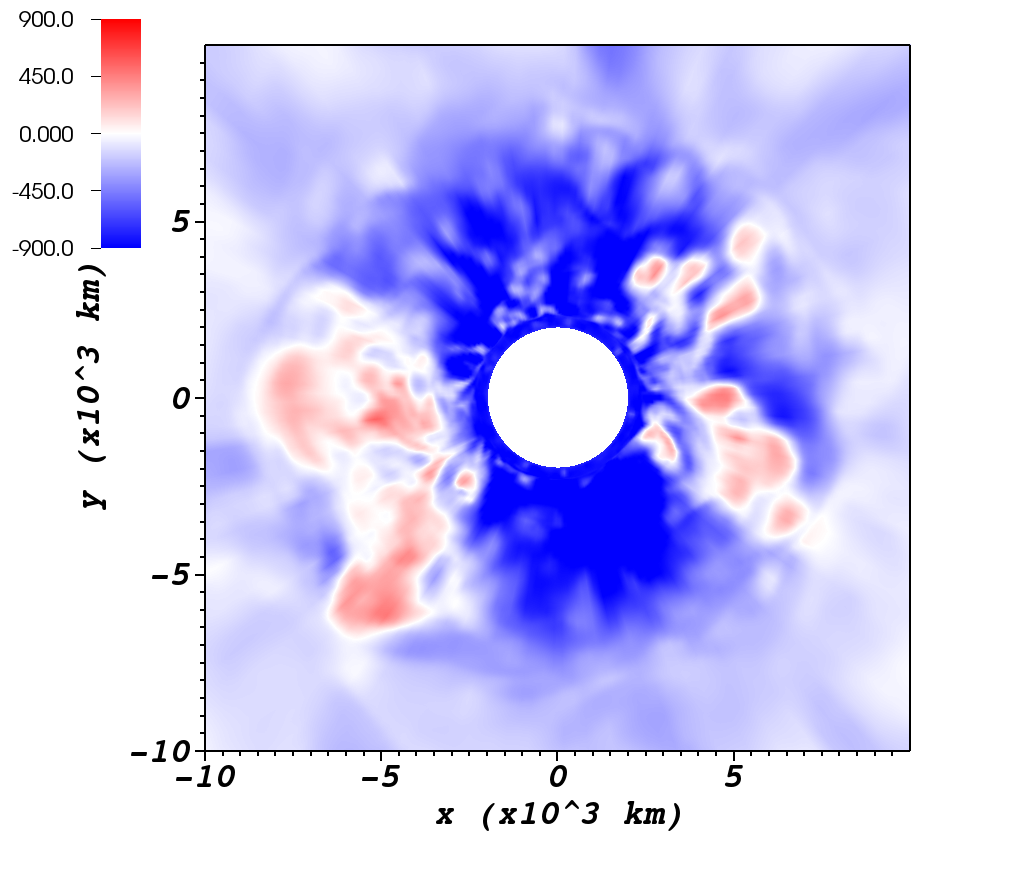}
  \caption{Slices showing the mass fraction $X_\mathrm{Si}$ of silicon
    (left column) and the radial velocity $v_r$ (right column) at
    times of $270 \, \mathrm{s}$, $286 \, \mathrm{s}$, and $293.5
    \, \mathrm{s}$ (onset of collapse) after the beginning of the 3D
    simulation (top to bottom). $v_r$ is given in units of
    $\mathrm{km} \, \mathrm{s}^{-1}$. Note that wave breaking at the
    outer boundary of the oxygen shell and the global asymmetry of
    convective motions become more conspicuous at late times. At the
    onset of collapse, a bipolar flow pattern emerges (bottom row).
\label{fig:snap2d_b}}
\end{figure*}

\begin{figure*}
  \plotone{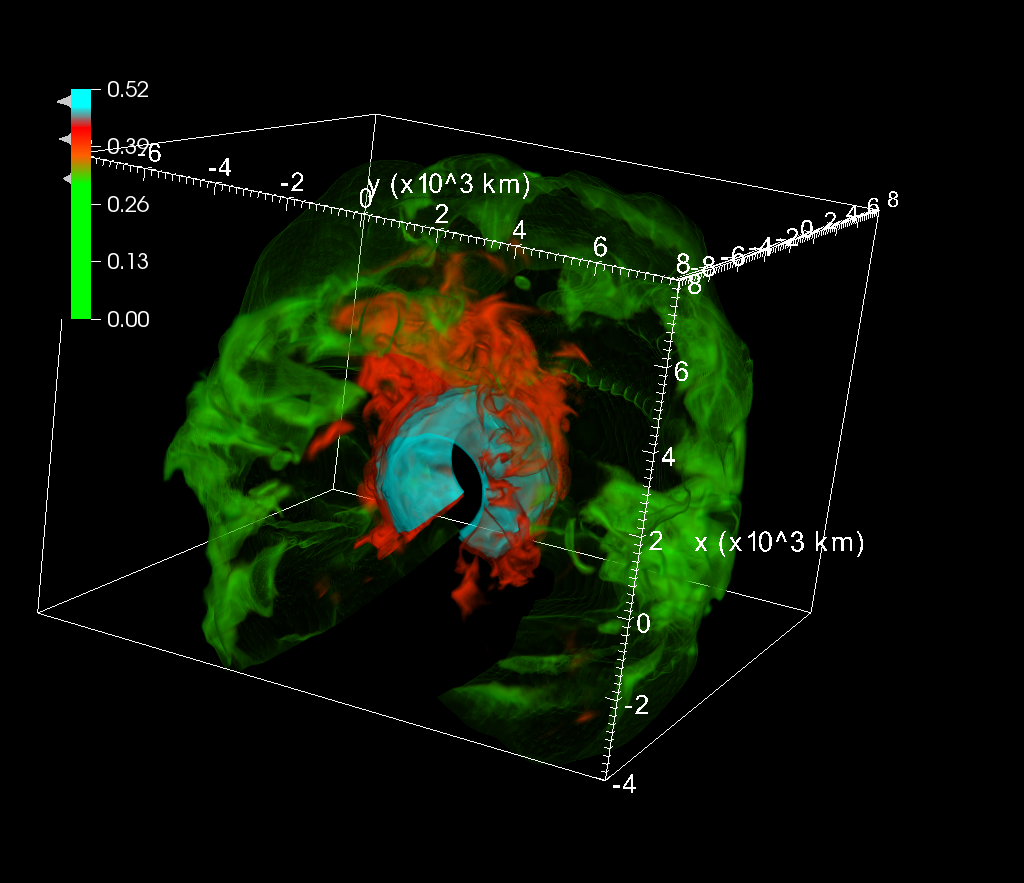}
  \caption{Volume rendering of the mass fraction of silicon at the end
    of the 3D simulation at $293.5 \, \mathrm{s}$ (onset of collapse)
    on one patch of the Yin-Yang grid, showing fuzzy silicon-rich
    updrafts of hot ashes (red) and silicon-poor downdrafts of
    fresh fuel. A global asymmetry in the updrafts is clearly
    visible. The inner boundary of the oxygen shell (cyan) is
    relatively ``hard'' due to the strong buoyancy jump between the
    silicon and oxygen shell and therefore remains almost spherical.
  \label{fig:vol3d}}
\end{figure*}

\begin{figure}
  \includegraphics[width=\linewidth]{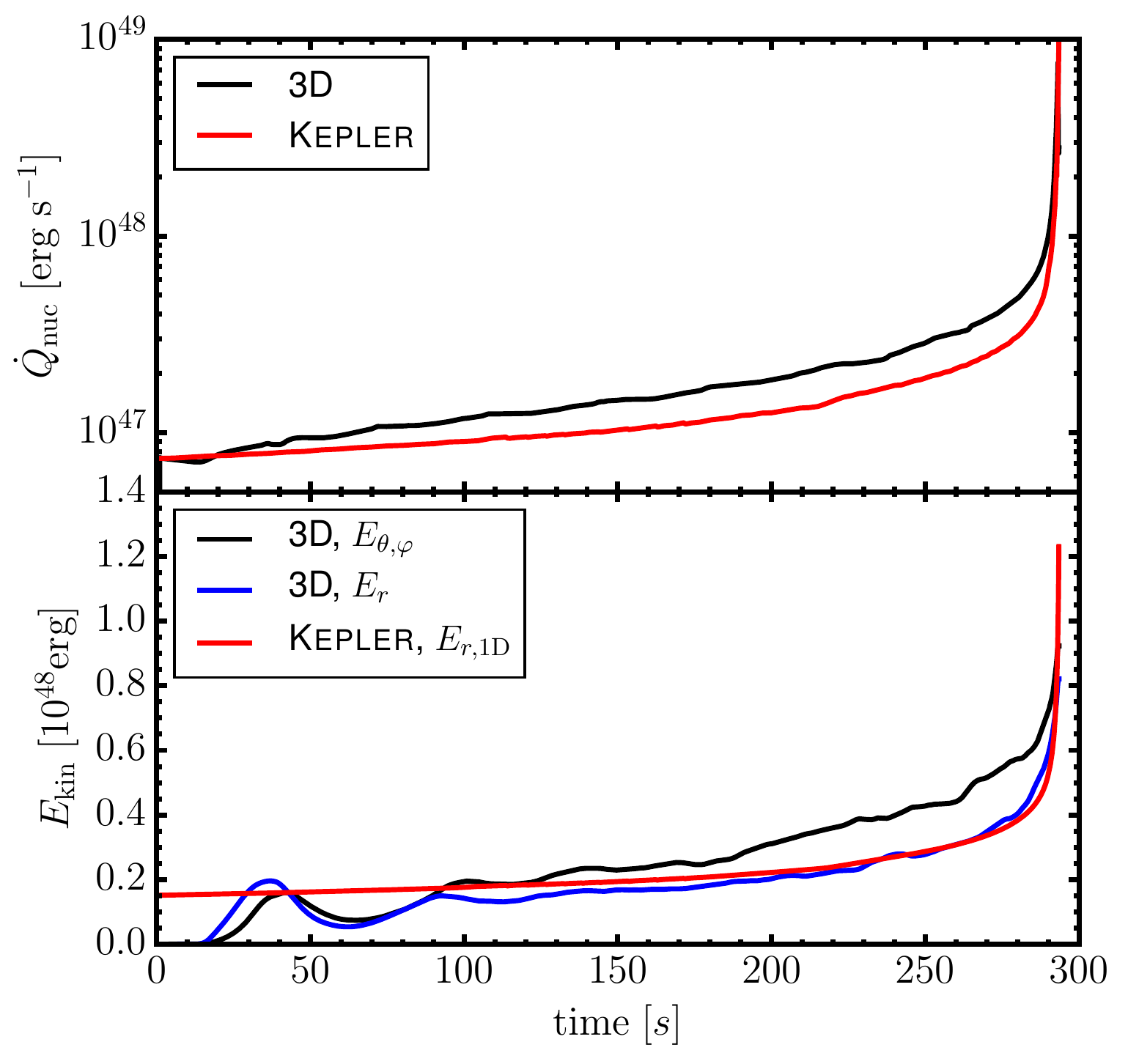}
  \caption{Top: Volume-integrated net nuclear energy
generation rate $\dot{Q}_\mathrm{nuc}$ in the oxygen shell
in the 3D simulation (black) and in \textsc{Kepler} (red).
Bottom: Kinetic energies $E_{\theta,\varphi}$ (black)
and $E_r$ (blue) contained in fluctuating non-radial and
radial motions in the 3D simulation; see
Equations~(\ref{eq:ekinr},\ref{eq:ekinl}) for definitions.
The MLT estimate of the volume-integrated kinetic energy $E_{r,\mathrm{1D}}$ in radial convective
motions in the oxygen shell for the \textsc{Kepler} model (red) is computed
by using Equation~(\ref{eq:vconv})
for the convective velocity assuming $\alpha_1=1$.
    \label{fig:qnuc_and_ekin}}
\end{figure}

\begin{figure}
  \includegraphics[width=\linewidth]{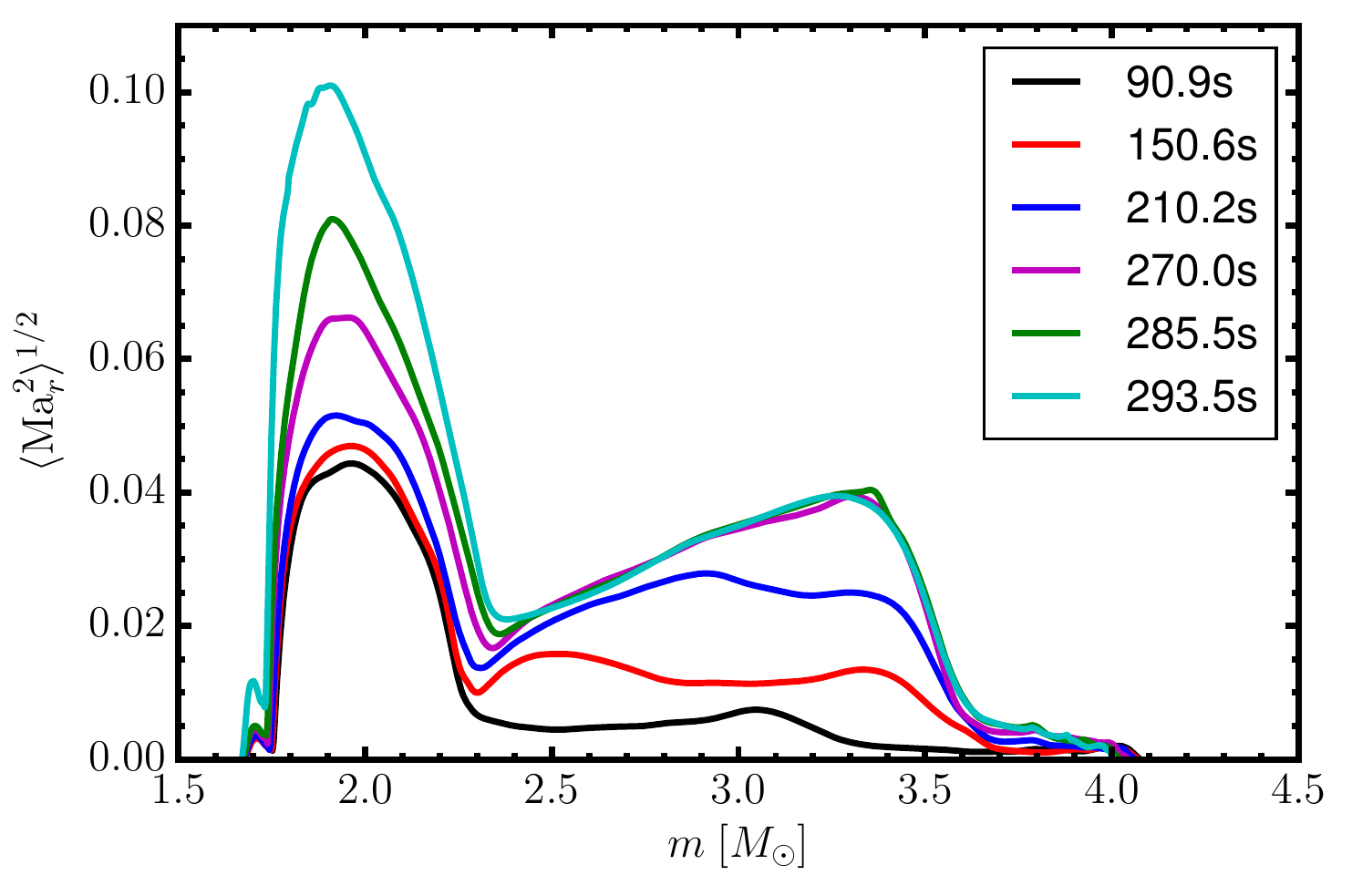}
  \caption{Profiles of the turbulent Mach number $\sqrt{\langle
      \mathrm{Ma}_r^2 \rangle}$ of radial velocity
    fluctuations in the oxygen and carbon shell at
    different times during the 3D simulation. Note that there is a
    secular increase in the Mach number in the oxygen shell even after
    convection has reached a quasi-stationary state due to the
    contraction of the inner boundary.  By contrast, the turbulent
    Mach number in the carbon shell merely increases because
    convection has not reached a quasi-stationary state in that shell.
    \label{fig:mach}}
\end{figure}

\begin{figure}
  \includegraphics[width=\linewidth]{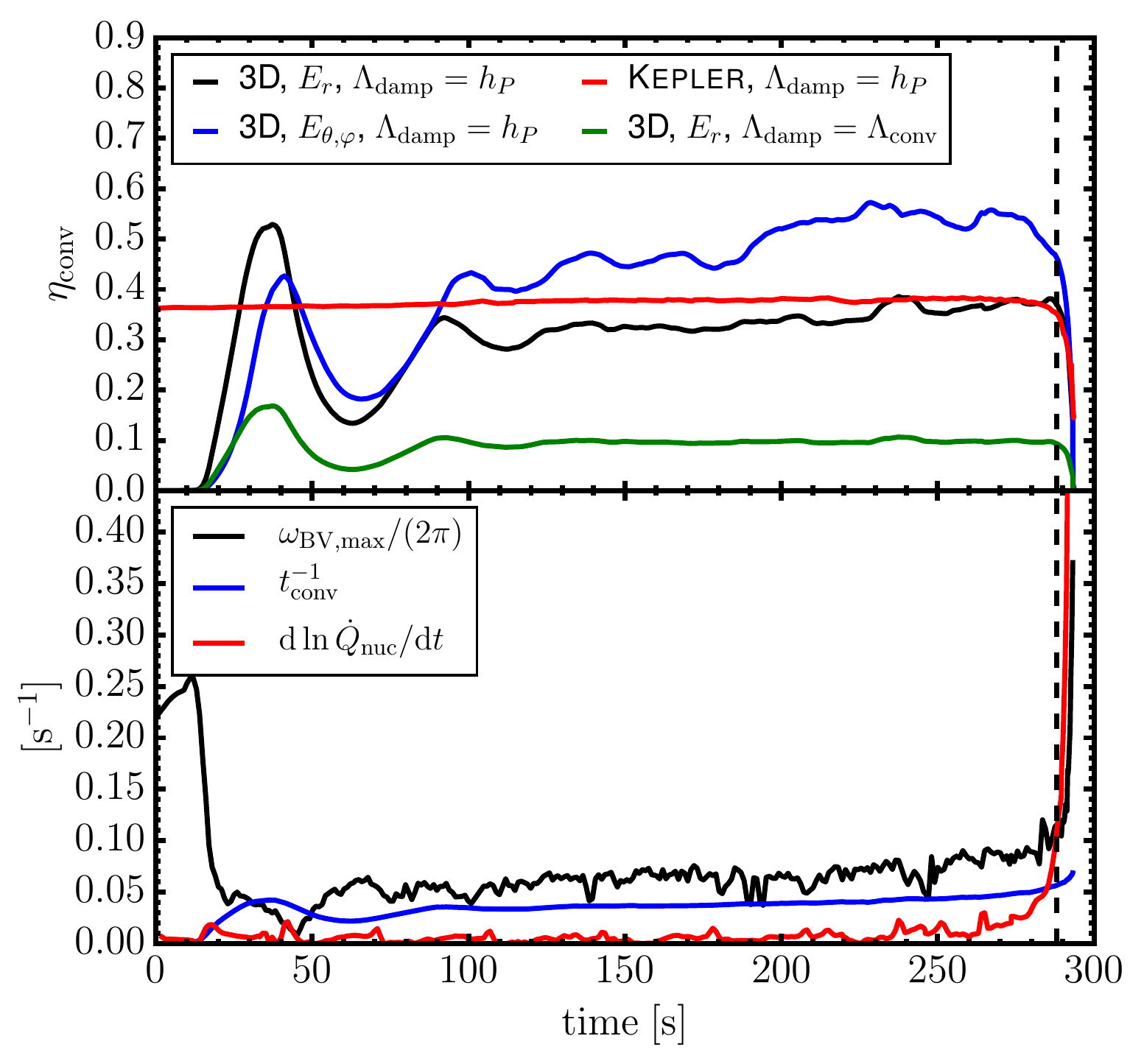}
  \caption{Top panel: Efficiency $\eta_\mathrm{conv}$ for the
    conversion of nuclear energy generation into convective kinetic
    energy as defined in Equation~(\ref{eq:eta_conv}) in the 3D run
    and the 1D \textsc{Kepler} model.  In the 3D case, we compute
    $\eta_\mathrm{conv}$ both for the kinetic energy in radial motions
    (Equation~\ref{eq:ekinr}, black curve) and transverse motions
    (Equation~\ref{eq:ekinl}, blue); for the \textsc{Kepler} run
    (red), we use the energy contained in radial convective motions
    computed according to Equation~(\ref{eq:ek_in_mlt}). By default, we use
    the pressure scale height in Equation~(\ref{eq:eta_conv}) as the
    mixing or damping length $\Lambda_\mathrm{damp}$.  The efficiency
    is much lower if $\Lambda_\mathrm{damp}$ is identified with the
    radial extent of the convective zone $\Lambda_\mathrm{conv}=r_+ -
    r_-$.  Bottom panel: Comparison of the Brunt-V\"ais\"al\"a
    frequency $\omega_\mathrm{BV,max}$ at the base of the oxygen shell (black), the reciprocal
    of the convective turnover time $t_\mathrm{conv}$ (blue), and the
    logarithmic derivative $\ud \ln \dot{Q}_\mathrm{nuc}/\ud t$ of the
    volume-integrated nuclear energy generation rate (red). The
    freeze-out of convection (denoted by a dashed vertical line)
    occurs roughly when $\omega_\mathrm{BV,max}/(2\pi) \approx
    t_\mathrm{conv}^{-1} \approx \ud \ln \dot{Q}_\mathrm{nuc}/\ud t$.
    \label{fig:qconv}}
\end{figure}

\section{Simulation Results}
\label{sec:results}

In Figures~\ref{fig:snap2d_a} and \ref{fig:snap2d_b}, we show 2D
slices depicting the evolution of the mass fraction $X_\mathrm{Si}$ of
silicon and the radial velocity $v_r$ to provide a rough impression of
the multi-D flow dynamics in our 3D simulation. Convective plumes
initially develop on small angular scales in the inner part of the
oxygen shell (where the burning rate is high). After about $100
\, \mathrm{s}$ we see fully developed convection with maximum plume
velocities of $\mathord{\sim} 500 \, \mathrm{km} \, \mathrm{s}^{-1}$
that increase towards collapse, and large-scale modes dominate the
flow. The latest snapshots at $286 \, \mathrm{s}$ and $293.5
\, \mathrm{s}$ suggest the
emergence of a bipolar flow structure right before
collapse. Large-scale structures are more clearly visible in the
velocity field than in $X_\mathrm{Si}$. Indeed, the rising plumes
enriched in silicon and the sinking plumes containing fresh fuel
appear rather ``wispy'', an impression which is reinforced by the 3D
volume rendering of $X_\mathrm{Si}$ at the onset of collapse in
Figure~\ref{fig:vol3d}.

Convection also develops in the overlying carbon shell. However,
since the convective velocities in the carbon shell are lower,
and since this shell extends out
to a radius of $27,000 \, \mathrm{km}$, convection never reaches a quasi-steady
state within the simulation time. We therefore do not address
convection in the carbon shell in our analysis.

As in earlier studies of mixing at convective boundaries
\citep{meakin_07}, the interface between the carbon and oxygen layer
proves unstable to the Holmb\"oe/Kelvin-Helmholtz
instability\footnote{We do not attempt to classify the precise type at
  instability at play, since this is immaterial for our purpose.}
with wave breaking leading to the entrainment of material from the
carbon shell. The snapshots suggest that such entrainment events
become more frequent and violent shortly before collapse.

The convective velocities and eddy scales thus fall roughly into
the regime where the parametric study of \citet{mueller_15a}
suggests a significant impact of pre-collapse asphericities
on shock revival (a convective Mach number
of order $0.05$ or higher, corresponding to velocities
of a few $100 \, \mathrm{km} \, \mathrm{s}^{-1}$, and dominant
$\ell=1$ or $\ell=2$ modes).

\begin{figure}
  \includegraphics[width=\linewidth]{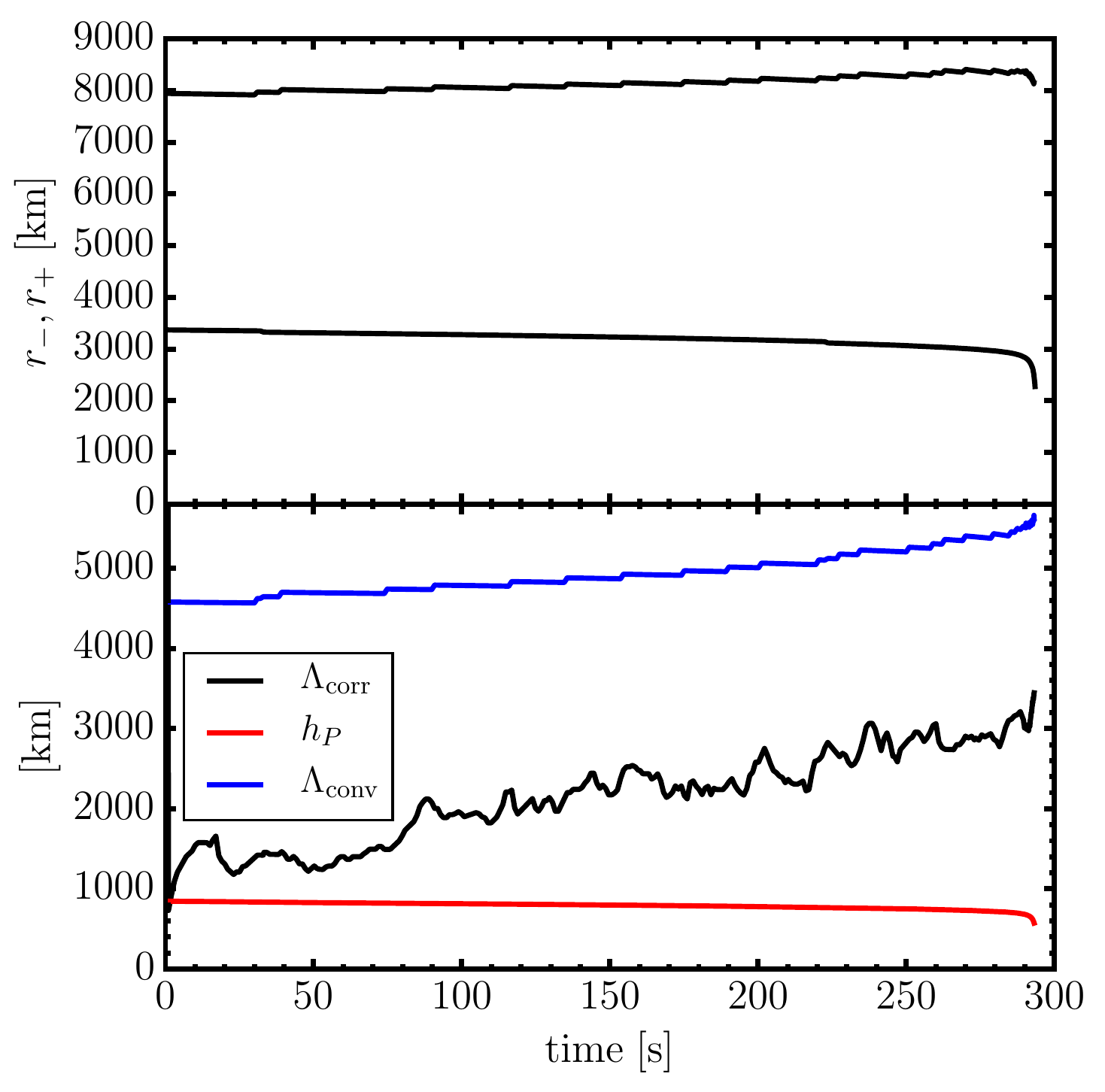}
  \caption{Top: Outer and inner boundary
radius $r_+$ and $r_-$ of the oxygen shell as functiosn
of time.
Bottom:
Correlation length $\Lambda_\mathrm{corr}$ for the radial velocity
computed at $r=4000 \, \mathrm{km}$ (black),
pressure scale height $h_P$ at the base of the oxygen shell (red)
and radial extent $\Lambda_\mathrm{conv}=r_+ - r_-$ of the oxygen shell (blue)
as a function of time.
    \label{fig:correlation_length}}
\end{figure}

\begin{figure}
  \includegraphics[width=\linewidth]{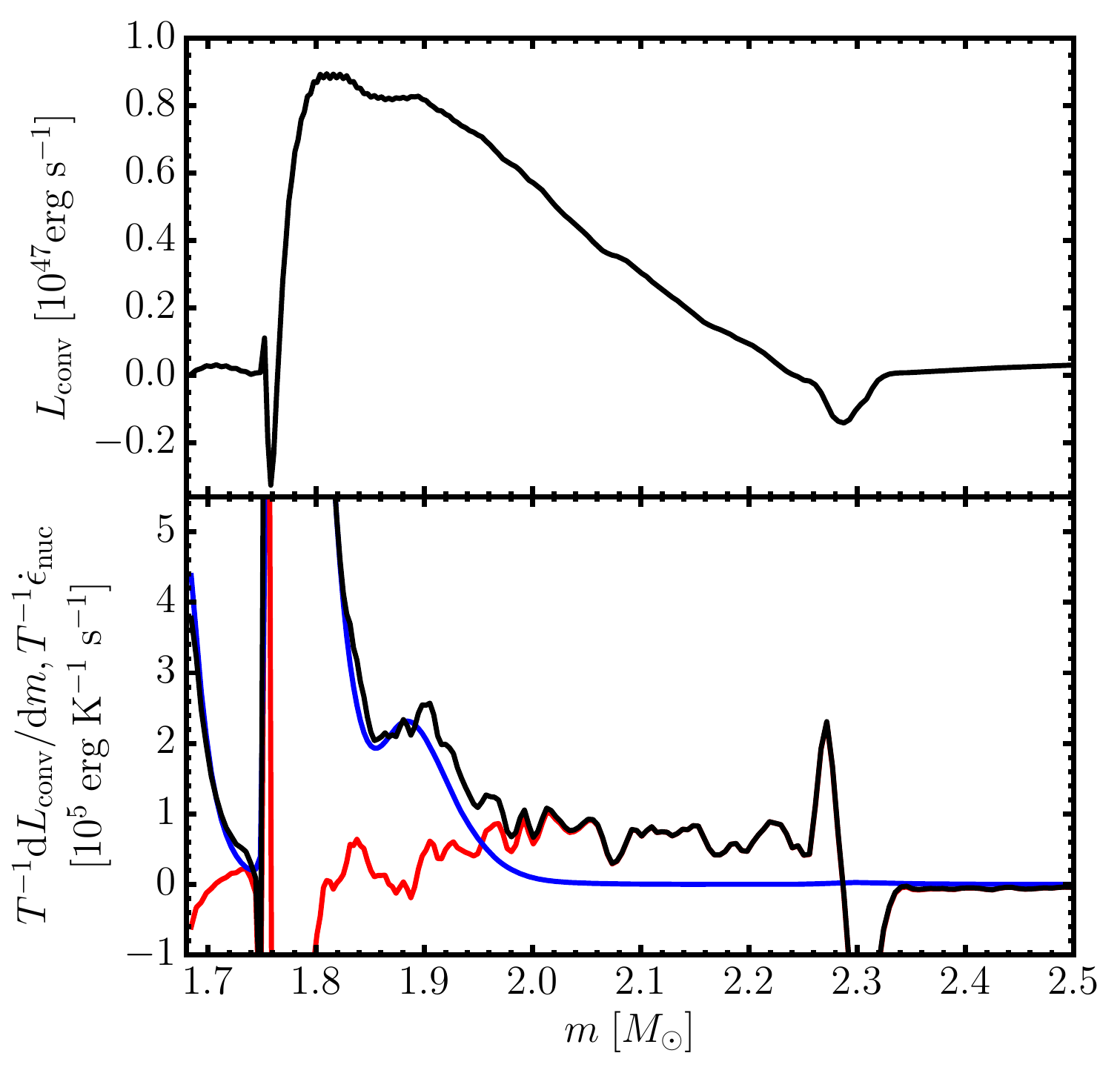}
  \caption{Top: Total convective luminosity
(including the kinetic energy flux) at $210 \, \mathrm{ms}$
in the 3D simulation
as a function of enclosed mass $m$. Bottom: Quantities determining
the spherically averaged entropy production.
The term $T^{-1} \ud L_\mathrm{conv}/\ud m$ stemming
from the divergence of the total convective
luminosity is shown in red, the entropy production
due to the nuclear source term $\dot{\epsilon}_\mathrm{nuc}/T$
(neglecting terms in the chemical potential of the
different nuclear species) is shown
in blue, and the black curve denotes the sum of both terms.
The curves are computed from averages over several time steps.
    \label{fig:lconv}}
\end{figure}

\begin{figure}
  \includegraphics[width=\linewidth]{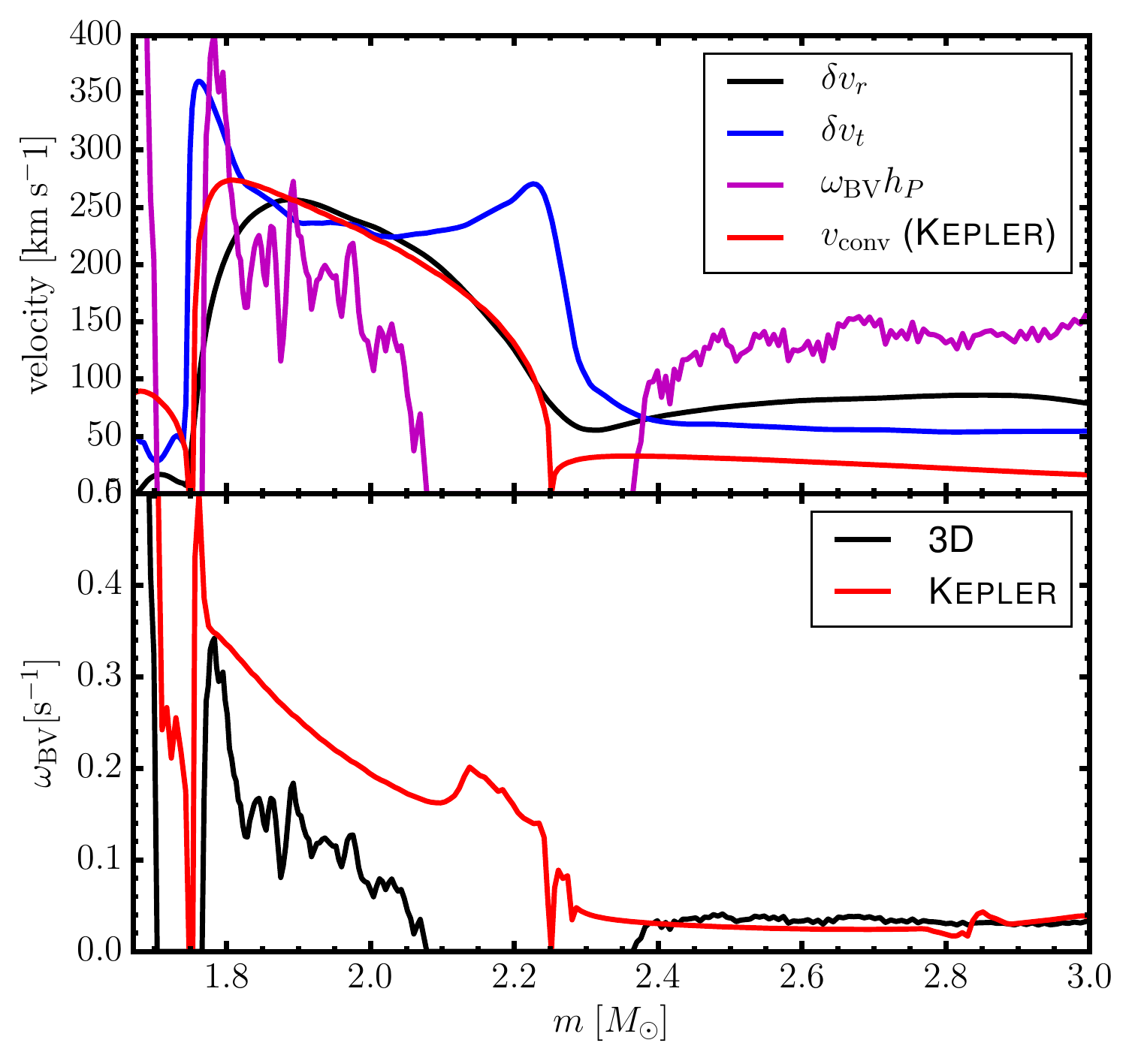}
  \caption{
Top panel: comparison of radial and transverse
RMS velocity fluctuations $\delta v_r$ (black)
and $\delta v_t$ (blue) in the 3D model
at $210 \, \mathrm{s}$
to the convective velocity 
$v_\mathrm{conv}$ computed in the \textsc{Kepler}
model using MLT (red), and to
$\omega_\mathrm{BV} h_P$ (violet).
Bottom panel: Comparison of the Brunt-V\"ais\"al\"a frequency computed
from spherical averages of the pressure, density, and sound
speed in the 3D run (black) and in the 1D \textsc{Kepler} model.
Note that there is a formally stable region around
the boundary between the oxygen and carbon shell in the 3D model
due to the aspherical deformation of the shell
boundary and the entrainment of material from the carbon shell,
which increases the spherically averaged entropy in the outer parts
of the oxygen shell.
  \label{fig:comp_vel}}
\end{figure}

\begin{figure}
  \includegraphics[width=\linewidth]{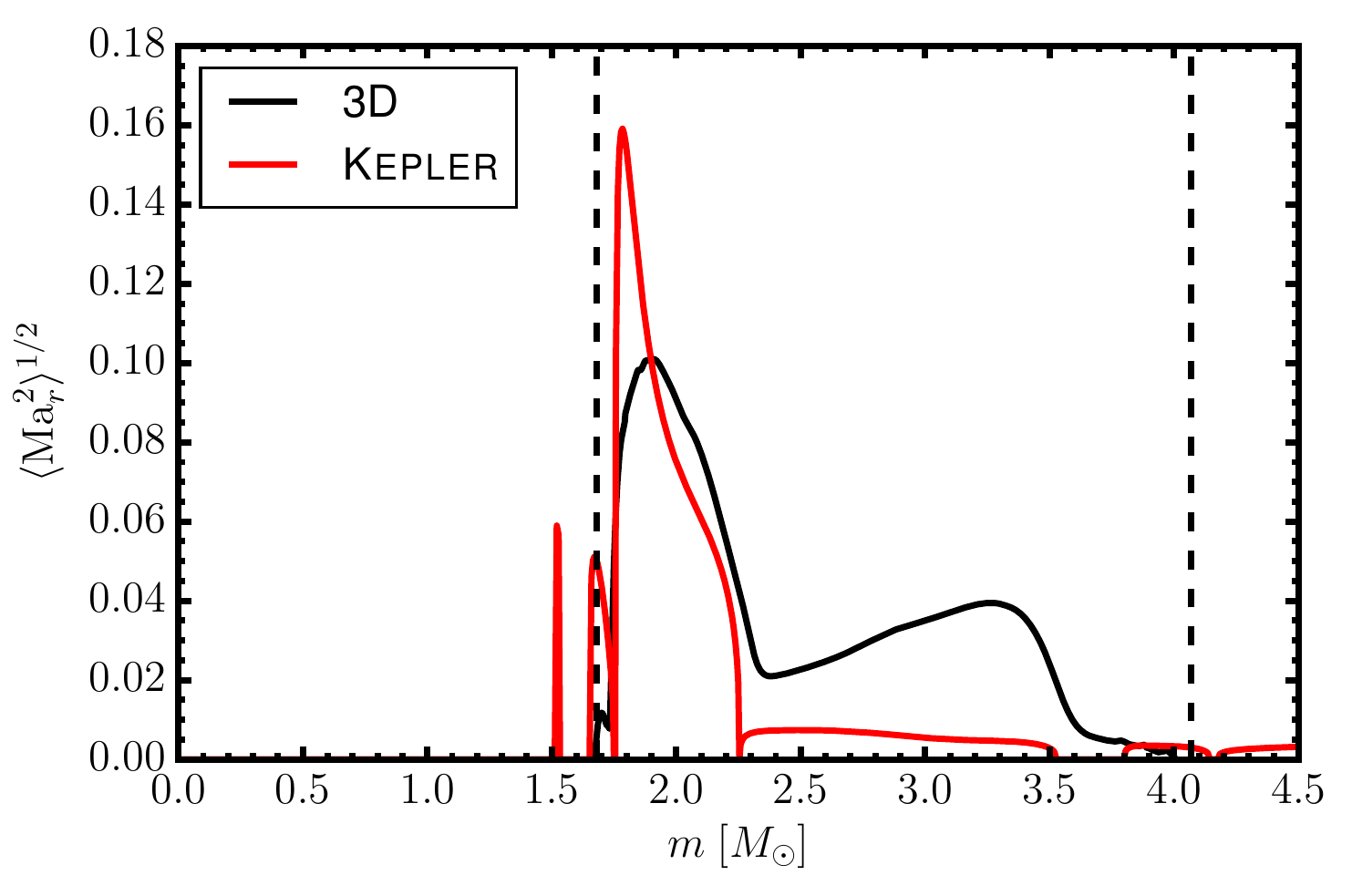}
  \caption{ Profiles of the RMS-averaged turbulent Mach number $\langle
      \mathrm{Ma}_r^2 \rangle^{1/2}$ of radial velocity fluctuations at the onset of collapse in the 3D model
    (black) and the 1D \textsc{Kepler} model (red).
Dashed lines denote the boundaries of the domain simulated
in 3D in \textsc{Prometheus}. The turbulent Mach number
in 1D and 3D agrees well in the bulk of the oxygen shell,
but the acceleration of nuclear burning and the concomitant
increase of the Brunt-V\"ais\"al\"a frequency
artificially increases the convective velocities
close to the base of the shell in \textsc{Kepler},
as MLT immediately translates the increase in $\omega_\mathrm{BV}$
into an increase in convective velocity.
Note that high nuclear burning rates in individual zones
inside the silicon core produce formally unstable
zones in the \textsc{Kepler} run shortly
before collapse, which do not affect the evolution
of the model to any significant degree.
    \label{fig:mach_collapse}}
\end{figure}

\subsection{Flow Dynamics for Quasi-Stationary Convection-- Quantitative Analysis and
Comparison with MLT}
\label{sec:comparison}

To analyze the flow dynamics more quantitatively, we consider the
volume-integrated net nuclear energy generation rate (including
neutrino losses) in the oxygen shell, $\dot{Q}_\mathrm{nuc}$, the
volume-integrated turbulent kinetic energy $E_r$ and $E_{\theta,\varphi}$
contained in the fluctuating components of radial and non-radial
velocity components, and profiles of the root-mean-square (RMS)
averaged turbulent Mach number $\langle \mathrm{Ma}_r^2\rangle^{1/2}$
of the radial velocity fluctuations in
Figures \ref{fig:qnuc_and_ekin} and \ref{fig:mach}.  $E_r$,
$E_{\theta,\varphi}$, and $\langle \mathrm{Ma}^2\rangle^{1/2}$, are computed
from the velocity field as follows,
\begin{eqnarray}
\label{eq:ekinr}
E_r
&=&
\frac{1}{2}\int\limits_{r_- \leq r \leq r_+}
\rho (v_r -\langle v_r\rangle )^2 \, \ud V,
\\
\label{eq:ekinl}
E_{\theta,\varphi}
&=&
\frac{1}{2}\int\limits_{r_- \leq r \leq r_+}
\rho (v_\theta^2+v_\varphi^2) \, \ud V,
\\
\langle \mathrm{Ma}_r^2\rangle^{1/2}
&=&
\left[\frac{\int \rho (v_r-\langle v_r\rangle)^2 \, \ud \Omega}{\int \rho c_s^2 \, \ud \Omega}\right]^{1/2},
\end{eqnarray}
where the domain of integration
in Equations~(\ref{eq:ekinr}) and (\ref{eq:ekinl}) extends
from the inner boundary radius $r_-$ to the outer boundary radius $r_+$
of the oxygen shell.
Angled brackets denote mass-weighted spherical Favre averages
for quantity $X$,
\begin{equation}
\langle X \rangle
=
\frac{\int \rho X\, \ud \Omega}{\int \rho \, \ud \Omega}.
\end{equation}
We note that one does not expect any mean flow in the non-radial
directions in the absence of rotation; therefore only
$v_\theta$ and $v_\varphi$ appear in Equation~(\ref{eq:ekinl}).
In Figure~\ref{fig:qnuc_and_ekin}, we also show the results
for $\dot{Q}_\mathrm{nuc}$ and the kinetic
energy in convective motions from the 1D \textsc{Kepler}
run for comparison. MLT only predicts the radial velocities
of rising and sinking convective plumes, so we only
compute the 1D analog to $E_r$,
\begin{equation}
\label{eq:ek_in_mlt}
E_{r,\mathrm{1D}}
=
\frac{1}{2}\int\limits_{r_-}^{r_+}
\rho v_\mathrm{conv}^2 \, \ud V,
\end{equation}
where $v_\mathrm{conv}$ is calculated according to Equation~(\ref{eq:vconv}).

The volume-integrated nuclear energy generation rate
$\dot{Q}_\mathrm{nuc}$ increases by more than two orders of magnitude
during the evolution towards collapse. Due to slight structural
adjustments after the initial transient and slightly different mixing
in the 3D model, $\dot{Q}_\mathrm{nuc}$ is roughly $30\ldots 50\%$
higher in 3D than in the \textsc{Kepler} for most of the run
(see discussion in Section~\ref{sec:mixing}), but
still parallels the \textsc{Kepler} run quite nicely and perhaps as
closely as can be expected given the extreme dependence of
the local energy generation $\dot{\epsilon}_\mathrm{nuc} \propto T^{30}$
on the temperature $T$ during oxygen burning.

The convective kinetic energy oscillates considerably during the first
$120 \, \mathrm{s}$, but exhibits a smooth secular increase reflecting
the acceleration of nuclear burning.  Equipartition between the radial
and non-radial kinetic energy in convective motions as suggested
by \citet{arnett_09} does not hold exactly,
instead we observe $E_\mathrm{\theta,\varphi}>E_r$ for most of the
simulation, suggesting that
there may not be a universal ratio between the
non-radial and radial kinetic energy and that this ratio is
instead somewhat dependent on the shell geometry (width-to-radius ratio,
ratio of width and pressure scale height), which can vary across
different burning shells, progenitors, and evolutionary phases.
There may also be stochastic variations in the eddy geometry
that the convective flow selects (see Appendix~\ref{app:res}) .
Anisotropic numerical dissipation might also account for different
results in different numerical simulations.
The
turbulent Mach number in the oxygen shell (Figure~\ref{fig:mach}) also
increases steadily from about $0.04 \ldots 0.05$ after the initial
transient to $0.1$ at collapse.

Again, there is reasonable agreement between the MLT prediction
$E_{r,\mathrm{1D}}$ for the convective kinetic energy and $E_r$ in the
3D simulation (Figure~\ref{fig:qnuc_and_ekin}).  $E_{r,\mathrm{1D}}$ and $E_r$ are in fact closer to
each other than $E_\mathrm{\theta,\varphi}$ and $E_r$ in 3D. Somewhat
larger deviations arise immediately prior to collapse when
convection is no longer fast enough to adjust to the acceleration
of nuclear burning as we shall discuss in Section~\ref{sec:freezeout}.

Except for the last few seconds, the kinetic energy in convection
scales nicely with the nuclear energy generation rate both in 1D and 3D.
For a case where the convective luminosity $L_\mathrm{conv}$ and
$\dot{Q}_\mathrm{nuc}$ balance each other in the case of steady-state
convection, MLT implies $v_\mathrm{conv}^3 \sim \dot{Q}_\mathrm{nuc}
\Lambda_\mathrm{mix} /M_\mathrm{conv}$, where $M_\mathrm{conv}$
is the mass contained in the convective shell \citep[note that only
  the form of the equations is slightly different in these
  references]{biermann_32,arnett_09}. In Figure~\ref{fig:qconv}, we
show the efficiency factors $\eta_\mathrm{conv}$ for the conversion of
nuclear energy generation into turbulent kinetic energy\footnote{Note that $\eta_\mathrm{conv}$
does not correspond to the ``convective efficiency'' as often used in stellar evolution, i.e.\
it is not the ratio of the convective luminosity to the radiative luminosity. }
$E_\mathrm{turb}$,
\begin{equation}
\label{eq:eta_conv}
\eta_\mathrm{conv}
=\frac{E_\mathrm{turb}/M_\mathrm{conv}}{(\dot{Q}_\mathrm{nuc} \Lambda_\mathrm{mix}/
M_\mathrm{conv})^{2/3}},
\end{equation}
for both the 3D model (using either the component $E_r$ or
$E_{\theta,\varphi}$ for $E_\mathrm{turb}$) and the \textsc{Kepler}
model (using $E_\mathrm{turb}=E_{r,\mathrm{1D}}$), with
$\Lambda_\mathrm{mix}$ set to the pressure scale height at the inner
boundary of the oxygen shell.  Between $130 \, \mathrm{s}$ and $290
\, \mathrm{s}$, $\eta_\mathrm{conv}$ shows only small fluctuations
around $0.35$ and $0.5$ for the kinetic energy in radial and
non-radial convective motions in 3D. For the \textsc{Kepler} model, we
find similar values around $\eta_\mathrm{conv}\approx 0.37$.

The scaling law $v_\mathrm{conv}^3 \sim \dot{Q}_\mathrm{nuc}
\Lambda_\mathrm{mix} /M_\mathrm{conv}$ can also be understood
as resulting from a balance of buoyant driving (or, equivalently,
kinetic energy generation by a heat engine) and turbulent dissipation
(see, e.g.\, \citealt{arnett_09} and in a different context
\citealt{mueller_15a}). In this picture, the scaling law emerges if
the mixing length is identified with the damping length
$\Lambda_\mathrm{damp}$. This identification
($\Lambda_\mathrm{damp}=\Lambda_\mathrm{mix}=h_P$), however, has
been criticized on the ground that $\Lambda_\mathrm{damp}$
should correspond to the largest eddy scale, which can be considerably
larger than $h_P$ if low-$\ell$ modes dominate the flow and the
updrafts and downdrafts traverse the entire convection zone,
which is precisely the situation that is realized in our 3D model.
The disparity of the pressure scale height
and the eddy scale can be quantified more rigorously by considering
the radial correlation length $\Lambda_\mathrm{corr}$ for fluctuations
in the radial velocity, $v'_r=v_r-\langle v_r\rangle$.
Following \citet{meakin_07} and \citet{viallet_13}
we compute the vertical correlation length $\Lambda_\mathrm{corr}$ as the full width
at half maximum of the correlation function $C(r, \delta r)$,
\begin{equation}
\label{eq:lcorr}
 C (r,\delta r)
  =
\frac{\langle v_r' (r,\theta,\varphi) v_r'(r+\delta r,\theta,\varphi)\rangle}
{\sqrt{\langle {v'}_r^2 (r,\theta,\varphi) \rangle \langle {v'}_r^2 (r+\delta r,\theta,\varphi) \rangle}}.
\end{equation}
The correlation function is computed at a radius of $r=4000 \, \mathrm{km}$
in the inner half of the oxygen shell.
$\Lambda_\mathrm{corr}$ is shown in Figure~\ref{fig:correlation_length}
and compared to the pressure scale height $\Lambda_\mathrm{mix}=h_P$ at the inner boundary
of the oxygen shell and the extent $\Lambda_\mathrm{conv}$
of the convective region. Once convection is fully developed,
we clearly have $\Lambda_\mathrm{corr}>\Lambda_\mathrm{mix}$
and $\Lambda_\mathrm{corr}\approx \Lambda_\mathrm{conv}/2$
(as expected for updrafts and downdrafts reaching over
the entire zone).

\citet{arnett_09} argued that the damping length should be of the
order of the width $\Lambda_\mathrm{conv}$ of the convective zone under
such circumstance.  If we compute the efficiency factor
$\eta_\mathrm{conv}$ based on $\Lambda_\mathrm{conv}$,
\begin{equation}
\eta_\mathrm{conv}
=\frac{E_\mathrm{turb}/M_\mathrm{conv}}{(\dot{Q}_\mathrm{nuc} \Lambda_\mathrm{conv}/M_\mathrm{conv})^{2/3}},
\end{equation}
we obtain suspiciously low values $\eta_\mathrm{conv}\lesssim 0.1$,
however. This suggests that the effective damping length is set by the
pressure scale height (or a multiple thereof) after all.  One could
opine that the energetics of the flow might still be described
adequately by $\Lambda_\mathrm{damp}=\Lambda_\mathrm{corr}$,
and that the efficiency factor $\eta_\mathrm{conv}$ merely
happens to be relatively low.

We argue, however, that there is a deeper reason for identifying
$\Lambda_\mathrm{damp}$ with a multiple of the pressure scale height
in the final phases of shell convection when neutrino cooling can no
longer balance nuclear energy generation. The crucial point is that
the average distance after which
buoyant convective blobs have to return their excess enthalpy $h'$ to
their surroundings cannot become arbitrarily large in a steady-state
situation, and since enthalpy and velocity fluctuations $v'$
are correlated ($h'\sim v'^2$), this also limits the damping length.

During the final stages, nuclear
energy generation, convective transport, and turbulent dissipation
must balance each other in such a way as to avoid both a secular build-up
of an ever-growing unstable entropy/composition gradient (although
the spherically average stratification always remains \emph{slightly}
unstable) and a complete erasure of the superadiabatic
gradient. Assuming that the Brunt-V\"ais\"al\"a frequency is
primarily set by the gradient of the entropy $s$,
this implies $\pd^2 s/\pd r \pd t \approx 0$, and hence
roughly constant entropy generation,
\begin{equation}
  \dot{s}=\frac{\pd s}{\pd t }\approx \mathrm{const.}
\end{equation}
throughout the convective region. In the late pre-collapse
stages, we can relate $\dot{s}$ to the local nuclear energy
generation rate $\dot{\epsilon}_\mathrm{nuc}$ and the derivative
of the ``total'' convective  luminosity $L_\mathrm{conv}$,
\begin{equation}
  \label{eq:dot_s}
\dot{s} \approx
  \frac{\dot{\epsilon}_\mathrm{nuc}}{T}
  +  \frac{1}{T}
  \frac{\pd L_\mathrm{conv}}{\pd m}
  =
  \frac{\dot{\epsilon}_\mathrm{nuc}}{T}
  +  \frac{1}{4 \pi r^2 \rho T}
  \frac{\pd L_\mathrm{conv}}{\pd r}.
  \end{equation}
Here, $L_\mathrm{conv}$ denotes the net total energy flux
resulting from fluctuations (denoted by primes) in the total energy density
and velocity around their spherical Favre average,
\begin{equation}
  L_\mathrm{conv}
  =
  r^2 \int
  \left[\rho e+P+\rho \frac{v^2}{2}\right]' v'_r \, \ud \Omega,
\end{equation}
where $e$ is the specific internal energy, and $P$ the pressure.  Note
that in formulating Equation~(\ref{eq:dot_s}), we implicitly assumed
that $\pd{L_\mathrm{conv}}/\pd m$ is equal to the rate of change of
the Favre average of the internal energy $e$ (instead of the total
energy density, which includes the contribution of the turbulent
kinetic energy). This assumption is justified for steady-state
convection in the late pre-collapse phase because of moderate Mach
numbers and the minor role of neutrino cooling. These two factors imply
that the energy that is generated by nuclear reactions and distributed
throughout the unstable region by convection mostly goes into internal
energy (whereas our argument cannot be applied to earlier phases
where neutrino cooling and nuclear energy generation balance each other).

Figure~\ref{fig:lconv} shows the two terms contributing to $\dot{s}$
in Equation~(\ref{eq:dot_s}) based on Favre averages over a few time
steps around $210 \, \mathrm{s}$, and demonstrates that $\dot{s}$ is
indeed roughly constant throughout the convective region. Since strong
nuclear burning is confined to a narrow layer at the bottom of the
convective shell, we even have $T^{-1} \pd L_\mathrm{conv}/\pd m
\approx \mathrm{const.}$ throughout a large part of the shell,
and for a stratification with roughly $\rho \propto r^{-3}$
and $T \propto r^{-1}$, this leads to
\begin{equation}
  \frac{\pd L_\mathrm{conv}}{\pd r}
  \propto  r^2 \rho T \propto r^{-2}.
\end{equation}
For such an idealized case, one can directly compute
that the energy transported by convective blobs from the lower boundary
must be dissipated after an average distance of
\begin{eqnarray}
  \nonumber
  \Lambda_\mathrm{damp}
  &=&
  -\frac{1}{L_\mathrm{conv}(r_-)-L_\mathrm{conv}(r_+)}
  \int\limits_{r_-}^{r_+}  (r-r_-) \frac{\pd L_\mathrm{conv}}{\pd r}\ud r
  \\
\label{eq:ldamp}
  &=&r_- \left(\frac{\xi \ln \xi}{\xi -1 }-1\right),
\end{eqnarray}
where $\xi=r_+/r_-$ is the ratio of the outer and inner boundary radius.
Evidently, $\Lambda_\mathrm{damp}$ grows only moderately at large
$\xi$, and is always smaller than $(r_+-r_-)/2$.

It thus appears unlikely that large damping lengths
$\Lambda_\mathrm{damp} \approx r_+-r_- \gg h_P$
can be realized in very extended convection zones in the final
pre-collapse stage. This is decidedly different to earlier
stages with strong neutrino cooling in the outer part
of the convective zone, for which
\citet{arnett_09} found high values of $\Lambda_\mathrm{damp} =0.85
(r_+-r_-)$. As outlined before, the different behavior is likely
due to the specific physical conditions right before collapse;
in the absence of strong cooling, the self-regulatory mechanism
that we outlined above automatically ensures that
$\Lambda_\mathrm{damp}$ cannot be considerably larger than
the pressure scale height. Thus, the implicit identification
of $\Lambda_\mathrm{damp}$ and $h_P$ (or a multiple thereof)
in MLT is likely less critical for shell convection right
before collapse than for earlier phases.

However, it still remains to be determined whether the damping length can
reach considerably higher values in deep convection zones with $\xi \gg 1$
during earlier stages when nuclear energy generation and neutrino cooling
balance each other. Since neutrino cooling generally decreases with radius
within a shell, it can still be argued that the convective luminosity must
decay not too far away from the burning region. Thus, an analog to
Equation~(\ref{eq:ldamp}) could still hold, and the damping length would only
increase slowly with the width of the shell in the limit of large $\xi$. In
that case, the difference between our simulation and the results of
\citet{arnett_09} would merely be due to a different depth of the convective
zone, which is much deeper in our model ($4\ldots 5$ pressure scale heights as
opposed to $\mathord{\sim} 2$ pressure scale heights in \citealt{arnett_09})
so that we approach a ``saturation limit'' for the damping length and can more
conveniently distinguish the damping length from the width of the convective
zone since the different length scale are sufficiently dissimilar.

Radial profiles of the convective velocities also point to reasonable
agreement between MLT and the 3D simulation.
In the upper panel of Figure~\ref{fig:comp_vel}, we compare the
convective velocities from \textsc{Kepler} to RMS
averages of the fluctuations
of the radial velocity ($\delta v_r$)
and the transverse velocity component
($\delta v_t$)
at $210 \, \mathrm{s}$,
\begin{eqnarray}
\delta v_r
&=&
\left(
\frac{\int \rho (v_r-\langle v_r\rangle)^2 \, \ud \Omega}
{\int \rho \, \ud \Omega}
\right)^{1/2},
\\
\delta v_t
&=&
\left(
\frac{\int \rho (v_\theta^2+v_\varphi^2) \, \ud \Omega}
{\int \rho \, \ud \Omega}
\right)^{1/2}.
\end{eqnarray}
We also compare these to the MLT estimate
$v_\mathrm{conv}=\omega_\mathrm{BV} \Lambda_\mathrm{mix}$ computed
from the Brunt-V\"ais\"al\"a frequency for the spherically averaged
stratification of the 3D model. It is evident that the agreement
especially between $\delta v_r$ and the convective velocity in
\textsc{Kepler} is very good in the oxygen shell.  In large parts of
the shell, $v_\mathrm{conv}=\omega_\mathrm{BV} \Lambda_\mathrm{mix}$
is also in very good agreement with $\delta v_r$, which again
demonstrates that the choice of the pressure scale height as the
acceleration and damping length for convective blobs is a reasonable
choice. However, no reasonable comparison can be made in the outer
part of the oxygen shell, where $\omega_\mathrm{BV}$ is formally
negative. This is due to the strong aspherical deformation of
the shell boundary and the entrainment of light, buoyant
material from the carbon shell; the fact that the outer part of the
oxygen shell is formally stable if $\omega_\mathrm{BV}$ is
computed from spherical averages of the density and pressure
is thus merely a boundary effect and has no bearing on the
validity of MLT in the interior of the shell.

The good agreement between the 3D simulation and the \textsc{Kepler}
model may seem all the more astonishing considering the rescaling
of the Brunt-V\"ais\"al\"a frequency according to
Equation~(\ref{eq:rescaling}) for stability reasons.  However, this
procedure is justified by the fact that the convective luminosity
automatically adjusts itself in such a way as to avoid a secular
build-up of $\omega_\mathrm{BV}$ as discussed before. In a steady
state, the convective luminosity in MLT in a shell will roughly
balance the nuclear energy generation rate, $L_\mathrm{conv}=4 \pi r^2
F_\mathrm{conv}\sim \dot{Q}_\mathrm{nuc}$, regardless of whether
$\omega_\mathrm{BV}$ is rescaled or not.  If a rescaling factor is
introduced in Equation~(\ref{eq:mlt2}), the result is simply that a
larger $\omega_\mathrm{BV}$ is maintained under steady state
conditions to balance the rescaling factor. Except for pathological
situations, the convective energy flux and the convective velocities
are thus essentially unaffected by this procedure. The
superadiabaticity of the stratification is changed, however. For
convection at low Mach number, it will be systematically
overestimated. This trend is evident from the lower panel of
Figure~\ref{fig:comp_vel}, which compares $\omega_\mathrm{BV}$
in \textsc{Kepler} and the 3D simulation. Since convection is not
extremely subsonic in our case, the rescaling factor is only slightly
smaller than unity, and the superadiabaticity in the 1D and 3D model
remains quite similar.

\subsection{Freeze-Out of Convection}
\label{sec:freezeout}

MLT in \textsc{Kepler} thus provides good estimates for the typical
convective velocities in the final stages of oxygen shell burning
as long as a steady-state balance between nuclear energy generation,
convective energy transport, and turbulent dissipation is maintained.
However, steady-state conditions are not maintained up to collapse.
Figure~\ref{fig:qconv} shows that the growth of the turbulent
kinetic energy can no longer keep pace with the acceleration
of nuclear burning in the last few seconds before collapse,
where $\eta_\mathrm{conv}$ drops dramatically.

The time at which convection ``freezes out'' can be nicely
determined by appealing to a time-scale argument: Freeze-out
is expected once the nuclear energy generation rate
(which sets the Brunt-V\"ais\"al\"a frequency and the convective velocity
under steady-state conditions) changes significantly over
a turnover time-scale. More quantitatively, the
efficiency factor $\eta_\mathrm{conv}$ drops abruptly once
the freeze-out condition
\begin{equation}
  \label{eq:freezeout_omega}
\frac{1}{\dot{Q}_\mathrm{nuc}}\frac{\ud \dot{Q}_\mathrm{nuc}}{\ud t}=\frac{\omega_\mathrm{BV,max}}{2\pi}
\end{equation}
is met as shown in the bottom panel of Figure~\ref{fig:qconv}. Equivalently, the freeze-out condition can be
expressed in terms of the convective turnover time
$t_\mathrm{conv}$,
\begin{equation}
  t_\mathrm{conv}=\frac{\Lambda_\mathrm{conv}}{\bar{v}_\mathrm{conv}},
\end{equation}
where $\bar{v}_\mathrm{conv}$ is an appropriate global average
of the convective velocity, e.g.,
\begin{equation}
  \bar{v}_\mathrm{conv}=(2 E_{\mathrm{kin},r}/M_\mathrm{conv})^{1/2}.
\end{equation}
Using these definitions, we find that freeze-out occurs roughly
when
\begin{equation}
\frac{1}{\dot{Q}_\mathrm{nuc}}
  \frac{\ud \dot{Q}_\mathrm{nuc}}{\ud t}=  t_\mathrm{conv}^{-1},
\end{equation}
which may be even more intuitive than
Equation~(\ref{eq:freezeout_omega})

Somewhat astonishingly, the \textsc{Kepler} run shows a
similar drop of $\eta_\mathrm{conv}$ in the last seconds,
although MLT implicitly assumes steady-state conditions
when estimating the density contrast and the convective
velocity. \textsc{Kepler} still overestimates
the volume-integrated turbulent kinetic energy somewhat
after freeze-out (Figure~\ref{fig:qnuc_and_ekin}), but
the discrepancy between the 1D and 3D models is not inordinate.

The key to the relatively moderate differences can be found in
profiles of the turbulent convective Mach number $v_\mathrm{conv}/c_s$
in \textsc{Kepler} and $\langle \mathrm{Ma}_r^2\rangle^{1/2}$ in 3D at
the onset of collapse in Figure~\ref{fig:mach_collapse}. Evidently,
MLT only overestimates the convective velocities in a narrow layer at
the lower boundary of the oxygen shell, where the acceleration of
nuclear burning greatly amplifies the superadiabaticity of the
stratification (as quantified by $\omega_\mathrm{BV}$). This
\emph{immediately} increases $v_\mathrm{conv}$, whereas the convective
velocity field adjusts only on a longer time-scale ($\mathord{\gtrsim}
\omega_\mathrm{BV}^{-1}$) in 3D. However, even in the \textsc{Kepler}
run, the convective velocities in the middle and outer region of the
oxygen shell remain unaffected by the increase of the
$\omega_\mathrm{BV}$ close to the inner shell boundary. Different from
the innermost region, where $\omega_\mathrm{BV}$ reacts
instantaneously to the nuclear source term, $\omega_\mathrm{BV}$ (and
hence the convective velocity in the outer region) responds to the
accelerated burning on a diffusion time-scale, which is again
of order
$\omega_\mathrm{BV}^{-1}$. For a slightly different reason
(insufficient time for convective diffusion vs.\, insufficient time for
the growth of plumes), the \textsc{Kepler} run therefore exhibits a
similar freeze-out of convection as the 3D model. We thus
conclude that the volume-integrated turbulent kinetic energy
and the average convective Mach number in 1D
stellar evolution codes still provide a reasonable estimate
for the state of convection \emph{even right at collapse}.
The spatial distribution of the turbulent kinetic energy,
on the other hand, appears more problematic; it will be somewhat
overestimated in the shell source at collapse due to the instantaneous
reaction of $\omega_\mathrm{BV}$ to the increasing burning rate.

The rescaling of $\omega_\mathrm{BV}$ in \textsc{Kepler}
according to Equation~(\ref{eq:rescaling}) can also affect the
time of freeze-out at a minor level. For a convective Mach number
of $\mathord{\sim} 0.1$, the rescaling procedure changes
$\omega_\mathrm{BV}$ only by $\mathord{\sim} 30 \%$, and given
the very rapid increase of $\ud \ln Q_\mathrm{nuc}/\ud t$, this will not
shift the time of freeze-out appreciably.

\begin{figure}
  \includegraphics[width=\linewidth]{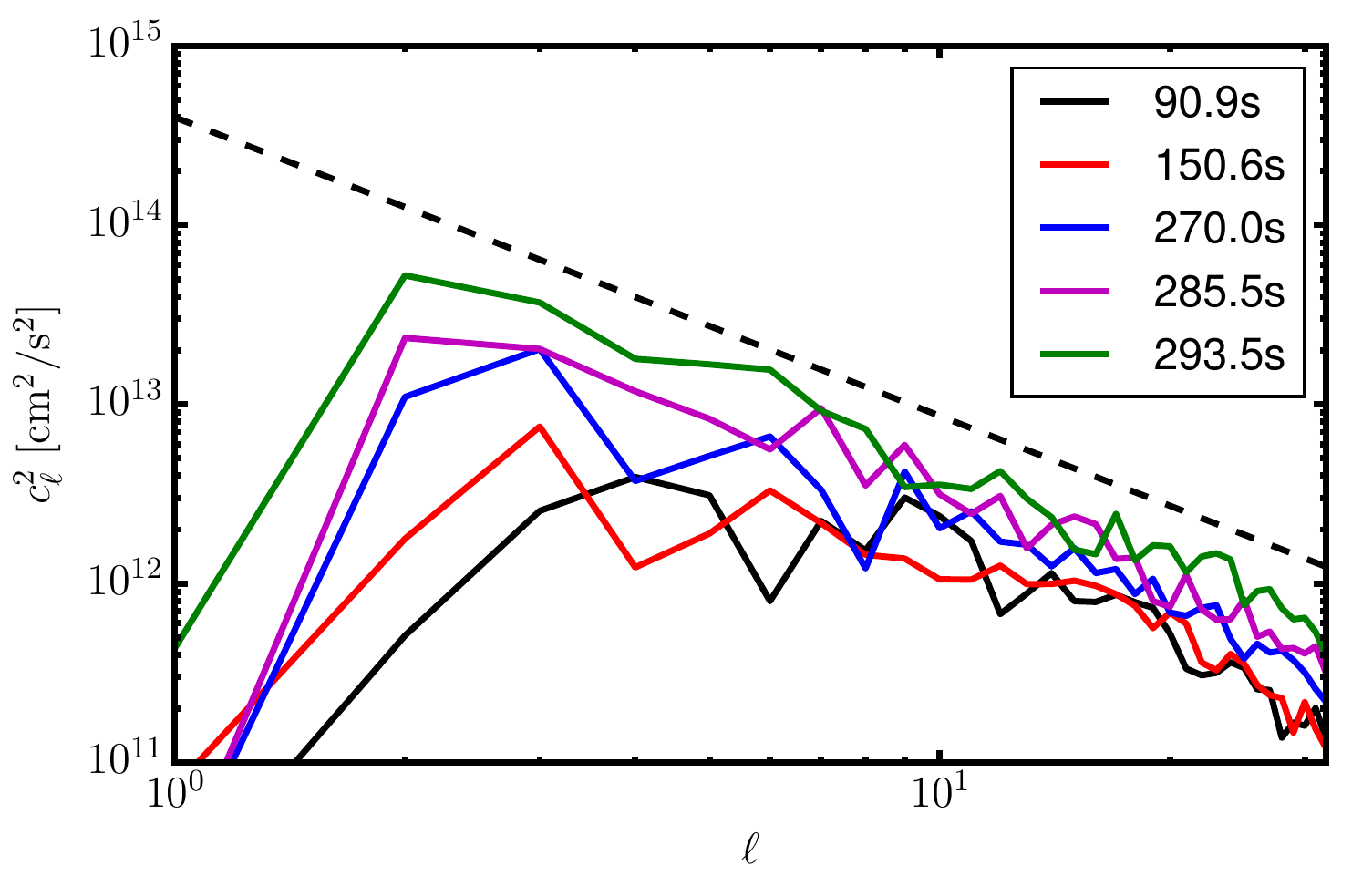}
  \caption{Power $c_\ell^2$ in different multipoles $\ell$
    for the decomposition of the radial velocity
    at $r=4000 \, \mathrm{km}$ into spherical
harmonics $Y_{\ell m}$ in the 3D model at different times
computed according to Equation~(\ref{eq:multipoles}).
The dominant angular
wave number shifts
from $\ell=3\ldots 5$ to $\ell=2$ over the course
of the simulation. The dashed line
indicates a slope of $\ell^{-5/3}$, which is roughly
expected for a Kolmogorov spectrum above the injection
scale in wave number  (i.e.\, at smaller spatial scales).
    \label{fig:vel_spectra}}
\end{figure}

\begin{figure}
  \includegraphics[width=\linewidth]{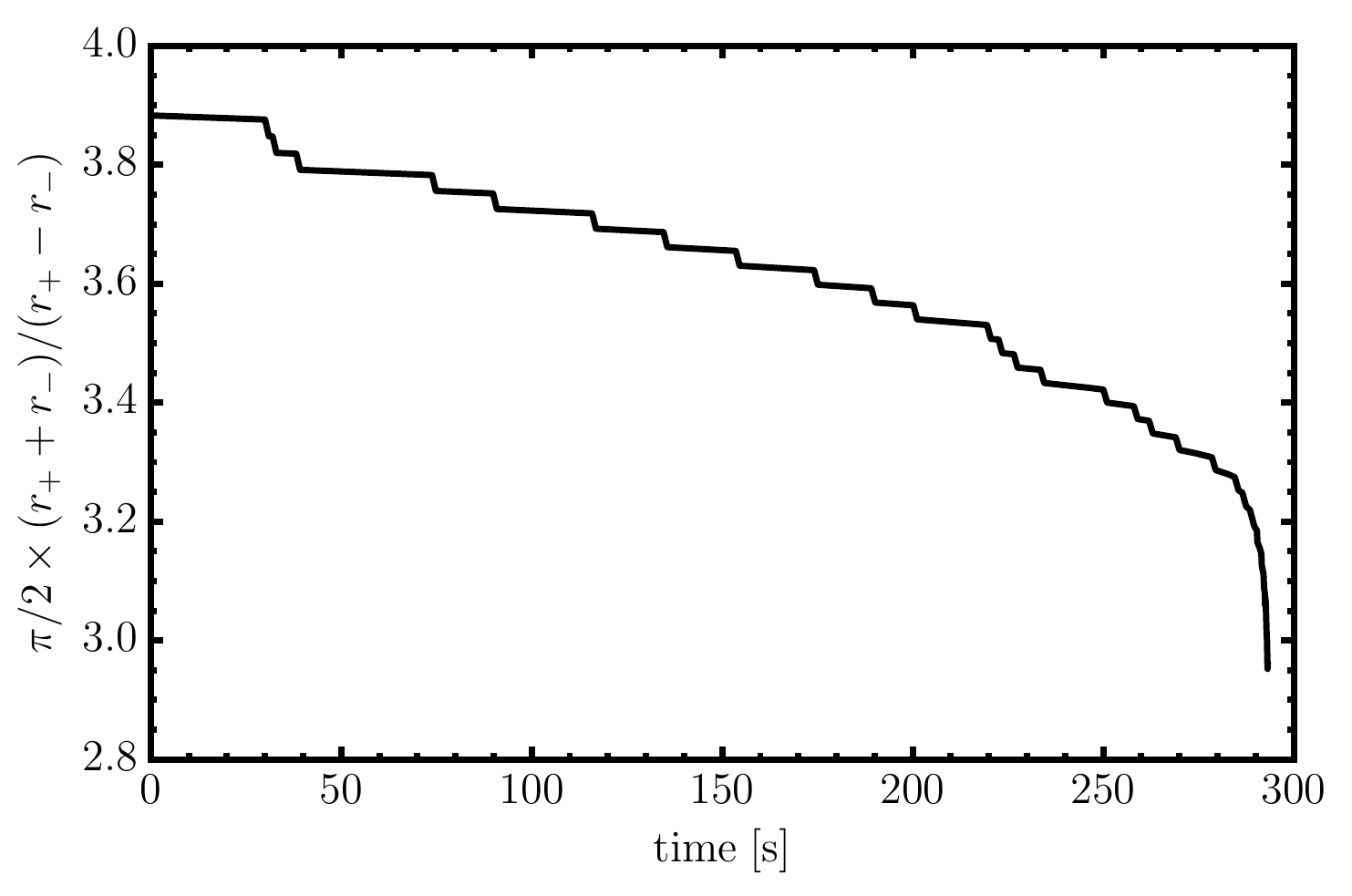}
  \caption{Estimate of the typical angular wave number $\ell=\pi/2
    \times (r_+ + r_-) /(r_+ - r_-)$ of convection in the linear
    regime from the inner and outer boundary radius of the oxygen
    shell according to Equation~(\ref{eq:dominant_l}) and
    \citet{foglizzo_06}. The rapid drop at the end
of the simulation is evident and suggests that
the emergence of a global $\ell=2$ mode is due to the
rapid contraction of the iron and silicon core shortly
before collapse.
    \label{fig:scale}}
\end{figure}

\subsection{Scale of Convective Eddies}
The role of progenitor asphericities in the explosion mechanism
depends not only on the magnitude of the convective velocities in the
burning shells, but also on the \emph{angular} scale of the infalling
eddies.  MLT does not make any strong assumptions about the eddy
scale; it assumes a radial correlation length for entropy and velocity
perturbations, but such a correlation length can in principle
be realized with very different flow geometries. Empirically,
simulations of buoyancy-driven convection in well-mixed
shells are usually characterized by eddies of
similar radial and angular extent $d$ that reach
across the entire unstable zone \citep[e.g.][]{arnett_09}.
The dominant modes are also typically close in scale
to the most unstable modes in the linear regime
\citep{chandrasekhar_61,foglizzo_06}, which
have $d \sim r_+ - r_-$.
This correspondence between
the linear and non-linear regime has sometimes
been justified by heuristic principles
for the selection of the eddy scale based on maximum kinetic
energy or maximum entropy production \citep{malkus_58,martyushev_06}.
Expressing the balance of kinetic energy
generation due to the growth of an instability
with a scale-dependent growth rate $\omega (d)$
and turbulent dissipation for the dominant mode
in a shell with mass $M$ yields
\begin{equation}
\dot{E}_\mathrm{kin}
\sim
 \omega(d) E_\mathrm{kin}
-v\frac{E_\mathrm{kin}}{d}
\sim
\omega (d) E_\mathrm{kin}
-\frac{\sqrt{2}E_\mathrm{kin}^{3/2}}{d M^{1/2}}
=0
\end{equation}
for the change of the kinetic energy $E_\mathrm{kin}$ in a given  mode.
The dominant mode(s) in the non-linear regime will be the one(s)
for which
\begin{equation}
E_\mathrm{kin}
\propto
M d^2 \omega(d)^2
\end{equation}
is maximal, which actually suggests a bias
towards slightly larger scales than in the linear regime.

A superficial inspection of Figures~\ref{fig:snap2d_a} and \ref{fig:snap2d_b}
already reveals that our 3D models conform to the typical
picture with $d \sim r_+ - r_-$. More quantitatively, the dominance
of large-scale modes is shown by a decomposition of the
radial velocity in the inner half of the oxygen shell
(at a radius of $4000 \, \mathrm{km}$) into spherical harmonics
(for more sophisticated decompositions of
the flow field see \citealt{fernandez_14,chatzopoulos_14}).
In Figure~\ref{fig:vel_spectra},
we plot the total power $c_{\ell}^2$ for each multipole
order $\ell$,
\begin{equation}
\label{eq:multipoles}
c_{\ell}^2
=
\sum_{m=-\ell}^\ell
\left |\int Y^*_{\ell m} (\theta,\varphi)
v_r (4000 \, \mathrm{km},\theta,\varphi) \, \ud \Omega\right|^2,
\end{equation}
which shows a clear peak at low $\ell$ that slowly moves from $\ell=4$ down to $\ell=2$
over the course of the simulation. The tail at high $\ell$ above the typical eddy
scale roughly exhibits an $\ell^{-5/3}$ slope as expected for a
Kolmogorov-like turbulent cascade \citep{kolmogorov_41} because of the
rough proportionality between $\ell$ and the wave number
\citep{peebles_93}.

The dominant eddy scale is consistent with the
crude estimate that the dominant $\ell$ is given by the
number of convective eddies of diameter $d=
r_+ - r_-$ that
can be fitted into one hemisphere of the
convective shell \citep{foglizzo_06},
\begin{equation}
\label{eq:dominant_l}
\ell
= \frac{\pi (r_+ + r_-)}{2 (r_+ - r_-)}.
\end{equation}
This estimate for the dominant multipole order is plotted in
Figure~\ref{fig:scale}. It agrees well with spectra of the radial
velocity, although it may not clearly predict the emergence of the
dominant quadrupole at the end (which is compatible with our argument
that the dominant angular scale for fully developed convection is
slightly larger than in the linear regime).  The slowly changing
geometry of the shell evidently accounts nicely for the secular trend
towards modes of lower $\ell$.  Figure~\ref{fig:correlation_length}
reveals that both the contraction of the inner boundary of the shell
by about one third in radius and a secular expansion of the (somewhat
ill-defined) outer shell boundary contribute to this trend. The fast
change of $\ell$ right before collapse is clearly due to the
contraction, however, as the outer boundary radius \emph{decreases}
again shortly before collapse.

The
expansion of the outer boundary is not seen in the \textsc{Kepler}
model and is the result of entrainment of material from the carbon
shell (see Section~\ref{sec:mixing} below).  If the amount of entrainment
is physical, this is another reason to suspect that estimates of
the dominant angular scale based on stellar evolution models
using Equation~(\ref{eq:dominant_l}) will slightly overestimate
the dominant $\ell$. Considering uncertainties and progenitor variations
in the shell structure,   Equation~(\ref{eq:dominant_l})
nonetheless furnishes a reasonable zeroth-order estimate
of the typical eddy scale.

\begin{figure}
  \includegraphics[width=\linewidth]{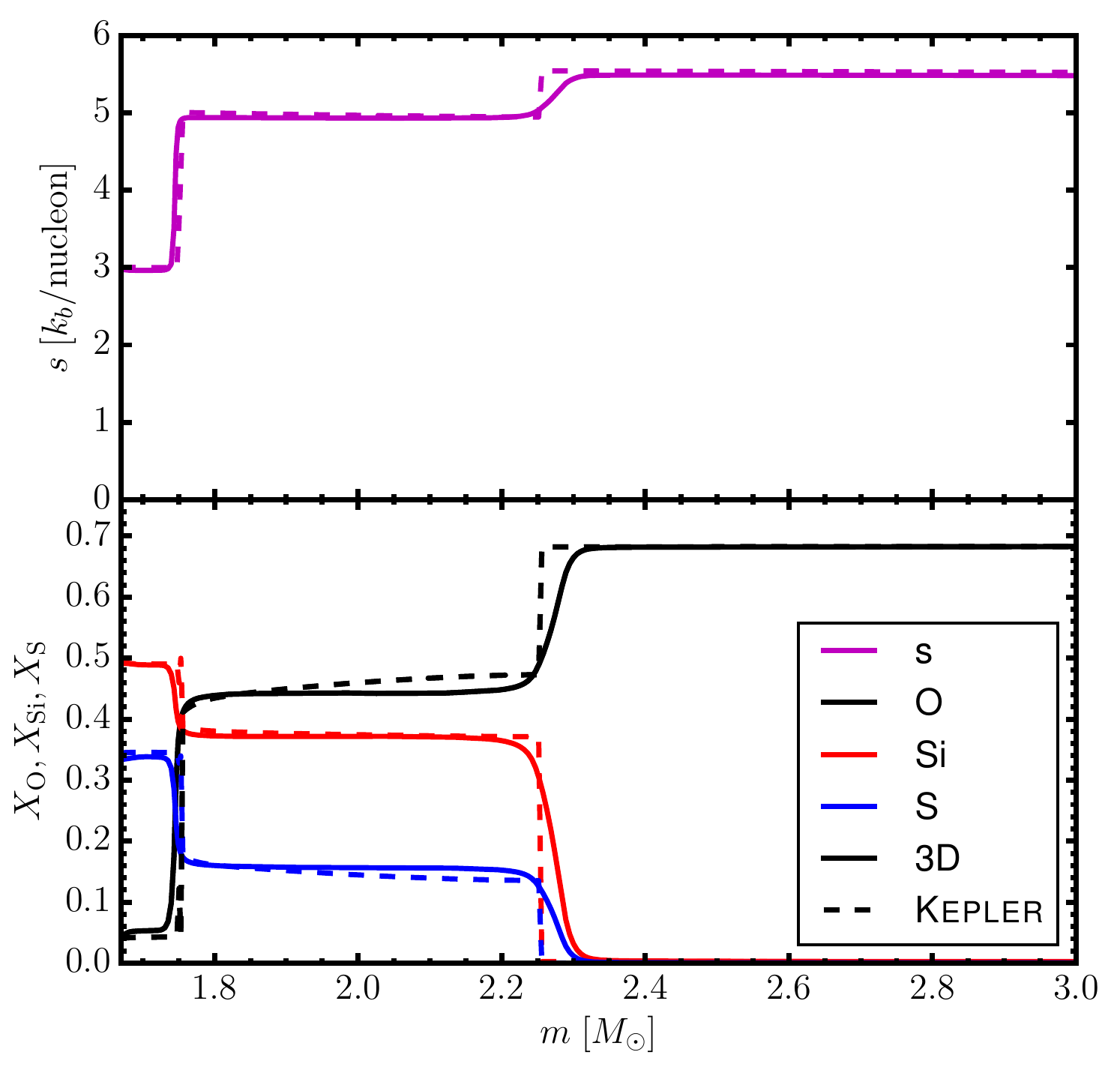}
  \caption{
Spherically averaged profiles of the
 entropy (violet curves, top panel) and the mass fractions
of oxygen (black), silicon (red), and sulfur (blue)
in the 3D run (solid curves) at $210 \, \mathrm{s}$ compared
to profiles from the 1D \textsc{Kepler} model (dashed) at the same time.
Note that the slope in the mass fractions is somewhat steeper
in the \textsc{Kepler} model, which we ascribe to the use of
an extra factor of $1/3$ in the diffusion equation
for compositional mixing.
  \label{fig:comp_mix}}
\end{figure}

\begin{figure}
  \includegraphics[width=\linewidth]{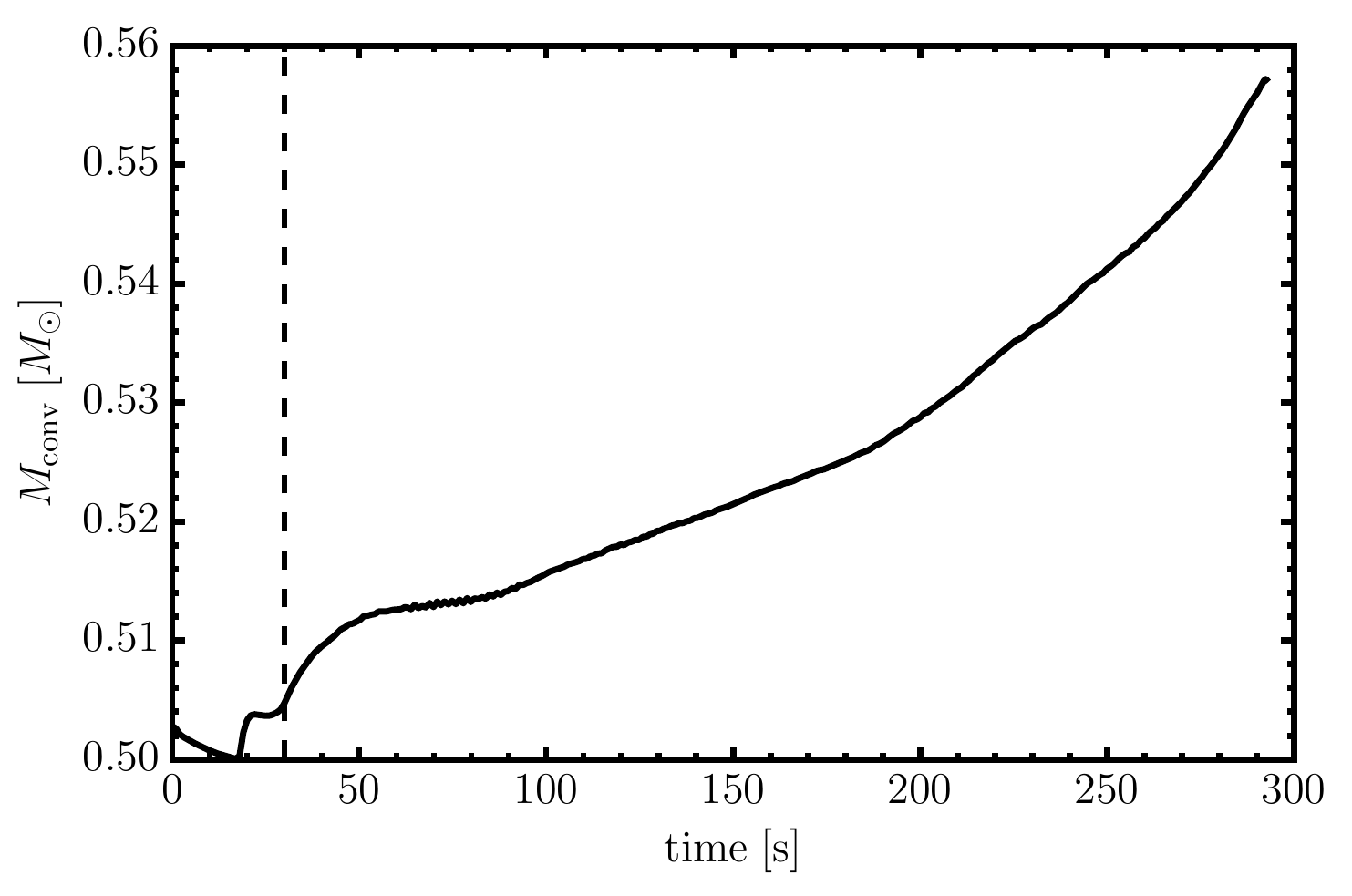}
  \caption{Mass $M_\mathrm{conv}$ contained in the convective oxygen shell in
the 3D simulation as a function of time. The mass increases by about
$0.05 M_\odot$ due to entrainment, which accelerates slightly
towards the end of the simulation as a result of higher
convective velocities and Mach numbers.
Note that the small changes in the first $\mathord{\sim} 30 \, \mathrm{s}$
are simply due to the advection of the entropy discontinuities
over the grid in the wake of hydrostatic adjustment, as a result
of which cells can jump around the threshold entropies
of
$3.6 k_b / \mathrm{nucleon}$ and $5.2 k_b /
\mathrm{nucleon}$ that we use to define the shell boundaries.
``Physical'' entrainment begins once the first convective plumes
reach the boundary between the carbon and oxygen shell
around $\mathord{\sim} 30 \, \mathrm{s}$ (denoted by a vertical
line).
    \label{fig:mconv}}
\end{figure}

\begin{figure}
  \includegraphics[width=\linewidth]{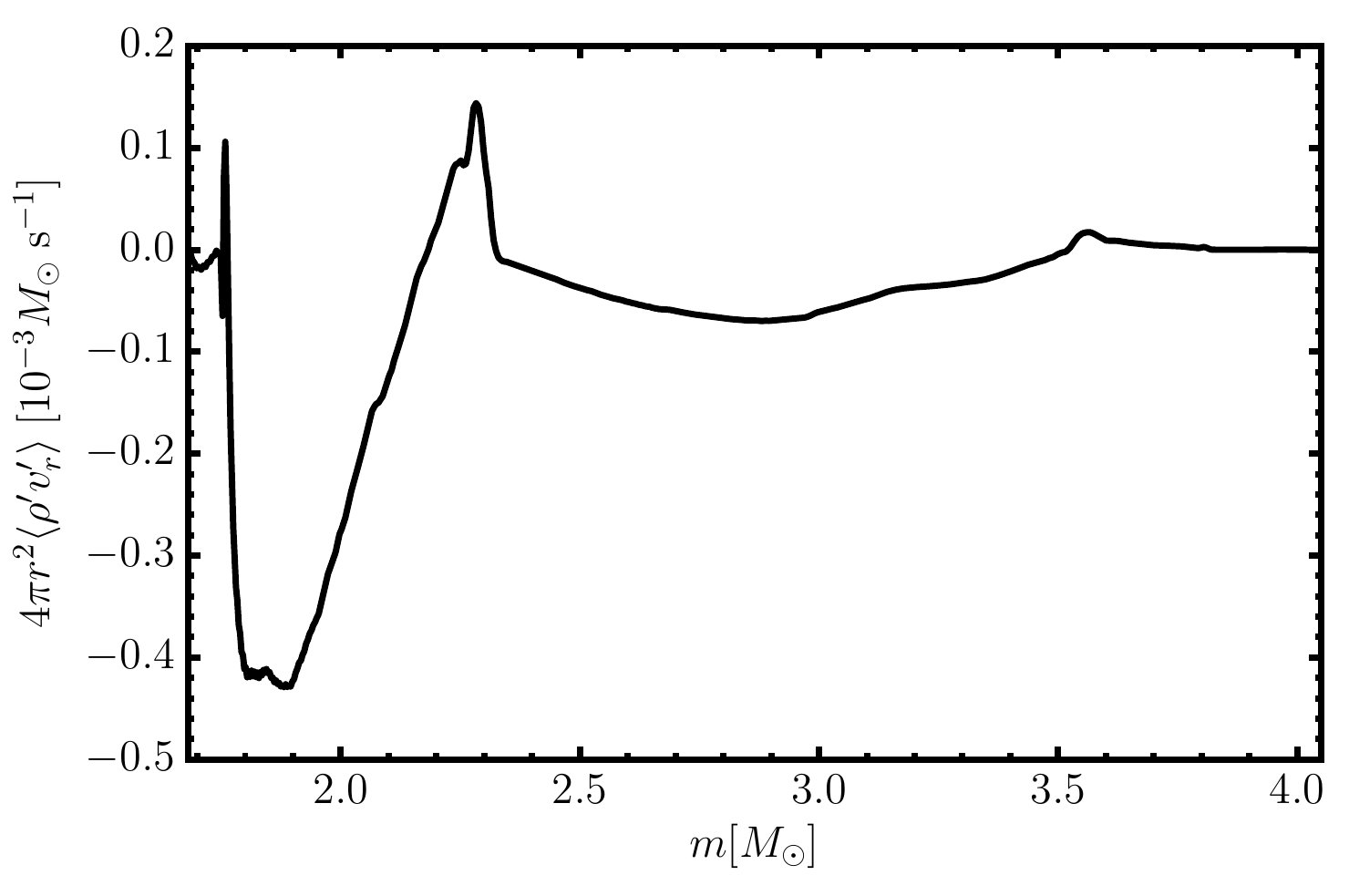}
  \caption{Turbulent mass flux
$4 \pi r^2 \langle \rho' v'_r \rangle$ in the 3D model
at a time of $210 \, \mathrm{s}$ as a function of enclosed
mass $m$. Positive values around the outer boundary
of the oxygen shell  at $m \approx 2.3 M_\odot$ indicate entrainment of
material from the carbon shell. The peak value
of $1.4 \times 10^{-4} M_\odot \, \mathrm{s}^{-1}$ roughly
corresponds to the average entrainment rate
over the course of the simulation.    \label{fig:turb_mflx}}
\end{figure}

\begin{figure}
  \includegraphics[width=\linewidth]{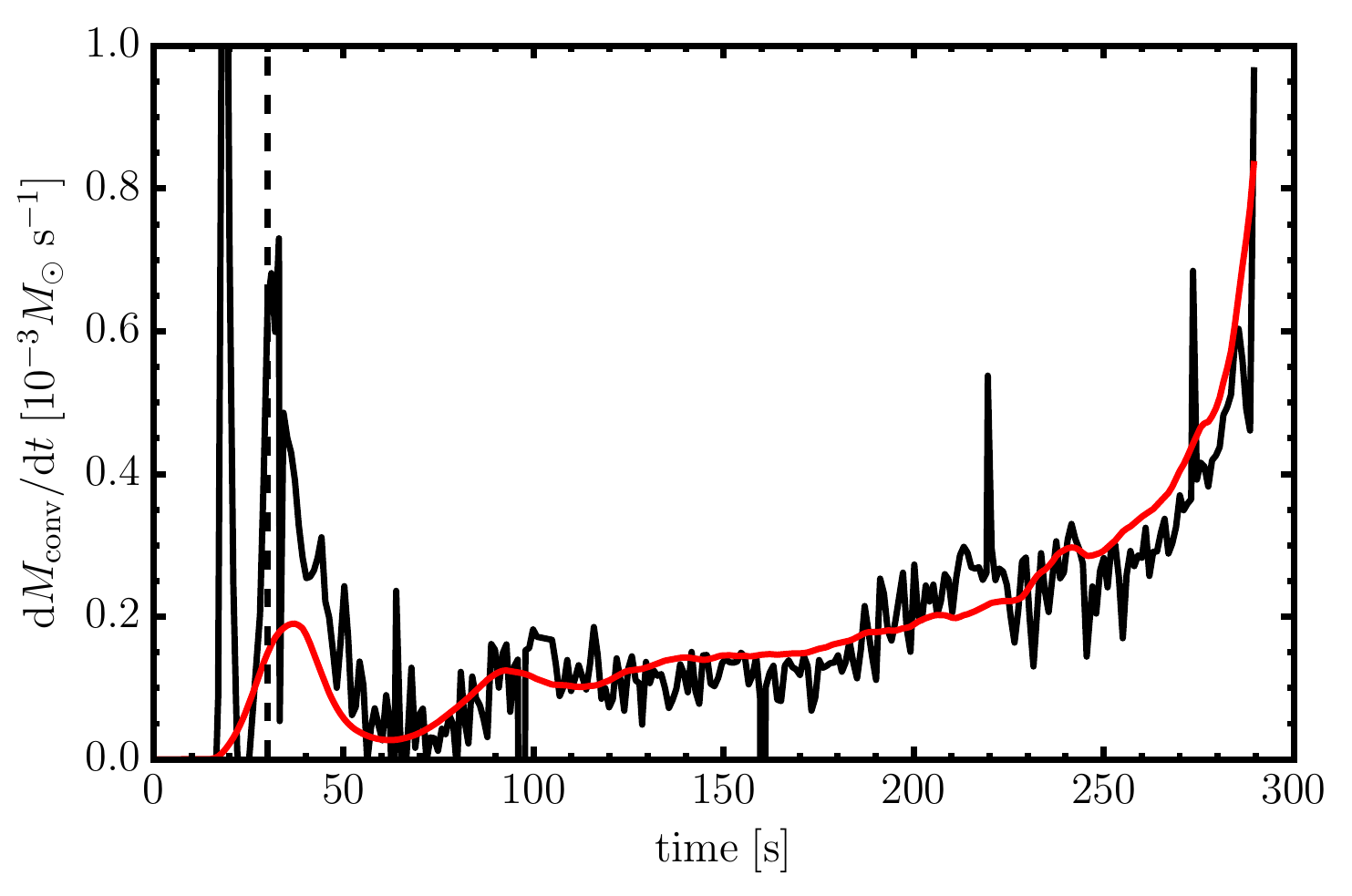}
  \caption{Comparison of the measured entrainment rate
    $\dot{M}_\mathrm{conv}$ in the 3D simulation (black) and a fit
    based on Equation~(\ref{eq:entrainment}) (red) computed using
    $A=0.37$ and a global average for the convective velocity (see
    text for details). Overall,  the time-dependent entrainment
    rate nicely follows Equation~(\ref{eq:entrainment}). Note
    that no data is shown later than $290 \, \mathrm{s}$, as the
    detection of the outer boundary of the oxygen shell becomes problematic
    due to increasingly violent boundary mixing shortly before collapse.
 As in Figure~\ref{fig:mconv},
the dashed vertical line indicates the time when convective
plumes first encounter the outer boundary and physical
entrainment begins.
    \label{fig:dotmconv}}
\end{figure}

\subsection{Comparison of Convective Mixing in 1D and 3D}
\label{sec:mixing}
Although the properties of the velocity field are more
directly relevant for the potential effect of progenitor asphericities
on supernova shock revival, some remarks about convective
mixing in our 3D model are still in order.

In Figure~\ref{fig:comp_mix}, we compare spherically averaged profiles
of the entropy $s$, and the mass fractions of oxygen, silicon, and
sulphur ($X_\mathrm{O}$, $X_\mathrm{Si}$, and $X_\mathrm{S}$) from the
3D model to the \textsc{Kepler} run at a time of $210 \,
\mathrm{ms}$. Although the treatment of convective mixing as a
diffusive process in 1D has sometimes been criticized
\citep{arnett_09}, the differences in the interior of the oxygen shell
remain minute; the most conspicuous among them are the somewhat
steeper gradients in the mass fractions in \textsc{Kepler}.  These
could potentially contribute (on a very modest level) to the lower
total nuclear energy generation rate in \textsc{Kepler}, since the
nuclear energy generation rate is roughly proportional to the square
of the mass fraction $X_\mathrm{O}$ of oxygen in the burning
region. Even if we account for spatial fluctuations in the composition
by computing $\langle X_\mathrm{O}^2 \rangle$, the compositional
differences do no appear to be sufficiently large to explain the
different burning rates; temperature changes due to hydrostatic
adjustment thus seem to be the major cause of the somewhat higher
total nuclear energy generation rate in \textsc{Prometheus}.

It is unclear whether the
composition gradients are really an artifact of MLT; we find it
equally plausible that they simply stem from the choices of different
coefficients $\alpha_2$ and $\alpha_3$ for energy transport and
compositional mixing in Equation~(\ref{eq:mlt2}) and (\ref{eq:mlt3}).
The introduction of an additional factor of $1/3$ in
Equation~(\ref{eq:mlt3}) is typically justified by interpreting
turbulent mixing as a random walk process of convective
blobs with a
mean free path $\Lambda_\mathrm{mix}$ and a \emph{total} velocity
$v_\mathrm{conv}$ with random orientation,
which translates into a \emph{radial} correlation length
$\Lambda_\mathrm{mix}/\sqrt{3}$ and an
RMS-averaged radial velocity of
$\langle v_r ^2\rangle^{1/2}=v_\mathrm{conv}/\sqrt{3}$. However,
the mixing length and MLT velocity are implicitly
identified with
the radial correlation length and
$\langle v_r ^2\rangle^{1/2}$ in Equation~(\ref{eq:mlt2}) already,
so that the choice $\alpha_3=\alpha_2$
rather than $\alpha_3=\alpha_2/3$ is arguably more appropriate.
With such a (more parsimonious) choice of parameters, the
composition gradients would be flattened considerably.

Figure~\ref{fig:comp_mix} also shows evidence of boundary mixing
(entrainment; \citealp{fernando_91,strang_01,meakin_07}) that is not
captured in the \textsc{Kepler} run. The fact that the entropy and
composition gradients are smeared out at the boundaries (especially at
the outer boundary) is mostly due to the aspherical deformation of the
shell interface by Kelvin-Helmholtz/Holmb\"oe waves; the shell
boundary remains relatively well defined in the multi-D snapshots in
Figures~\ref{fig:snap2d_a} and \ref{fig:snap2d_b}. However, the oxygen
shell is clearly expanding in $m$ at the outer boundary. To capture
the increase of the total mass $M_\mathrm{conv}$ in the convective
oxygen shell, we integrate the mass in all zones with entropies
between $3.6 k_b / \mathrm{nucleon}$ and $5.2 k_b / \mathrm{nucleon}$
(Figure~\ref{fig:mconv}).  $M_\mathrm{conv}$ increases by about $0.05
M_\odot$ over the course of the simulation with some evidence for
higher $\dot{M}_\mathrm{conv}$ towards the end, corresponding to an
entrainment rate of $1.4 \times 10^{-4} M_\odot$, which is also 
roughly the maximum value of the turbulent mass flux $4 \pi r^2
\langle \rho' v_r'\rangle$ that is reached in the formally stable
region around the outer boundary (Figure~\ref{fig:turb_mflx}).

Higher resolution is ultimately required to decide whether this
entrainment rate is physical or partially due to numerical
diffusion, which could lead to an overestimation of the amount
of entrained mass in wave breaking events (see Appendix~\ref{app:res}) . Our simulations
are, however, consistent with semi-empirical entrainment
laws found in the literature. Laboratory experiments
and simulations
\citep{fernando_91,strang_01,meakin_07}
suggest
\begin{equation}
  \label{eq:entrainment}
  \dot{M}_\mathrm{conv}
  =4 \pi r^2 \rho v_\mathrm{conv} A \, \mathrm{Ri}_\mathrm{B}^{-1},
\end{equation}
for the entrainment rate in the relevant regime of
the bulk Richardson number $\mathrm{Ri}_\mathrm{B}$ and a
dimensionless proportionality constant $A$.
$\mathrm{Ri}_\mathrm{B}$ is defined in terms of the density
contrast $\delta \rho /\rho$ at the interface,
the gravitational acceleration $g$, the typical
convective velocity $v_\mathrm{conv}$, and the eddy scale
$\Lambda$ as
\begin{equation}
  \mathrm{Ri}_\mathrm{B}=
  \frac{\delta \rho}{\rho} \frac{g \Lambda}{v_\mathrm{conv}^2}.
\end{equation}
If we identify $\Lambda$ with the pressure scale height, this
amounts to
\begin{equation}
  \label{eq:rib}
  \mathrm{Ri}_\mathrm{B}
  =
  \frac{\delta \rho}{\rho} \frac{P}{\rho v_\mathrm{conv}^2}.
\end{equation}

In our case, we have $\delta \rho/\rho =0.1$, and with
$v_\mathrm{conv}= 2.5 \times 10^7 \, \mathrm{cm} \, \mathrm{s}^{-1}$
(corresponding to the non-radial velocities near the boundary, which
are relevant for the dynamics of interfacial
Holmb\"oe/Kelvin-Helmholtz waves), we obtain $\mathrm{Ri}_\mathrm{B}
=17$, indicating a very soft boundary.
Together with an average convective velocity
of $\mathord{\sim} 200 \, \mathrm{km} \, \mathrm{s}^{-1}$
and an average entrainment rate of $1.4 \times 10^{-4} M_\odot$, this
points to a low $A \sim
0.1$ in the entrainment law~(\ref{eq:entrainment}), although the
ambiguities inherent in the definition of $\mathrm{Ri}_\mathrm{B}$ can
easily shift this by an order of magnitude, which may account for the
higher value $A\approx 1$ obtained by \citet{meakin_07}.
It is obvious that the calibration of the entrainment
law is fraught with ambiguities:
If we calibrate Equation~(\ref{eq:entrainment}) by using
a global average for $v_\mathrm{conv}$,
\begin{equation}
  v_\mathrm{conv}=\sqrt{\frac{2 E_{\mathrm{kin},r}}{M_\mathrm{conv}}},
\end{equation}
and the initial values for $\delta\rho /\rho$,
and the density $\rho$ at
the outer boundary radius $r$ in
(\ref{eq:entrainment}) and (\ref{eq:rib}),
the time-dependent entrainment rate is well fitted
by $A=0.37$ (Figure~\ref{fig:dotmconv}). If anything, relatively low values of $A$ merely demonstrate that entrainment in our
3D model is no more affected by numerical diffusion than
in comparable simulations. Considering the low value
of the bulk Richardson number and the small
entropy jump of $\sim 0.5 k_b/\mathrm{nucleon}$, which should be
conducive to entrainment effects, the dynamical impact of
boundary mixing in our simulation is remarkably small, but
its long-term effect warrants further investigation.

\section{Requirements for 3D Pre-Supernova Simulations}
\label{sec:requirements}
If 3D simulations of shell burning in massive stars are to be used as
input for core-collapse simulations, it is essential that the typical
convective velocities and eddy scales are captured accurately. The
analysis of our model in the preceding section provides guidelines about the
approximations that can (or cannot) be justified in such simulations.

The emergence of large-scale motions ($\ell=2$ modes) during the final
phase of our model implies that pre-SN model \emph{generally need to
  cover the full solid angle}  (which has been done previously
for oxygen shell burning only by \citealt{kuhlen_03}, albeit for
an earlier phase). However, for sufficiently narrow
convective shells, simulations restricted to a wedge or octant may
still cover the flow geometry accurately notwithstanding that such
symmetry assumptions remain questionable in the ensuing SN
phase. Thus, for the pre-SN phase, the assumption of octant
symmetry in \citet{couch_15} may be adequate for their model of
silicon shell burning, which has $r_+/r_- \approx 2$ towards the end
of the simulation. The eddies should then remain of a moderate scale
with a preferred $\ell$ of $\ell \approx \pi/2
(r_++r_-)/(r_+-r_-)=4.71$.

An accurate treatment of nuclear burning is even more critical because
of the scaling of convective velocities with
$(\dot{Q}_\mathrm{nuc}/M_\mathrm{conv})^{1/3}$. Since the nuclear
generation rates in the silicon and oxygen shell are sensitive to the
contraction of the deleptonizing iron core, this not only applies to
the burning shell in question itself, but also to the treatment
applied for the iron core. If the contraction of the core is
artificially accelerated as in \citet{couch_15}, this considerably
reduces the nuclear time-scale in the outer shells as well. For example
$\mathord{\sim}0.2 M_\odot$ of intermediate mass-elements in the
silicon shell are burned to iron group elements within $160
\, \mathrm{s}$ in the 3D model of \citet{couch_15}, i.e.\, silicon
burning on average proceeds $6.25$ times faster than in the
corresponding stellar evolution model, where this takes $1000
\, \mathrm{s}$. This suggests an artificial increase of the convective
velocities by $84 \%$ in their 3D model.

Approximations that affect the nuclear burning time-scale are also
problematic because they change the ratio
$\tau_\mathrm{conv}/\tau_\mathrm{nuc}\propto
\tau_\mathrm{nuc}^{-2/3}$, which plays a crucial role in the
freeze-out of convective motions shortly before the onset of collapse
(see Section~\ref{sec:freezeout}). If the nuclear burning is
artificially accelerated and continues until collapse, then the
freeze-out will occur somewhat earlier, which may compensate the
overestimation of convective velocities discussed before. However,
the simulation of \citet{couch_15} suggests that the opposite
may also occur: In their 3D model, silicon burning slows
down towards the end of their simulation as the shell almost
runs out of fuel. In the corresponding 1D stellar evolution
model, convection in the original silicon shell has already died
down completely as can be seen from their Figure~2, which
shows non-zero convective velocities only in regions with $Y_e=0.5$. While it is conceivable that convection
subsides more gradually in 3D as the available fuel is
nearly consumed -- probably over a few turnover time-scales --
increasing the ratio $\tau_\mathrm{conv}/\tau_\mathrm{nuc}$
by more than a factor of $\gtrsim 3$ evidently introduces
the risk of artificially \emph{prolonging} convective
activity in almost fully burned shells.

Other worries about the feasibility of multi-D simulations of
supernova progenitors include the problem of thermal adjustment after
mapping from a 1D stellar evolution model as well as artificial
boundary mixing. We have largely circumvented the problem of thermal
adjustment in this study by focusing on the final stages. The somewhat
higher nuclear burning rate in the 3D model (by up to $\mathord{\sim}
50\%$ compared to \textsc{Kepler}), which may be due to physical
multi-D effects or transients after the mapping such as an adjustment
to a new hydrostatic equilibrium, suggests that even for a setup where
the problem of hydrostatic and thermal adjustment is rather benign, we
still face uncertainties of the order of $15\%$ -- because of
$v_\mathrm{conv} \propto
(\dot{Q}_\mathrm{nuc}/\mathrm{M}_\mathrm{conv})^{1/3}$ -- in the final
convective velocity field at collapse. The slight expansion of the
outer boundary of the oxygen shell, which may be the result of an
adjustment effect or driven by (physical) entrainment, also deserves
attention because it plays some role in fostering the emergence of an
$\ell=2$ mode right before collapse. It appears less worrisome,
however, since there are natural variations in shell geometry anyway,
and since the emergence of the $\ell=2$ mode may still be primarily
driven by the contraction of inner shell boundary. There is no
evidence for artificial boundary mixing at this stage, although
further high-resolution tests remain desirable.

\section{Effect of Convective Seed Perturbations on Supernova Shock Revival}
\label{sec:theory}

With typical convective Mach numbers of $\sim 0.1$
and a dominant $\ell=2$ mode at collapse, the progenitor
asphericities fall in the regime where they may be able
to affect shock revival in the ensuing core-collapse supernova,
as has been established by the parameter study of \citet{mueller_15a}.
Considering that several recent works have shown that
the conditions for shock revival in multi-D can be captured
with good accuracy by surprisingly simple scaling laws
\citep{mueller_15a,summa_16,janka_16,mueller_16}
that generalize the concept of the critical luminosity
\citep{burrows_93} to multi-D, it is reasonable to ask
whether the effect of progenitor asphericities
can also be predicted more quantitatively by
simple analytic arguments. Given the
good agreement between our 3D model and MLT, such
a theory could help to better identify progenitors
for which convective seed asphericities play a major
role in the explosion before investing considerable
computer time into multi-D simulations.

The key ingredient to accomplish this consists in a first quantitative
theory for the interaction of asymmetries in the supersonic infall
region with the shock, which \citet{mueller_15a} only described
qualitatively as ``forced shock deformation''. The starting
point is the translation of initial radial velocity
perturbations into density perturbations at the shock
due to differential infall \citep{mueller_15a},
\begin{equation}
  \delta \rho_\mathrm{pre}/\rho_\mathrm{pre}
  \approx \mathrm{Ma},
\end{equation}
which can also be understood more rigorously using linear perturbation
theory \citep{goldreich_97,takahashi_14}.  Note that we now designate
the typical convective Mach number during convective shell burning
simply as $\mathrm{Ma}$ to avoid cluttered notation. The perturbations
in the transverse velocity components are amplified
as $r^{-1}$ \citep{goldreich_97} and are roughly given by
\begin{equation}
\delta v_t \approx \mathrm{Ma}\, c_\mathrm{s,ini} (r_\mathrm{ini}/r_\mathrm{sh}),
\end{equation}
where $c_\mathrm{s,ini}$ and $r_\mathrm{ini}$ are the initial sound
speed and radius of the shell before collapse and $r_\mathrm{sh}$
is the shock radius. Radial velocity perturbations only grow
with $r^{-1/2}$ \citep{goldreich_97} and can therefore be neglected.

\subsection{Generation of Turbulent Kinetic Energy by Infalling Perturbations}
The interaction of the pre-shock perturbations with the shock can then
be interpreted as an injection of additional turbulent kinetic energy
into the post-shock region. While this problem has not yet been
addressed in the context of spherical accretion onto a neutron star,
the interaction of planar shocks with incident velocity and density
perturbations has received some attention in fluid dynamics
\citep{ribner_87,andreopoulos_00,wouchuk_09,huete_10,huete_12}.  The
perturbative techniques that allow a relatively rigorous treatment in
the planar case cannot be replicated here, and we confine ourselves to
simple rule-of-thumb estimate for the generation of turbulent energy
by the infalling perturbations: If we neglect the deformation of the
shock initially, we can assume transverse velocity perturbations
$\delta v_\mathrm{t}$ and density fluctuations $\delta \rho /\rho$
compared to the spherically averaged flow are conserved across the
shock as a first-order approximation. The anisotropy of the ram
pressure will also induce pressure fluctuations $\delta P/P \sim
\delta \rho /\rho$ downstream of the shock. In a more self-consistent
solution, these pressure fluctuations would induce lateral flows and
modify the shape of the shock, and larger vorticity perturbations
would arise if the shock is asymmetric to begin with (which is
important in the context of the SASI
\citep{foglizzo_07,guilet_12}. As a crude first-order estimate such a
rough estimate is sufficient for our purpose; it is not incompatible
with recent results about shocks traveling in inhomogeneous
media \citep{huete_10,huete_12}.

From the density and pressure perturbations $\delta \rho/\rho \sim \delta P/P \mathord{\sim} \mathrm{Ma}$
and transverse velocity perturbations
$\delta v_t \sim \mathrm{Ma} \, c_\mathrm{s,ini} (r_\mathrm{ini}/r_\mathrm{sh})$
downstream of the shock, we can estimate fluxes of
transverse kinetic energy ($F_\mathrm{t}$),
acoustic energy ($F_\mathrm{ac}$), and
an injection rate of kinetic energy due to the work
done by buoyancy during the advection of the accreted
material through down to the gain radius $r_\mathrm{g}$.
$F_\mathrm{t}$ is roughly given by,
\begin{eqnarray}
F_\mathrm{t}
&=&
\frac{\dot{M}}{2} \delta v_\mathrm{t}^2
=
\frac{\dot{M}}{2} \mathrm{Ma}^2 c_\mathrm{s,ini}^2 \left(\frac{r_\mathrm{ini}}{r_\mathrm{sh}}\right)^2
\\
\nonumber
&\approx&
\mathrm{Ma}^2 \frac{GM \dot{M}}{6r_\mathrm{ini}} \left(\frac{2r_\mathrm{ini}^2}{3r_\mathrm{g} r_\mathrm{sh}}\right)
=
\mathrm{Ma}^2 \frac{GM \dot{M}}{9r_\mathrm{g}} \left(\frac{r_\mathrm{ini}}{r_\mathrm{sh}}\right),
\end{eqnarray}
where we approximated the initial sound speed as $c_\mathrm{s,ini}^2 \approx GM/(3r_\mathrm{ini})$,
which is a good approximation for the shells outside the iron core.
Note that we use a typical ratio $r_\mathrm{sh}/r_\mathrm{g}=3/2$
during the pre-explosion phase to express $F_\mathrm{t}$ in terms of the gravitational
potential of the gain radius; the reason for this will become apparent when
we compare the injection rate of turbulent kinetic energy at the shock
to the contribution from neutrino heating.

Following \citet{landau_fluid}, the acoustic energy flux can be
estimated by assuming that the velocity fluctuations in acoustic waves
are roughly $\delta v \sim c_s \delta P/P$ (where $c_s$ is the sound
speed behind the shock and $\delta P\approx  \mathrm{Ma}\, P$).
The post-shock pressure $P$ can be determined from
the jump conditions,
\begin{equation}
P=\rho_\mathrm{pre} \frac{\beta-1}{\beta} v_\mathrm{pre}^2
=\rho \frac{\beta-1}{\beta^2}\frac{GM }{r_\mathrm{sh}},
\end{equation}
where $\rho_\mathrm{pre}$ and $\rho$ are the pre- and
post-shock density, $\beta \approx 7$ is the compression
ratio in the shock, and $v_\mathrm{pre}$
is the pre-shock velocity, which we approximate as $v_\mathrm{pre}
=\sqrt{GM/r_\mathrm{sh}}$. The acoustic energy flux
is thus,
\begin{eqnarray}
\nonumber
F_\mathrm{ac}
&=&4 \pi r_\mathrm{sh}^2\delta P\, \delta v
=4 \pi r_\mathrm{sh}^2 \frac{\delta P^2 c_s}{P}
=4 \pi \rho  |v_r| r_\mathrm{sh}^2 \frac{\delta P^2 c_s}{\rho P |v_r|}
\\
\nonumber
&=&
4 \pi \rho  |v_r| r_\mathrm{sh}^2 \frac{\mathrm{Ma}^2 c_s P}{\rho |v_r|}
= \dot{M}\, \mathrm{Ma}^2 \frac{\beta-1}{\beta^2} \frac{GM}{r_\mathrm{sh}}
\frac{\beta}{\sqrt{3}}
\\
&\approx& 0.49\mathrm{Ma}^2 \frac{G M \dot{M}}{r_\mathrm{g}}.
\end{eqnarray}
Here, $|v_r|=v_\mathrm{pre}/\beta$ is the spherical average of the
post-shock velocity, and $c_s^2\approx
GM/(3 r_\mathrm{sh})$ has been used following \citet{mueller_15a}.

Finally, the gravitational potential energy corresponding
to density fluctuations $\delta \rho$ will be converted
into kinetic energy by buoyancy forces at a rate of\footnote{If
we assume that the density perturbations in the post-shock region
adjust on a dynamical time-scale as pressure equilibrium between over- and underdensities
is established, then this estimate might be lower, but pressure
adjustment itself would involve the generation of lateral
flows and hence generate turbulent kinetic energy, so that our estimate
is probably not too far off.}
\begin{eqnarray}
\nonumber
F_\mathrm{pot}
&=&
  \frac{\dot{M }\delta \rho}{\rho}
  \left(\frac{GM}{r_\mathrm{sh}}-\frac{GM}{r_\mathrm{g}}\right)
  =
  \mathrm{Ma}\, \dot{M}  \left(\frac{GM}{r_\mathrm{g}}-\frac{GM}{r_\mathrm{sh}}\right)
\\
  &\approx&
 \mathrm{Ma} \frac{G M \dot{M}}{3r_\mathrm{g}}.
\end{eqnarray}
Especially for moderate Mach numbers,
$F_\mathrm{pot}$ is clearly the dominating term, as the
flux of acoustic and transverse kinetic energy scale with $\mathrm{Ma}^2$.

In the absence of infalling perturbations,
\citet{mueller_15a} established
a semi-empirical scaling law that
relates transverse kinetic energy $E_\mathrm{kin,t}$ stored in the post-shock
region to the volume-integrated neutrino heating rate $\dot{Q}_\nu$,
the mass in gain region, $M_\mathrm{g}$, and the shock and gain radius,
\begin{equation}
\label{eq:etrans}
\frac{E_\mathrm{kin,t}}{M_\mathrm{g}}
\approx
\frac{1}{2}
\left[\frac{(r_\mathrm{sh}-r_\mathrm{g}) \dot{Q}_\nu}{M_\mathrm{g}}\right]^{2/3}.
\end{equation}
At least for convection-dominated models, this scaling law can be understood
as the result of a balance between kinetic energy generation by buoyancy
and turbulent dissipation with a dissipation length $\Lambda=r_\mathrm{sh}-r_\mathrm{g}$
(cf.\, also \citealt{murphy_12}).
Assuming a local dissipation rate of $v^3/\Lambda=(2E_\mathrm{kin,t}/M_\mathrm{g})^{3/2}/\Lambda$,
this leads to
\begin{equation}
\label{eq:balance}
\dot{E}_{\mathrm{kin,t}}
=
\dot{Q}_\nu-\frac{1}{\Lambda}\left(\frac{2E_\mathrm{kin,t}}{M_\mathrm{g}}
\right)^{3/2} M_\mathrm{g}=0,
\end{equation}
from which  Equation~(\ref{eq:etrans}) immediately follows.

In the presence of infalling perturbations, it is natural
to add another source term to Equation~(\ref{eq:balance}),
\begin{equation}
\dot{Q}_\nu
+F_\mathrm{pot}
-\frac{1}{\Lambda}\left(\frac{2E_\mathrm{kin,t}}{M_\mathrm{g}}
\right)^{3/2} M_\mathrm{g}=0.
\end{equation}
To keep the calculation tractable, we only include
the dominant contribution  $F_\mathrm{pot}$ arising from infalling perturbations and discard $F_\mathrm{ac}$ and $F_\mathrm{t}$.

However, this obviously poses the question about the
appropriate choice for $\Lambda$, which can now no longer assumed
to be simply given by $r_\mathrm{sh}-r_\mathrm{g}$. To get
some guidance, we can consider the limit in which neutrino heating
is negligible; here the appropriate choice for $\Lambda$
is clearly given by the scale of the infalling perturbations, i.e.\
$\Lambda \approx \pi r_\mathrm{sh}/\ell$ in
terms of their typical angular wave number $\ell$.
Hence we find,
\begin{equation}
\label{eq:etrans_no_nu}
\frac{E_\mathrm{kin,t}}{M_\mathrm{g}}
=
\frac{1}{2}
\left[\frac{\pi r_\mathrm{sh} F_\mathrm{pot}}{\ell M_\mathrm{g}}\right]^{2/3},
\end{equation}
in this limit. The general case can be accommodated
by simply interpolating between the two limits,
\begin{equation}
\label{eq:etrans_new}
\frac{E_\mathrm{kin,t}}{M_\mathrm{g}}
=
\frac{1}{2}
\left[
\frac{(r_\mathrm{sh}-r_\mathrm{g}) \dot{Q}_\nu}{M_\mathrm{g}}
+
\frac{\pi r_\mathrm{sh} F_\mathrm{pot}}{\ell M_\mathrm{g}}\right]^{2/3}.
\end{equation}
We emphasize that a different dissipation length enters in both terms:
In the limit of neutrino-driven convection with small seed perturbations,
the dissipation length is given by the width of the gain layer, whereas
the dissipation length $\pi r_\mathrm{sh}/\ell$ can be considerably larger
for ``forced'' convection/shock deformation due to infalling perturbations
with small $\ell$.

For deriving the modification of the critical luminosity,
it will be convenient to express
$E_\mathrm{kin,t}$ in terms of its value
in the limit of small seed perturbations
(Equation~\ref{eq:etrans}) and a correction term
$\psi$,
\begin{equation}
\frac{E_\mathrm{kin,t}}{M_\mathrm{g}}
=
\frac{1}{2}
\left[
\frac{(r_\mathrm{sh}-r_\mathrm{g}) \dot{Q}_\nu}{M_\mathrm{g}}
\right]^{2/3}
(1+\psi)^{2/3},
\end{equation}
where $\psi$ is defined as
\begin{equation}
\psi=
\frac{\pi r_\mathrm{sh} F_\mathrm{pot}/(\ell M_\mathrm{g})}
{(r_\mathrm{sh}-r_\mathrm{g}) \dot{Q}_\nu/M_\mathrm{g}}
=
\frac{\pi r_\mathrm{sh} F_\mathrm{pot}}{\ell (r_\mathrm{sh}-r_g) \dot{Q}_\nu}.
\end{equation}

Different from the case of negligible seed
perturbations, it is hard to validate Equation~(\ref{eq:etrans_new})
in simulations. In the 2D study of \citet{mueller_15a},
the amplitudes of the infalling perturbations change significantly
over relatively short time-scales, and the phase during
which they have a significant impact on the turbulent kinetic
energy in the post-shock region but have not yet triggered
shock revival was therefore too short to detect
any deviations from Equation~(\ref{eq:etrans}), especially
since the turbulent kinetic energy fluctuates considerably
around its saturation value in 2D.

\subsection{Effect on the Heating Conditions
and the Critical Luminosity}
Conceptually, the steps from Equation~(\ref{eq:etrans_new})
to a modified critical luminosity are no different from
the original idea of \citet{mueller_15a}, i.e.\, one
can assume that the average shock radius can be obtained
by rescaling the shock radius $r_\mathrm{sh,1D}$ for the stationary 1D
accretion problem
with a correction factor that depends
on the average RMS Mach number $\langle \mathrm{Ma}_\mathrm{gain}^2\rangle$
in the gain region (Equation~42 in \citealt{mueller_15a}),
\begin{equation}
r_\mathrm{sh}
\approx
r_\mathrm{sh,1D}
\left(1+\frac{4}{3} \langle \mathrm{Ma}_\mathrm{gain}^2\rangle \right)^{2/3},
\end{equation}
which then leads to a similar correction factor for the critical values for
the neutrino luminosity and mean energy $L_\nu$ and $E_\nu$
(Equation 41 in \citealt{mueller_15a}).

In the presence of strong seed perturbations, we can express
$\langle \mathrm{Ma}_\mathrm{gain}^2 \rangle$ at
the onset of an explosive runaway in terms
of its value  $\langle \mathrm{Ma}_\nu^2\rangle$ at shock revival
in the case of small seed perturbations
and a correction factor $(1+\psi)^{2/3}$
as in Equation~(\ref{eq:etrans_new}),
\begin{equation}
\label{eq:lcrit}
L_\nu E_\nu^2
\propto (\dot M M)^{3/5}
r_\mathrm{g}^{-2/5}
\left[1+\frac{4}{3} \langle \mathrm{Ma}_\nu^2\rangle
(1+\psi)^{2/3} \right]^{-3/5}.
\end{equation}

Equation~(\ref{eq:lcrit}) obviously hinges on the
proper calibration (and validation) of Equation~(\ref{eq:etrans_new}),
which needs to be provided by future core-collapse supernova
simulations. Nonetheless, it already allows some crude estimates.

The ratio of the critical luminosity with strong
seed perturbation $(L_\nu E_\nu^2)_\mathrm{pert}$
to the critical luminosity
$(L_\nu E_\nu^2)_\mathrm{3D}$ value in multi-D for small seed perturbations
is found to be
\begin{eqnarray}
\nonumber
\frac{(L_\nu E_\nu^2)_\mathrm{pert}}
{(L_\nu E_\nu^2)_\mathrm{3D}}
&=&
\left(\frac{1+4/3 \langle\mathrm{Ma}_\nu^2\rangle (1+\psi)^{2/3}}{1+4/3
\langle\mathrm{Ma}_\nu^2\rangle}\right)^{-3/5}
\\
&\approx&
1-\frac{8 \langle \mathrm{Ma}_\nu^2 \rangle \psi}{15 \big(1+4/3
\langle\mathrm{Ma}_\nu^2\rangle \big)},
\end{eqnarray}
where we linearized in $\psi$.
In order not to rely on an increasingly long chain of uncertain
estimates, it is advisable to use the known
multi-D effects without strong seed perturbations
as a yardstick; they bring about a reduction
of the critical luminosity by about $25 \%$ compared
to 1D \citep{murphy_08b,hanke_12,couch_12b,dolence_13,mueller_15a}.
This reduction is obtained by setting
$\mathrm{Ma}_\nu^2=0.4649$ at the onset
of runaway shock expansion, which is also the value
derived by \citet{mueller_15a}
based on analytic arguments.

Using this value, we
estimate a reduction of the critical luminosity by $\mathord{\sim}0.15 \psi$
relative to the the critical luminosity in multi-D without perturbations,
which remains only a very rough indicator for the importance
of perturbations in shock revival barring any further
calibration and a precise definition of how and where
$\mathrm{Ma}$ is to be measured.

It is illustrative to express $\psi$ in terms of
the heating efficiency $\eta_\mathrm{heat}$, which
is defined as the ratio of the volume-integrated
neutrino heating rate and the
sum of the electron neutrino and antineutrino luminosities
$L_{\nu_e}$ and $L_{\bar{\nu}_e}$,
\begin{equation}
\eta_\mathrm{heat}=\frac{\dot{Q}_\nu}{L_{\nu_e}+L_{\bar{\nu}_e}},
\end{equation}
and the accretion efficiency $\eta_\mathrm{acc}$,
\begin{equation}
\eta_\mathrm{acc}=\frac{L_{\nu_e}+L_{\bar{\nu}_e}}{GM \dot{M}/r_\mathrm{g}}.
\end{equation}
We then obtain
\begin{equation}
\label{eq:psi}
\psi=
\frac{\pi r_\mathrm{sh} \mathrm{Ma}}{3\ell (r_\mathrm{sh}-r_\mathrm{g})
\eta_\mathrm{acc} \eta_\mathrm{heat}}
\approx
\frac{\pi \mathrm{Ma}}{\ell \eta_\mathrm{acc} \eta_\mathrm{heat}}.
\end{equation}

Using Equation~(\ref{eq:psi}), we can verify that the estimated
reduction of the critical luminosity due to seed perturbations by
$\mathord{\sim} 0.15 \psi$ is in the ballpark: If we take
$\mathrm{Ma}$ to be half the maximum value of the Mach number in the
infalling shells in the models of \citet{mueller_15a} and work with
reasonable average values of $\eta_\mathrm{acc}=2$ and
$\eta_\mathrm{heat}=0.05$, we obtain a reduction of 11\% for their
model p2La0.25 ($\mathrm{Ma}=0.045$), 24\% for p2La1
($\mathrm{Ma}=0.1$), and 36\% for p2La2 ($\mathrm{Ma}=0.15$), which
agrees surprisingly well with their inferred reduction of the critical
luminosity (Figure~12 in their paper). It also explains why their
models with $\ell=4$ require twice the convective Mach numbers in the
oxygen shell to explode at the same time as their corresponding
$\ell=2$ models. For the models of \citet{couch_13,couch_14} with
$\eta_\mathrm{heat} \approx 0.1$ and $\ell=4$, our estimate would
suggest a reduction in critical luminosity by 6\%.  This prediction
  cannot be compared quantitatively to the results of
  \citet{couch_13,couch_14} since an analysis of
  the effect on the critical luminosity in the
  vein of \citet{mueller_15a} and \citet{summa_16}
  would require additional data (e.g., trajectories
  of the gain radius). Qualitatively, such a moderate
  reduction of the critical luminosity seems consistent with their results:
  The effect of infalling perturbations corresponds to a change
  in the critical heating factor\footnote{The change in the critical
    heating factor is related to but not necessarily identical
    to the change in the generalized critical luminosity as
    introduced by \citet{mueller_15a} and \citet{summa_16}, which also depends, e.g., on the relative change of the gain radius and the specific binding
  energy in the gain region.} (by which they multiply the
  critical luminosity to compute the neutrino heating terms)
  by only $2 \ldots 3\%$, and their inferred reduction
  of the ``critical heating efficiency'' by $\mathord{\sim} 10\%$ due to
  infalling perturbations is much smaller than the reduction
  of the critical heating efficiency by a factor of $\mathord{\sim} 2$ in 3D compared to 1D. 
 For the
simulations of \citet{couch_15}, for which we estimate the convective
Mach number in the silicon shell as roughly $0.02$, the expected
reduction in the critical luminosity (again for $\eta_\mathrm{heat}
\approx 0.1$ and a dominant $\ell=4$ mode) is roughly $1\%$ ,
which is consistent with the development of an explosion in both the
perturbed and the unperturbed model.

For our $18 M_\odot$ progenitor model with a typical convective Mach
number $\mathrm{Ma} \approx 0.1$ in the middle of the oxygen shell, we
expect a much more sizable reduction of the critical luminosity by
$12\ldots 24\%$ if we assume $\eta_\mathrm{acc}=2$, $\ell=2$ and
$\eta_\mathrm{heat} =0.05 \ldots 0.1$, although this crude
estimate still needs to be borne out by a follow-up core-collapse
supernova simulation. Because the importance
of the infalling perturbations relative to the
contribution of neutrino heating to non-radial instabilities
is determined by
$\eta_\mathrm{acc}$ and $\eta_\mathrm{heat}$,
reasonably accurate multi-group transport is obviously required;
the inaccuracy of leakage-based models
like \citet{couch_13} and \citet{couch_15} that has been
pointed out by  \citet{janka_16} evidently does not
permit anything more than a proof of principle.

\section{Summary and Conclusions}
\label{sec:summary}
In this paper, we presented the first 3D simulation of the last
minutes of oxygen shell burning outside
a contracting iron and silicon core in a massive star (ZAMS mass $18
M_\odot$) up to the onset of collapse.  Our simulation was conducted using a
19-species $\alpha$-network as in the stellar evolution code
\textsc{Kepler} \citep{weaver_78} and an axis-free, overset Yin-Yang
grid \citep{kageyama_04,wongwathanarat_10a} to cover the full solid
angle and allow for the emergence of large-scale flow patterns. To
circumvent the problem of core deleptonization and nuclear
quasi-equilibrium in the silicon shell without degrading the accuracy
of the simulation by serious modifications of the core evolution, a
large part of the silicon core was excised and replaced by a
contracting inner boundary with a trajectory determined from the
corresponding \textsc{Kepler} run.  The model was evolved over almost
5~minutes, leaving ample time for transients to die down and roughly
3~minutes or 9~turnover time-scalse of steady-state convection for a
sufficiently trustworthy analysis of the final phase before collapse.

For the simulated progenitor, an $18 M_\odot$ star of
solar metalicity with an extended oxygen shell, our 3D simulation
shows the acceleration of convection from typical Mach numbers of
$\mathord{\sim} 0.05$ to $\mathord{\sim} 0.1$ at collapse due to the
increasing burning rate at the base of the shell. The contraction of
the core also leads to the emergence of larger scales in the flow,
which is initially dominated by $\ell=3$ and $\ell=4$ modes before a
pronounced quadrupolar ($\ell=2$) mode develops shortly before
collapse. As a result of a small buoyancy jump between the oxygen and
carbon shell, the oxygen shell grows from $0.51 M_\odot$ to $0.56
M_\odot$  due to the entrainment of material from the
overlying carbon shell over the course of the simulation, which
appears compatible with empirical scaling laws for the entrainment
rate at convective boundaries \citep{fernando_91,strang_01,meakin_07}.

The comparison with the corresponding \textsc{Kepler} model shows that
-- aside from entrainment at the boundaries -- convection is well
described by mixing length theory (MLT) in the final stage before
  collapse in the model studied here.  MLT at least captures the bulk
  properties of the convective flow that matter for the subsequent
  collapse phase quite accurately:  If properly ``gauged'', the
convective velocities predicted by MLT in \textsc{Kepler} agree well
with the 3D simulation, and the time-dependent implementation of MLT
even does a reasonable job right before collapse when the nuclear
energy generation rate changes significantly within a turnover
time-scale, which results in a ``freeze-out'' of convection.  The good
agreement with MLT is also reflected by the fact that the kinetic
energy in convective motions obeys a scaling law of the expected form
\citep{biermann_32,arnett_09}. The kinetic energy $E_{\mathrm{kin},r}$
in radial convective motions can be described to good accuracy in
terms of the average nuclear energy generation rate per unit mass
$\dot{q}_\mathrm{nuc}$, the pressure scale height $h_P$ at the base of
the shell, and the mass $M_\mathrm{conv}$ in the shell as,
\begin{equation}
\label{eq:ekinr_scaling}
E_{\mathrm{kin},r}
\approx 0.35  M_\mathrm{conv} (\dot{q}_\mathrm{nuc} h_P)^{2/3},
\end{equation}
and the convective velocities are not too far
from
\begin{equation}
\delta v_r \approx
\omega_\mathrm{BV} h_P,
\end{equation}
where $\omega_\mathrm{BV}$ is the Brunt-V\"ais\"al\"a frequency and
$h_P$ is the \emph{local} value of the pressure scale height. Our
results are consistent with the assumption that convective blobs are
accelerated only over roughly one pressure scale height, and there
appears to be no need to replace the pressure scale height
in Equation~(\ref{eq:ekinr_scaling}) with the extent of the convective
zone as the arguments of \citet{arnett_09} suggest.  We surmise that
this may be a specific feature of the final phases of shell convection
before collapse that requires the dissipation of turbulent energy
within a limited distance: Since neutrino cooling no longer balances
nuclear energy generation, the convective flow will adjust such as to
maintain a constant rate of entropy generation throughout the shell to
avoid a secular build-up or decline of the unstable gradient.  During
earlier phases with appreciable neutrino cooling in the outer regions
of convective shells, Equation~(\ref{eq:ekinr_scaling}) may no longer
be adequate.

Similarly, the dominant scale of the convective eddies agrees well
with estimates based on linear perturbation theory
\citep{chandrasekhar_61,foglizzo_06}.
In terms of the radii $r_-$ and $r_+$ of the inner and outer
shell boundary, the dominant angular wave number $\ell$ is roughly
\begin{equation}
\label{eq:typical_l}
\ell
\approx \frac{\pi (r_+ + r_-)}{2 (r_+ - r_-)}.
\end{equation}

Our findings already allow some conclusions about one of the primary
questions that has driven the quest for 3D supernova progenitors,
i.e.\, whether progenitor asphericities can play a beneficial role for
shock revival.  We suggest that Equations~(\ref{eq:ekinr_scaling}) and
(\ref{eq:typical_l}) can be used to formulate an estimate for the
importance of convective seed perturbations for shock revival in the
ensuing supernova \citep{couch_13,mueller_15a,couch_15}.  To this end,
these two equations (or alternatively, the convective velocities
obtained via MLT in a stellar evolution code) need to be evaluated at
the time of freeze-out of convection to obtain the typical convective
Mach number $\mathrm{Ma}$ and angular wave number $\ell$ at collapse.
The time of freeze-out can be determined by equating the typical
time-scale for changes in the volume-integrated burning rate
$\dot{Q}_\mathrm{nuc}$ with the turnover time-scale
$t_\mathrm{conv}$, which results in
the condition
\begin{equation}
\frac{\ud \ln \dot{Q}_\mathrm{nuc}}{\ud t}=\frac{\omega_\mathrm{BV,max}}{2\pi},
\end{equation}
or,
\begin{equation}
\frac{\ud \ln \dot{Q}_\mathrm{nuc}}{\ud t}=t_\mathrm{conv}^{-1}.
\end{equation}
Relying on an estimate for the extra turbulent energy generated in the
post-shock region in the supernova core by the infall of seed
perturbations, and using the reduction of the
energy-weighted critical neutrino
luminosity $\mathcal{L}_\mathrm{crit}$ for explosion by $\mathord{\sim} 25 \%$ in multi-D
\citep{murphy_08b,hanke_12,mueller_15a} as a yardstick,
one finds that strong seed perturbations
should reduce $\mathcal{L}_\mathrm{crit}$ further by
\begin{equation}
\label{eq:dlcrit}
\frac{\Delta \mathcal{L}_\mathrm{crit}}{\mathcal{L}_\mathrm{crit}}
\approx
0.47 \frac{\mathrm{Ma}}{\ell \eta_\mathrm{acc} \eta_\mathrm{heat}},
\end{equation}
relative to the control value in multi-D simulations without
strong seed perturbations.
Here $\eta_\mathrm{acc}$ and $\eta_\mathrm{heat}$ are the accretion
and heating efficiency in the supernova core,
 $\mathrm{Ma}$ is the typical convective Mach number
in the infalling shell at the onset of collapse,
and $\mathcal{L}_\mathrm{crit}=(L_\nu E_\nu^2)_\mathrm{crit}$
includes the proper weighting of the neutrino luminosity
$L_\nu$ with the square of the neutrino mean energy $E_\nu^2$
\citep{janka_12,mueller_15a}. This estimate appears
to be roughly in line with recent multi-D studies of shock revival
with the help of strong seed perturbations and nicely accounts for
the range of effect sizes from \citet{couch_15} (no qualitative change in
shock revival) to \citet{mueller_15a} (reduction of $\mathcal{L}_\mathrm{crit}$
by tens of percent for $\ell=2$ modes with sufficiently strong
perturbations). For our 3D progenitor model, we expect
a reduction of the critical luminosity by $12 \ldots 24 \%$.

Considering these numbers, the prospects for a significant and
supportive role of progenitor asphericities in the supernova explosion
mechanism seem auspicious.  Yet caution is still in order. Because the
relative importance of seed perturbations is determined by the ratio
$\mathrm{Ma}/(\eta_\mathrm{acc} \eta_\mathrm{heat})$, a reliable
judgment needs to be based both on a self-consistent treatment of
convective burning in multi-D before collapse (which determines
$\mathrm{Ma}$) \emph{and} accurate multi-group neutrino transport
after bounce (which determines $\eta_\mathrm{acc}$ and
$\eta_\mathrm{heat}$).  Again, first-principle models of supernovae
face a curious coincidence: As one typically finds $\mathrm{Ma} \sim
\eta_\mathrm{heat}$, progenitor asphericities are just large enough to
play a significant role in the explosion mechanism, but not large
enough to provide a clear-cut solution for the problem of shock revival.
The danger of a simplified neutrino treatment has already been
emphasized repeatedly in the literature (see \citealt{janka_16} for a
recent summary), but pitfalls also abound in simulations of convective
burning: For example, the recipes employed by \citet{couch_15} can be shown to
considerably affect the convective Mach number at collapse by
appealing to scaling laws from MLT and time-scale considerations.

Our method of excising the core seems to be a viable avenue towards
obtaining 3D initial conditions in the oxygen shell (which is the
innermost active convective shell in many progenitor models) without
introducing inordinate artifacts due to initial transients or
artificial changes to the nuclear burning. Nonetheless, the model
presented here is only another step towards a better understanding of
the multi-D structure of supernova progenitors.  In particular, the
  effects of resolution and stochasticity on the convective flow need
  to be studied in greater depth, though a first
restricted resolution study (Appendix~\ref{app:res}) suggests
that the predicted convective velocities are already accurate
to within $10 \%$ or less. 

Future simulations will also  need to address progenitor variations in the
shell geometry, shell configuration, and the burning rate; in fact the
$18 M_\odot$ was deliberately chosen as an optimistic case with strong
oxygen burning at the base of a very extended convective shell, and
may not be representative of the generic situation (if there is
any). Moreover, massive stars with active convective silicon shells at
collapse also need to be explored even if they form just a subclass of
all supernova progenitors. Treating this phase adequately to avoid the
artifacts introduced by an approach like that of \citet{couch_15} is
bound to prove a harder challenge due to the complications of nuclear
quasi-equilibrium.  Finally, the long-term effects of entrainment and
other phenomena that cannot be captured by MLT need to be examined:
If such effects play a major role in the evolution of supernova
progenitors, capturing them with the help of exploratory 3D models and
improved recipes for 1D stellar evolution in the spirit of the 321D
approach \citep{arnett_2015} will be much more challenging than 3D
simulations of the immediate pre-collapse stage, where the problems of
extreme time-scale ratios (e.g.\ of the thermal adjustment and
turnover time-scale), numerical diffusion, and energy/entropy
conservation errors are relatively benign.  It is by no means certain
that supernova progenitor models will look fundamentally different
once this is accomplished; but there is little doubt
that groundbreaking discoveries will be made along the way.

\acknowledgements We thank T.~Foglizzo, E.~M\"uller, and S.~Woosley
for useful discussions and T.~Melson for support and discussions
concerning the Yin-Yang grid.  We acknowledge support by the
Australian Research Council through a Discovery Early Career
Researcher Award DE150101145 (BM) and an ARC Future Fellowship
FT120100363 (AH), by the Deutsche Forschungsgemeinschaft through the
Excellence Cluster Universe EXC 153 (TJ) and by the European Research
Council through grant ERC-AdG No.~341157-COCO2CASA (MV, TJ).  This
research was undertaken with the assistance of resources from the
National Computational Infrastructure (NCI), which is supported by the
Australian Government and was supported by resources provided by the
Pawsey Supercomputing Centre with funding from the Australian
Government and the Government of Western Australia.  This material is
based upon work supported by the National Science Foundation under
Grant No.~PHY-1430152 (JINA Center for the Evolution of the Elements).

\bibliography{paper}

\appendix

\section{Ledoux Criterion in Terms of Entropy and Composition
Gradients}
\label{app:ledoux}
The Ledoux and Schwarzschild criteria for convective instability
are often expressed in terms of entropy and composition gradients
in the stellar evolution literature. It is straightforward
to show that Equation~(\ref{eq:mlt1}) can equally be expressed
in terms of these gradients:
\begin{eqnarray}
\label{eq:mlt_s_yi}
\frac{\delta \rho}{\rho}
&=&
\Lambda_\mathrm{mix}
\left(
\frac{1}{\rho} \frac{\pd  \rho}{\pd r}
-\frac{1}{\rho c_s^2}\frac{\pd P}{\pd r}\right)
=
\frac{\Lambda_\mathrm{mix}}{\rho}
\left[
 \frac{\pd  \rho}{\pd r}
-\left(\frac{\pd \rho}{\pd P}\right)_{s,Y_i}
\frac{\pd P}{\pd r}\right]
\\
\nonumber
&=&
\frac{\Lambda_\mathrm{mix}}{\rho}
\left[
\left(\frac{\pd \rho}{\pd s}\right)_{P,Y_i}
 \frac{\pd  s}{\pd r}
+\sum_i \left(\frac{\pd \rho}{\pd Y_i}\right)_{P,s,Y_{j,j \neq i}}
 \frac{\pd  Y_i}{\pd r}
\right.\\
\nonumber
&&
\left.
+\left(\frac{\pd \rho}{\pd P}\right)_{s,Y_i}
\frac{\pd P}{\pd r}
-\left(\frac{\pd \rho}{\pd P}\right)_{s,Y_i}
\frac{\pd P}{\pd r}\right]
\\
\nonumber
&=&
\frac{\Lambda_\mathrm{mix}}{\rho}
\left[
\left(\frac{\pd \rho}{\pd s}\right)_{P,Y_i}
 \frac{\pd  s}{\pd r}
+\sum_i \left(\frac{\pd \rho}{\pd Y_i}\right)_{P,s,Y_{j,j \neq i}}
 \frac{\pd  Y_i}{\pd r}\right].
\end{eqnarray}

\begin{figure}
  \centering
\includegraphics[width=0.6\linewidth]{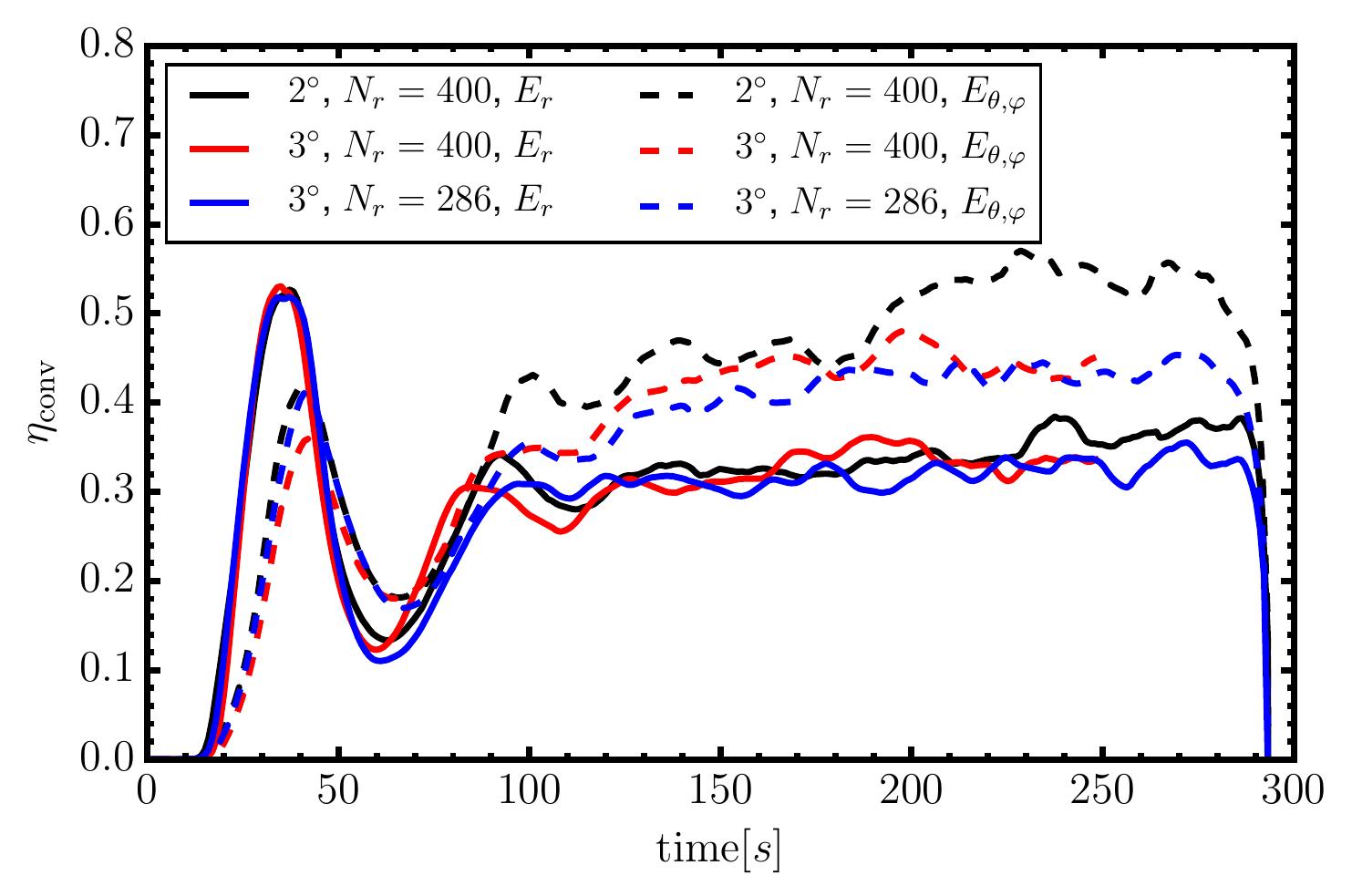}
\caption{Efficiency factors $\eta_\mathrm{conv}$  for the conversion
of the total nuclear energy generation rate
into turbulent energy in radial motions ($E_r$, solid lines)
and non-radial motions ($E_\theta,\varphi$, dashed lines)
for
the baseline run with an angular resolution of $3^\circ$
degrees and  $N_r=400$ radial zones (black curves) and for two
low-resolution runs with an angular resolution
of $2^\circ$ and a radial resolution of 400
zones  (red) and 286 zones (blue). 
\label{fig:qconv_res}
}
\end{figure}

\begin{figure}
  \centering
\includegraphics[width=0.6\linewidth]{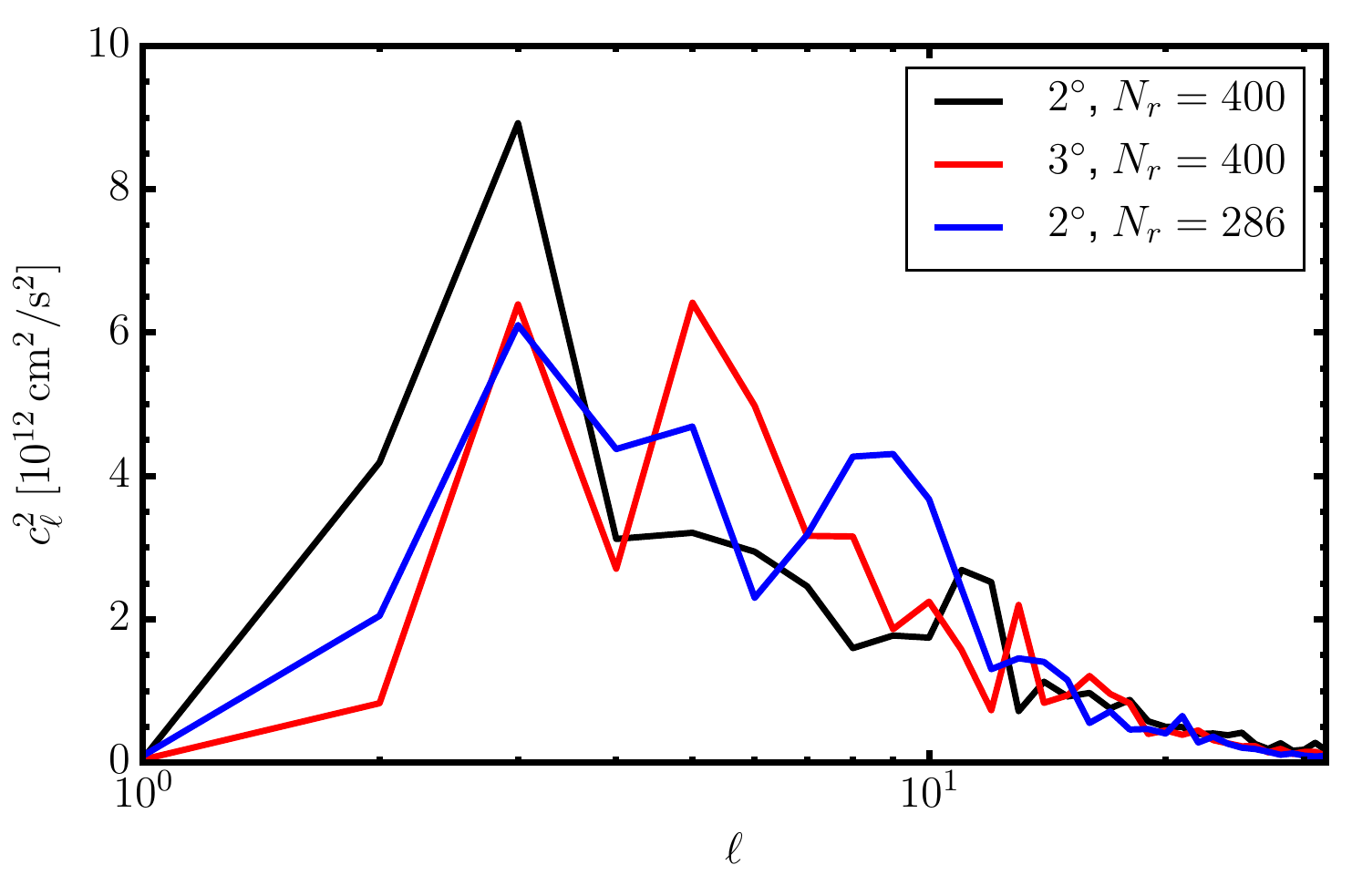}
\caption{Power $c_\ell^2$ in different multipoles $\ell$
    for the decomposition of the radial velocity
    at $r=4000 \, \mathrm{km}$ into spherical
harmonics $Y_{\ell m}$ at a time of $210 \, \mathrm{s}$
for
the baseline run with an angular resolution of $3^\circ$
degrees and  $N_r=400$ radial zones (black curve) and for two
low-resolution runs with an angular resolution
of $2^\circ$ and a radial resolution of 400
zones  (red) and 286 zones (blue).
Although the dominant angular wavenumbers are similar,
the baseline run shows more power in $\ell=2$ and $\ell=3$
modes. 
\label{fig:mpole_res}
}
\end{figure}

\begin{figure}
  \centering
\includegraphics[width=0.6\linewidth]{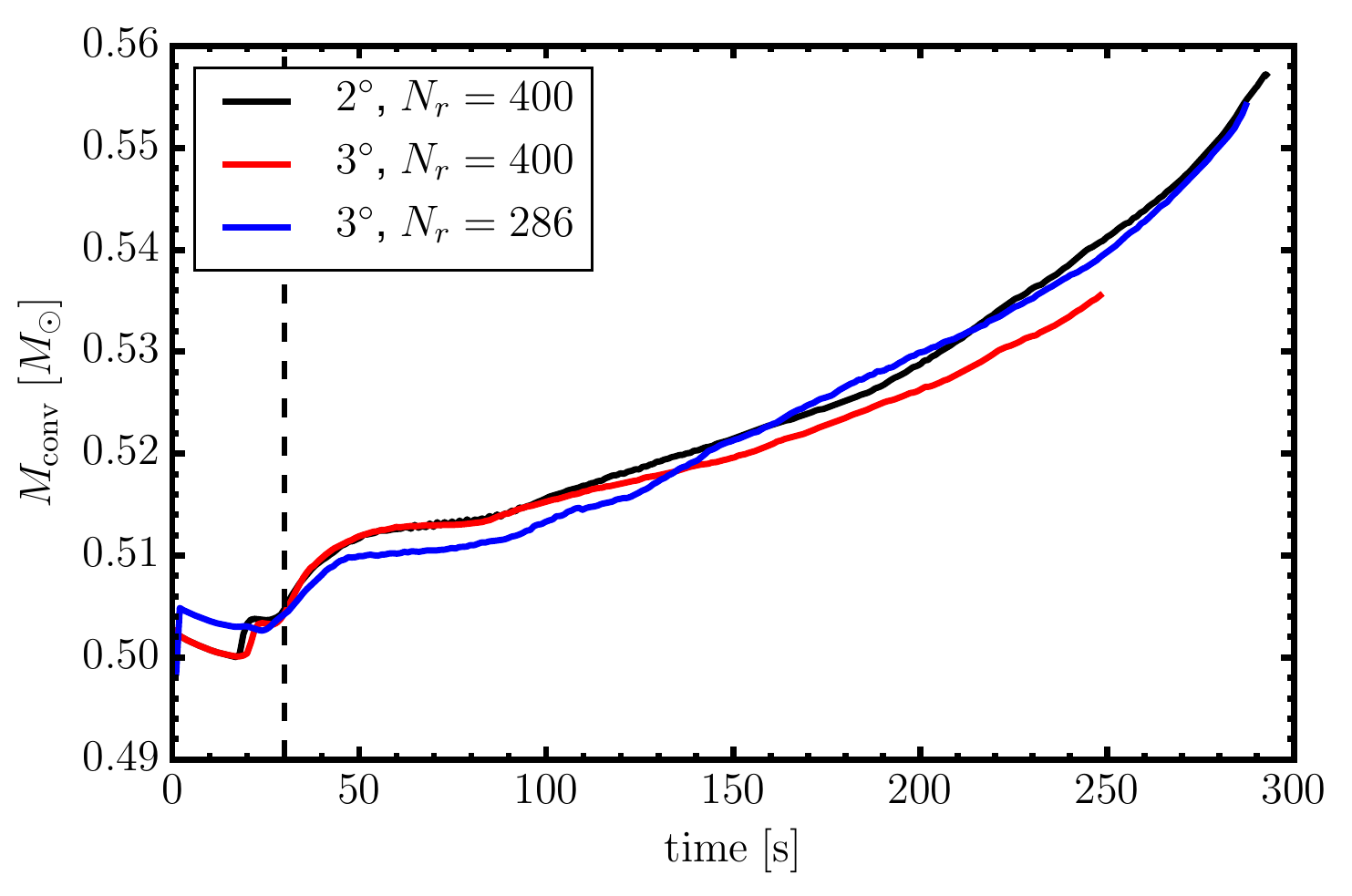}
\caption{Growth of the mass $M_\mathrm{conv}$ of the
oxygen shell as a function of time for
the baseline run with an angular resolution of $3^\circ$
degrees and  $N_r=400$ radial zones (black curve) and for two
low-resolution runs with an angular resolution
of $2^\circ$ and a radial resolution of 400
zones  (red) and 286 zones (blue).
Note that $M_\mathrm{conv}$ starts at a slightly
different value in the run with $N_r=286$ because the initial
model is mapped to a different grid than for $N_r=400$. 
\label{fig:mconv_res}
}
\end{figure}

\section{Effects of Resolution and Stochasticity}
\label{app:res}
Limited resolution is always a cause of concern
for turbulent flow as in convective shell burning. Current models of
shell burning during the final stages like \citet{couch_15} and ours
are limited to $\mathord{\sim} 1.5$~million zones to cover the region
of interest, and a significant increase in resolution is currently
not affordable. To gauge potential resolution effects, we have, however,
computed two additional runs with lower resolution, and compare
key quantities of those models to the baseline run. We both
consider the case of a reduced angular resolution of $3^\circ$
with the same radial grid as well as a simulation with
a resolution of $3^\circ$ and a coarser radial grid
of 286 zones (with $\Delta r/r =1\%$). 

The two complementary low-resolution simulations also help to
address another important effect to a limited extent, i.e.\
stochastic variations in the convective flow geometry. With
just three different models computed at different resolutions,
it is obviously impossible to distinguish cleanly between
the effects of resolution and stochasticity. The three
models merely define a vague ``band of uncertainty'',
and only permit limited conclusions about the underlying
cause for the variations between the models.

While all three models exhibit very similar nuclear energy generation
rates, the dynamics of the convective flow is slightly
different. Figure~\ref{fig:qconv} compares the efficiency factor
$\eta_\mathrm{conv}$ (Equation~\ref{eq:eta_conv}) for the conversion of nuclear energy into the
components $E_r$ and $E_{\theta,\varphi}$ of the turbulent kinetic
energy (which is essentially the same as comparing $E_r$ and
$E_{\theta,\varphi}$ directly because of the extremely similar burning
rate). $\eta_\mathrm{conv}$ does not differ substantially for the two
low-resolution runs.  Variations between the different simulations are
typically on the level of $\mathord{10} \%$ or less, i.e.\ of the same
order as the stochastic fluctuations within each run, and there is
thus no clear evidence for a strong resolution dependence. This in
line with findings from a different context (convection in
core-collapse supernova explosions), where \citet{handy_14} and
\citet{radice_16} found that the global energetics of the
flow is well captured even with a modest resolution of $2\ldots
3^\circ$ in angle and $\mathord{\sim}100$ radial zones across the
convective region.

The non-radial kinetic energies in the low-resolution runs differ more
strongly from the baseline run, especially after $190 \,
\mathrm{s}$. It is not clear, however, whether this is a resolution
effect or due to stochastic variations. The higher non-radial kinetic
energy in the baseline run appears to be connected to a slightly
different eddy geometry, which could suggest stochasticity rather than
differences in resolution as the culprit for the differences between
the runs. Figure~\ref{fig:mpole_res} shows the coefficients
$c_{\ell}^2$ for the decomposition of the radial velocity into
spherical harmonics (computed according to
Equation~\ref{eq:multipoles}) at a time of $210 \, \mathrm{s}$ and
demonstrates a slightly stronger preponderance of low-$\ell$ modes in
the baseline run compared to the low-resolution models. It is thus
conceivable that the lateral flows from the updrafts to the downdrafts at the top
and bottom of the convective shell must be slightly faster in the
baseline run merely to ensure mass conservation in a flow
that is (almost) anelastic: In the baseline run, the lateral flow must
transport mass from the updrafts to the downdrafts (and vice versa a the bottom
of the shell) at the same rate as in the low-resolution runs, but over a larger
distance; hence the ratio of non-radial to radial turbulent velocities is
larger.

Whatever the underlying cause, the lesson of our minimal resolution study
is that we should anticipate uncertainties in the convective velocities,
which scale with $\eta_\mathrm{conv}^{1/2}$,
of $\mathord{\lesssim} 10 \%$, and that the dominant angular wavenumber of
the convective eddies may also vary slightly.

Even the growth of the oxygen shell by entrainment is not too
dissimilar for the three different runs as shown by
Figure~\ref{fig:mconv_res}. Again it is unclear to what extent the
variations in entrainment are stochastic or due to resolution
effects. The resolution requirements for the problem at hand may be
mitigated by the ``softness'' of the convective boundary: Interfacial
waves develop on relatively large scales, so that the individual
breaking billows are typically covered by $\mathord{\sim} 10$ zones or
more. This is still well below the resolution of almost 50 zones in
the transition of the best-resolved global simulations of convective
boundary mixing in the context of hydrogen ingestion
\citep{herwig_14,woodward_15}, and hence more caution is in order
concerning the convergence of the entrainment rate than for the
convective velocities and eddy scales.

\end{document}